%% file: main.tex
\documentclass{article} %
\usepackage{iclr2027_conference,times}
\usepackage{fontawesome5}
\usepackage{graphicx}
\newcommand{\faHuggingface}{\raisebox{-0.15em}{\includegraphics[height=1em]{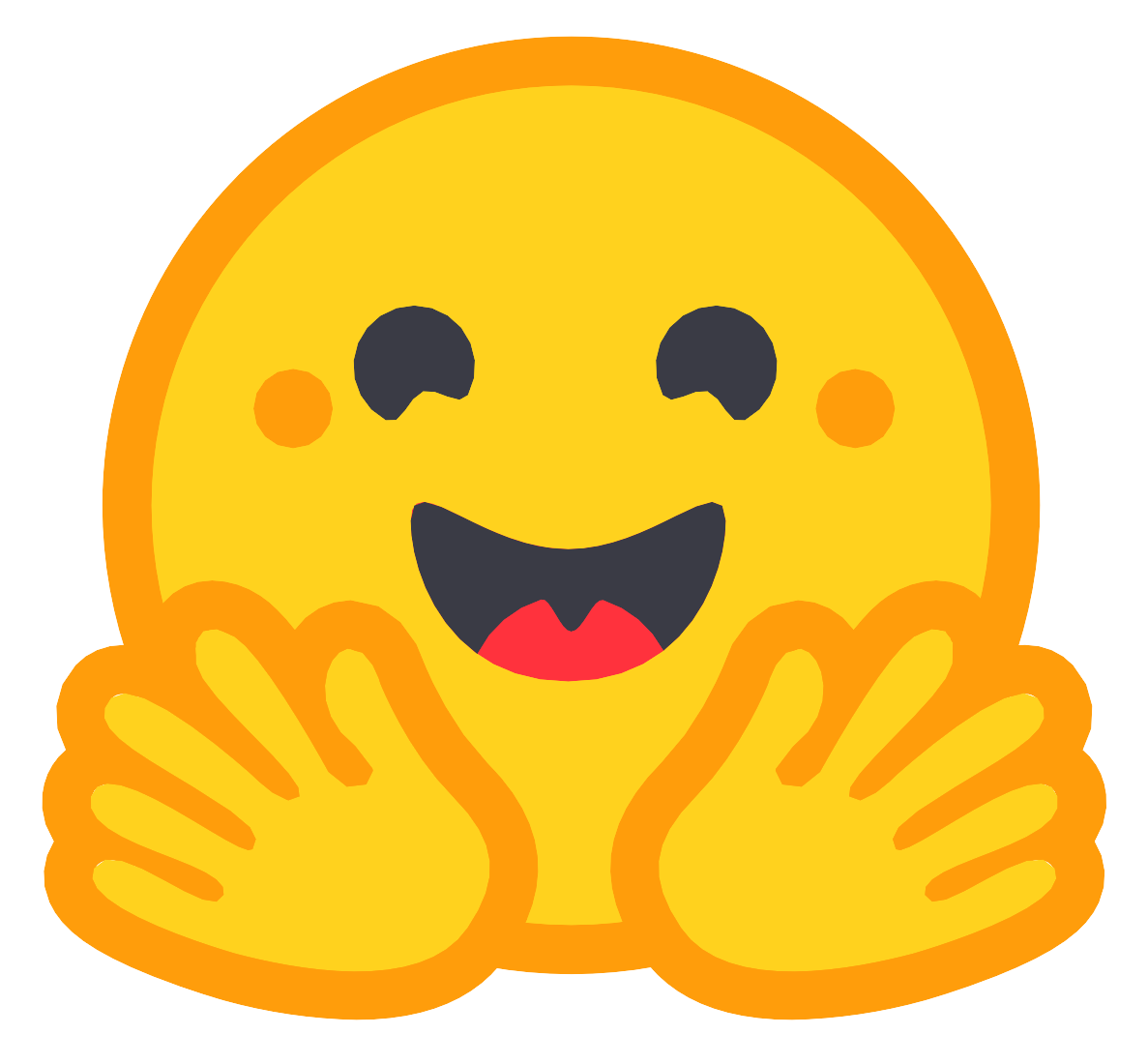}}}

\input{math_commands.tex}

\input{headers/header-lqa-head}

\input{headers/header}

\input{headers/header-lqa-tail}

\title{TestSuiteBench: \\Coding Agents Struggle to Generate \\Strong Test Suites from Specifications}

\author{
  Kazuki Egashira$^1$, Niels M\"{u}ndler-Sasahara$^1$, Mark Niklas M\"{u}ller$^2$, Martin Vechev$^1$ \\
  $^1$ ETH Zurich, $^2$ LogicStar.ai \\
  \texttt{\{kazuki.egashira, niels.muendler\}@inf.ethz.ch}
}

\iclrfinalcopy %
\begin{document}

\maketitle
\lhead{Preprint}

\vspace{-2.2em}
\begin{center}\small
  \faGlobe\enspace\href{https://testsuitebench.com}{\texttt{testsuitebench.com}}\hspace{2em}
  \faGithub\enspace\href{https://github.com/eth-sri/testsuitebench}{\texttt{eth-sri/testsuitebench}}\hspace{2em}
  \faHuggingface\enspace\href{https://huggingface.co/collections/eth-sri/testsuitebench}{\texttt{eth-sri/testsuitebench}}
\end{center}

\begin{abstract}
\input{abstract.tex}
\end{abstract}

\input{src/intro}

\input{src/overview.tex}
\input{src/construction.tex}

\input{src/experiment.tex}

\input{src/related_work}

\input{src/discussion.tex}

\input{src/conclusion.tex}

\bibliography{iclr2027_conference}
\bibliographystyle{iclr2027_conference}

\message{^^JLASTREFERENCESPAGE \thepage^^J}

\ifincludeappendixx
  \newpage
  \appendix
  \onecolumn
  \include{appendix}

\fi

\end{document}

%% file: math_commands.tex
\usepackage{amsmath,amsfonts,bm}

\def\1{\bm{1}}

\def\vr{{\bm{r}}}

\DeclareMathAlphabet{\mathsfit}{\encodingdefault}{\sfdefault}{m}{sl}
\SetMathAlphabet{\mathsfit}{bold}{\encodingdefault}{\sfdefault}{bx}{n}

%% file: headers/header-lqa-head.tex
\usepackage[utf8]{inputenc} %

\usepackage{graphicx}

\usepackage[T1]{fontenc}

\usepackage[hidelinks,colorlinks]{hyperref}       %
\usepackage{url}            %
\usepackage{booktabs}       %
\usepackage{amsfonts}       %
\usepackage{nicefrac}       %
\usepackage{microtype}      %
\usepackage{amssymb}
\usepackage{pifont}

\input{headers/lqa/overfull}

\input{headers/lqa/comments}

\input{headers/lqa/appendix}

\input{headers/lqa/abbreviations}

\input{headers/lqa/acronyms}

\input{headers/lqa/listing}

\input{headers/lqa/prompt.tex}

\newcommand{\cmark}{\ding{51}}
\newcommand{\xmark}{\ding{55}}

%% file: headers/lqa/overfull.tex
\IfFileExists{headers/config/showoverfull.config}{
	\overfullrule=1cm
}{
}

%% file: headers/lqa/comments.tex
\usepackage{marginnote}

\usepackage[backgroundcolor=none,linecolor=red,textsize=footnotesize]{todonotes}

%% file: headers/lqa/appendix.tex
\usepackage{etoolbox}

\newbool{includeappendix}
\setbool{includeappendix}{true} %
\IfFileExists{headers/config/noappendix.config}{
	\setbool{includeappendix}{false}
}{}

\newif\ifincludeappendixx
\ifbool{includeappendix}{
	\includeappendixxtrue
}{
	\includeappendixxfalse
}

\usepackage{xr} %
\usepackage{filecontents}

\ifbool{includeappendix}{}{
	\input{appendix-labels-loader}

	\externaldocument{appendix-labels}
}

%% file: headers/lqa/abbreviations.tex
\newcommand{\eg}{e.g., }
\newcommand{\ie}{i.e., }

%% file: headers/lqa/acronyms.tex
\usepackage{acro} %
\makeatletter
\@ifpackagelater{acro}{2020/06/01}{\acsetup{patch/longtable=false}}{}
\makeatother

\DeclareAcronym{cli} {
    short = CLI,
    long = Command Line Interface,
}

%% file: headers/lqa/listing.tex
\usepackage{listings}

\usepackage{textcomp}

\usepackage[scaled=0.8]{beramono}

\definecolor{ckeyword}{HTML}{7F0055}
\definecolor{ccomment}{HTML}{3F7F5F}
\definecolor{cstring}{HTML}{2A0099}

\lstdefinestyle{numbers}{
	numbers=left,
	framexleftmargin=20pt,
	numberstyle=\tiny,
	firstnumber=auto,
	numbersep=1em,
	xleftmargin=2em
}

\lstdefinestyle{layout}{
	frame=none,
	captionpos=b,
}

\lstdefinestyle{comment-style}{
	morecomment=[l]//,
	morecomment=[s]{/*}{*/},
	commentstyle={\color{ccomment}\itshape},
}

\lstdefinestyle{string-style}{
	morestring=[b]",%
	morestring=[b]',%
	stringstyle={\color{cstring}},
	showstringspaces=false,%
}

\lstdefinestyle{keyword-style}{
	keywordstyle={\ttfamily\bfseries},
	morekeywords={
		function,
		constructor,
		int,
		bool,
		return,
		returns,
		uint
	},
	morekeywords = [2]{},
	keywordstyle = [2]{\text},
	sensitive=true,
}

\lstdefinestyle{input-encoding}{
	inputencoding=utf8,
	extendedchars=true,
	literate=
	{ℝ}{$\reals$}1%
	{→}{$\rightarrow$}1%
	{α}{$\alpha$}1%
	{β}{$\beta$}1%
	{λ}{$\lambda$}1%
	{θ}{$\theta$}1%
	{ϕ}{$\phi$}1%
}

\lstdefinestyle{escaping}{
	moredelim={**[is][\color{blue}]{\%}{\%}},
	escapechar=|,
	mathescape=true
}

\lstdefinestyle{default-style}{
	basicstyle=\fontencoding{T1}\ttfamily\footnotesize,
	style=numbers,
	style=layout,
	style=comment-style,
	style=string-style,
	style=keyword-style,
	style=input-encoding,
	style=escaping,
	tabsize=2,
	upquote=true
}

\lstdefinelanguage{BASIC}{
	language=C++,
	style=default-style
}[keywords,comments,strings]%

%% file: headers/lqa/prompt.tex
\usepackage[most]{tcolorbox}

\definecolor{PromptAccent}{HTML}{51a35f}

\newcommand{\placeholder}[1]{%
  \tcbox[on line,
    colback=PromptAccent!9,
    colframe=PromptAccent!35,
    boxrule=0.35pt,
    arc=0.9mm,
    left=2.5pt,right=2.5pt,top=1pt,bottom=1pt
  ]{\ttfamily\footnotesize\{#1\}}%
}

\newtcolorbox{promptbox}[1][]{%
  enhanced,
  breakable,
  colback=white,
  colframe=black!16,
  boxrule=0.55pt,
  arc=1.6mm,
  outer arc=1.6mm,
  left=10pt,right=10pt,
  top=8pt,bottom=8pt,
  borderline west={2.2pt}{0pt}{PromptAccent},
  drop shadow={black!12},
  fontupper=\scriptsize,
  before upper={%
    \setlist[itemize]{leftmargin=1.0em}%
    \setlist[enumerate]{leftmargin=1.2em}%
  },
  fonttitle=\sffamily\bfseries\footnotesize,
  coltitle=PromptAccent!90!black,
  colbacktitle=PromptAccent!10,
  title={#1},
  attach boxed title to top left={xshift=10pt,yshift*=-2mm},
  boxed title style={
    enhanced,
    arc=1.2mm,
    boxrule=0.35pt,
    colframe=PromptAccent!35,
    colback=PromptAccent!10,
    left=6pt,right=6pt,top=2pt,bottom=2pt
  },
  before skip=10pt,
  after skip=10pt
}

\newtcblisting{schemabox}{%
  enhanced,
  breakable,
  colback=black!2,
  colframe=black!14,
  boxrule=0.4pt,
  arc=1.2mm,
  left=8pt,right=8pt,
  top=6pt,bottom=6pt,
  listing only,
  listing options={
    basicstyle=\ttfamily\footnotesize,
    breaklines=true,
    columns=fullflexible,
    keepspaces=true,
    showstringspaces=false
  }
}

%% file: headers/header.tex
\usepackage[utf8]{inputenc} %
\usepackage[T1]{fontenc}    %
\usepackage{algcompatible}
\usepackage{caption}
\usepackage{algpseudocode}
\usepackage{enumerate}
\usepackage{url}            %
\usepackage{booktabs}       %
\usepackage{amsfonts}       %
\usepackage{nicefrac}       %
\usepackage{microtype}      %
\usepackage{xcolor}         %
\usepackage{algpseudocode}
\usepackage{algorithm}
\usepackage{tikz,booktabs,multirow,enumitem,bm}
\usepackage{colortbl}
\usepackage{makecell}
\usepackage{tabularx}
\usepackage{longtable}      %
\usepackage{arydshln}
\usepackage{subcaption}
\usepackage{wrapfig}
\usepackage{tabulary}
\usepackage{amsmath,amsthm,amsfonts}
\usepackage{setspace}
\usepackage{xspace}
\usepackage{siunitx}

\newcommand{\ms}{\ensuremath{\mathcal{M}}\xspace} %
\renewcommand{\vr}{\ensuremath{\mathcal{V}}\xspace} %
\newcommand{\pr}{\ensuremath{\mathcal{R}}\xspace} %

\definecolor{hyperlinkblue}{HTML}{0000AA}
\hypersetup{allcolors=hyperlinkblue} %

\newcommand{\benchmark}{\textsc{TestSuiteBench}\xspace}
\newcommand{\programbench}{\textsc{ProgramBench}\xspace}

%% file: headers/header-lqa-tail.tex
\input{headers/lqa/references}

%% file: headers/lqa/references.tex
\usepackage[capitalize, noabbrev]{cleveref}

\crefrangeformat{section}{\S#3#1#4\crefrangeconjunction\S#5#2#6}

\crefmultiformat{section}{\S#2#1#3}{\crefpairconjunction\S#2#1#3}{\crefmiddleconjunction\S#2#1#3}{\creflastconjunction\S#2#1#3}

\newcommand{\crefrangeconjunction}{--}

\crefname{listing}{Lst.}{listings}
\crefname{line}{Lin.}{Lin.}
\crefname{appendix}{App.}{App.}

\newcommand{\appref}[1]{%
	\ifbool{includeappendix}{\cref{#1}}{the appendix}%
}
\newcommand{\Appref}[1]{%
	\ifbool{includeappendix}{\cref{#1}}{The appendix}%
}

%% file: abstract.tex
Coding agents are increasingly used to create large software projects. While code-generation benchmarks have scaled accordingly to full applications, test-generation benchmarks remain limited to individual functions or bug fixes. In particular, generating complete test suites for complex software from scratch remains unexplored.
To address this gap, we introduce TestSuiteBench, a benchmark of 165 open-source applications, each with a structured and detailed specification of intended behavior, a reference implementation, a reference test suite, and on average 74 mutants. Each mutant is generated from a single requirement of the specification and refined to isolate that behavior: 73\% are killed by exactly one reference test and, on average, by only 0.2\% of the remaining tests.
We find that even frontier coding agents struggle to generate strong test suites given our specifications, achieving mutation scores below 60\%, despite their suites being largely valid and 88\% of their tests passing on the reference implementation. We observe that generated test suites often lack relevant input combinations and are smaller than reference test suites. Meanwhile, implementations generated for the same specifications pass 82\% of the reference tests. These results show agents can build software they cannot verify, a gap that directly limits fully autonomous software development.

%% file: src/intro.tex
\section{Introduction}
\label{sec:intro}

Coding agents have rapidly progressed from resolving isolated issues \citep{sweagent} to being used across the software engineering lifecycle \citep{aitw,huang2026professionalsoftwaredevelopersdont}.
Increasingly, they re-implemented complex applications from scratch with little human involvement, including a C compiler and a web browser \citep{carliniBuildingCCompiler2026,linScalingLongRunningAutonomousCoding2026}.
Systems like these took human teams years to build, suggesting that fully autonomous development of genuinely new software is within reach.

However, such autonomous development depends on strong verification that checks the agent's work against the intended behavior \citep{carliniBuildingCCompiler2026}.
While re-implementations can rely on existing test suites and reference implementations for verification, those are not available for genuinely new software, which initially only has a natural language specification.
The agent must thus be able to verify their implementation against this specification by translating it into a strong test suite.

Existing benchmarks do not measure this capability.
Code generation benchmarks have scaled from resolving individual issues \citep{jimenez2024swebench,swebenchpro} to building complete applications \citep{ding2025nl2repo,yang2026programbench,lu2025webgenbenchevaluatingllmsgenerating,baxbench}.
Test generation benchmarks, in contrast, remain limited to individual issues \citep{mundler2024swtbench,sun2026swemutation}, few files \citep{jain2025testgeneval,wang2026multifiletest}, or single scenarios \citep{kogler2026restestbench}, while always providing the implementation to test.
Building a benchmark closing this gap is challenging: ambiguity, inconsistency, and incompleteness of natural language are a main source of software testing errors \citep{mustafa2021automated}, and measuring a test suite's strength requires faults that violate specific requirements, which generic rule-based mutations cannot guarantee \citep{jia2010analysis}.

\paragraph{Our work} We introduce \benchmark, a benchmark of 165 open-source applications in which agents must write a complete test suite given only a natural-language specification (\cref{fig:teaser}).
We derive each specification from the application's reference test suite, so that it describes every tested behavior, and iteratively refine it to remove ambiguities and contradictions.
To measure a suite's strength, we construct on average 74 mutants per application, each violating a single requirement. We refine mutants to isolate the violated requirement, with 73\% being killed by exactly one reference test.
We report a generated suite's validity score \vr as the fraction of tests that pass on the reference implementation and its strength as the mutation score \ms, the fraction of mutants killed by valid tests; since the reference suite kills every mutant, a perfect score is attainable by construction.

We evaluate coding agents with six recent frontier models and find that they struggle to write strong test suites: even the best agent's suites kill only 55\% of mutants, and four of the six agents stay below 40\%.
This is not due to a lack of understanding of the specification: on average, 88\% of generated tests are valid, and implementations built by the same agents from the same specifications pass up to 93\% of the reference tests.
Even when given the reference executable in addition to the specification, mutation scores remain unchanged, indicating that the bottleneck lies in test design rather than in the specification's precision.
Indeed, generated suites are often too small and miss the input combinations needed to expose violated requirements.

These results reveal a critical gap: agents can build software they cannot verify.
Until they can turn specifications into strong test suites, fully autonomous development of new software remains out of reach; \benchmark measures progress toward this goal.

\input{src/figure/teaser/teaser.tex}

\paragraph{Key Contributions} of our work are:
\begin{itemize}
    \item \benchmark, a benchmark for repository-scale test suite generation from natural-language specifications, comprising 165 applications with detailed specifications, reference implementations, reference test suites, and requirement-targeted mutants (\cref{sec:construction}).
    \item A scalable pipeline that derives detailed specifications from existing test suites, refines them by attributing agent failures to specification defects, and generates mutants that each violate a single requirement, enabling the benchmark to be extended to new applications (\cref{sec:construction}).
    \item An extensive evaluation of 6 coding agents, revealing that they can implement specifications they cannot test (\cref{sec:experiment}).
\end{itemize}

%% file: src/figure/teaser/teaser.tex
\begin{figure}
    \centering
    \includegraphics[width=\linewidth]{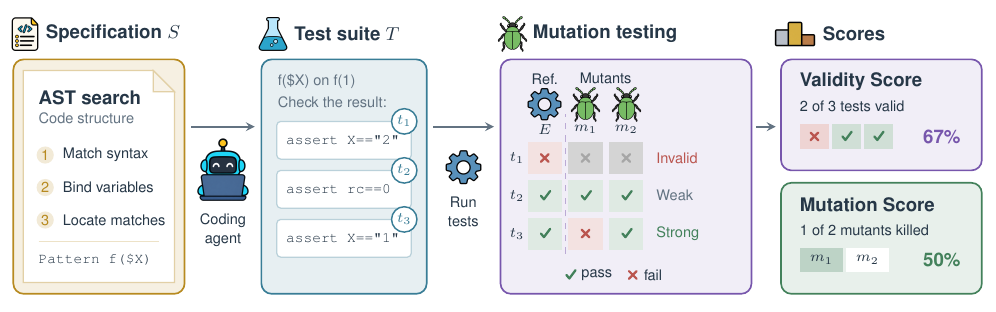}
    \vspace{-7mm}
    \caption{
        \benchmark:
        One instance consists of a specification $S$, a test suite $T$ asserting its requirements, and a set of mutants $M$ that break these requirements in targeted ways.
        An agent is tasked with generating a full test suite based solely on the natural-language specification $S$.
        We then evaluate the \emph{Validity Score} \vr, the fraction of tests that pass on the reference implementation, and the \emph{Mutation Score} \ms, the fraction of mutants killed by the generated test suite.
    }
    \label{fig:teaser}
    \vspace{-4mm}
\end{figure}

%% file: src/construction.tex
\section{Constructing \benchmark}
\label{sec:construction}
\paragraph{Notation}
We denote a repository as $R$, containing a reference implementation $I^*$ from which a build script produces a reference executable $E^*$, and a reference test suite $T^*$ that validates the correctness of $I^*$ and $E^*$. Throughout, we mark reference artifacts with a star (\eg $T^*$) and artifacts under evaluation with a hat (\eg $\hat{T}$). A test suite $T$ is a set of individual tests $t$. Typically, $T$ can be considered an executable verifier that checks the alignment between an executable $E$ and a specification $S$, which describes in natural language how $E$ should behave. Mutants $M$ are executables that have specific defects with respect to $S$ and can be used to assess the strength of $T$.
We say a test $t$ or a test suite $T$ containing $t$ \emph{kills} an executable $E$ if $t$ fails when processing $E$. A high quality test suite never kills the reference executable $E^*$, while killing all mutants $M$.

\paragraph{Benchmark Construction}
We build a \benchmark instance in three conceptual steps, described below. We first create a detailed draft specification from the reference test suite (\cref{subsec:initial_test_spec}). Then we systematically construct a set of mutants from this draft specification and the reference test suite that each get killed by a single test (\cref{subsec:mutant_construction}), before finally polishing the specification iteratively to remove any remaining ambiguities and contradictions (\cref{subsec:iterative_spec}).
\input{src/figure/construction.tex}

\subsection{Generating a Draft Specification}
\label{subsec:initial_test_spec}
We manually inspected developer-written documentations for representative repositories and found them severely underspecified and thus unsuitable for test generation (see \cref{tab:developer_spec}). Thus, we create richer specifications suitable for test and program generation. To this end, we first refine a test suite to remove low quality tests before generating a specification document from it.
Throughout, we use Codex GPT-5.6 Sol \citep{gpt56} with high reasoning.

\paragraph{Reference Implementation}We start by collecting a reference implementation $I^*$ building on the 200 instances constituting \programbench{} \citep{yang2026programbench}. We summarize their initial construction process below.

\begin{enumerate}[topsep=-2pt, itemsep=0pt, parsep=0pt, partopsep=0pt, leftmargin=12pt]
    \item Collect 200 open-source GitHub repositories $R$ with compiled standalone executables $E^*$, based primarily on Rust, Go or C++.
    \item Instruct a coding agent to construct an extensive test suite $T^*$ based on the test suite contained in $R$ and the observed behavior of $E^*$. All tests in $T^*$ are purely input-output behavior assertions.
    \item Manually and automatically audit the constructed test suite for underspecification or overspecification with respect to $E^*$.
\end{enumerate}

\paragraph{Test Suite Refinement}
We refine the \programbench test suites by filtering out tests that require internet access, excessive resources, or are flaky w.r.t. the reference implementation, i.e. produce inconsistent results across reruns. See \cref{appsec:construction_detail} for more details.
If, after this filtering, the size of $T^*$ has been significantly reduced, we drop the instance.
Overall, this leads to a set of 165 initial instances with reference implementation and test suite. Among those instances, the reference test suites are shrunken by $5.7\%$ on average.

\paragraph{Draft Specification Generation}
We use a coding agent  to generate a semi structured specification draft $S'$ describing every tested behavior of $T^*$ explicitly with at most 50 sections.
This initial draft will later be iteratively refined to remove ambiguities, contradiction, and under specification w.r.t. input-output behaviour.

\subsection{Mutant Construction}
\label{subsec:mutant_construction}
The set of mutants are at the core of \benchmark and their specificity and diversity determine its ability to measure a test suites strength.
We thus aim to create mutants with targeted changes that break only specific requirements, while being otherwise indistinguishable.
For each mutant, we select a target test function $t_{target}$ that is intended to kill it, and denote additional tests $T^* \setminus{} \{t_{target}\}$ that kill it as $\bar{T}$. We aim to construct mutants that are killed by exactly $t_{target}$ and have an empty $\bar{T}$.

As $T^*$ is often very large (median size of 679), we target only a subset of test functions, selected to uniformly cover $S'$. This is sufficient for strong mutation testing, since the later test generation can not know which requirements will be tested by the mutants and thus has to evenly cover the entire specification.
We generate the actual mutants using the following two-step approach.

\paragraph{Candidate Construction}
We create a bi-graph $G$ linking each section $s_i$ of $S'$ describing a set of requirements to the test functions $t_j \in T^*$ asserting them.
Based on $G$, we then pick 6 test functions $t_{j,target}$ per specification section $s_i$, such that each test is selected by as few sections as possible. This can be described and solved optimally as a min-cost flow problem. See \cref{appsubsec:minflow} for more details.

We then use a separate coding agent for every $t_{j,target}$ to modify $I^*$ into $I_j$ such that it compiles into a mutant executable $m_j$ killed by $t_{j,target}$ due to a violation of a specific behavior in $s_i$. The agent is instructed to retry the process in case $I_j$ does not compile up to three times.

\input{src/figure/method_spec_polish/spec_polish.tex}

\paragraph{Candidate Refinement}
For every mutant candidate, we determine the test functions that kill it and rank the 6 mutants $m_j$ associated with each $s_i$ by (i) whether $m_j$ is killed by $t_{j,target}$ and (ii) the size of $\bar{T}_j$. We select the top three mutants by these criteria. Then, we use a coding agent to inspect these mutant candidates and the test cases that kill it, and refine them to reduce $\bar{T}_j$ where needed.

Finally, we run the reference test suite against the created mutants and drop mutants that are either (i) not killed by the intended test function $t_{j,target}$, or (ii) killed by $>20\%$ of unrelated functions.
This refinement process significantly increases the fraction of mutants killed exactly by the targeted test ($32.4\% \rightarrow 73.0\%$) and reduces the average size of $\bar{T}_j$ ($1.45\% \rightarrow 0.17\%$).

\subsection{Specification Refinement}
\label{subsec:iterative_spec}

Finally, we refine the draft specifications to remove contradictions and ambiguities while ensuring that the specifications contain enough detail for test and program generation. To this end, we run an iterative process, creating tests and implementations based on the specification, analysing resulting failures, and remedying any errors attributed to the specification.

\paragraph{Trial Generations}
We use two coding agents to generate an implementation $\hat{I}$ and a test suite $\hat{T}$ based on the draft specification $S'$.
We then evaluate $\hat{I}$ against the reference tests $T^*$, and $\hat{T}$ against the reference executable $E^*$ and our mutant set $M$.

\paragraph{Error Attribution}
For each failure ($t \in T^*$ killing $\hat{I}$, $\hat{t} \in \hat{T}$ killing $I^*$, and $m \in M$ surviving $\hat{T}$), we classify whether the failure is due to a bad generation, or the underlying specification $S'$, e.g., underspecification (e.g., ambiguous, insufficient) or contradiction (e.g., misleading, wrong) with respect to $T$.

\paragraph{Specification Refinement}
For every failure attributed to $S'$, we refine the affected sections of the specification using an agent. Concretely, for contradictory information we expect a correction of the specification, and for an underspecification, we expect further details to be inserted based on the reference tests.

We iterate the above three steps until fewer than $5\%$ of generated test failures were attributed to the specification.
We visualize the improvement process in \cref{fig:spec-polish}.
We see that in particular the test validity rate increased significantly as ambiguities in expected behaviour were clarified.

\subsection{Manual Quality Control}
\label{subsec:manual_check}
Finally, we check the resulting specifications, mutants, and reference test suites manually, fixing minor issues such as tests from the original \programbench suite expecting crashes in specific lines. We discuss 34 instances where such manual intervention was required in \cref{appsec:construction_detail}.

\subsection{\benchmark}
\label{sec:overview}

\input{src/table/stats.tex}
Our final benchmark consists of 165 instances, spanning Rust, Go, and C/C++ and a diverse set of domains including text processing, file and disk utilities, and build tools. We show key statistics in \cref{tab:stats}, and a programming language breakdown in \cref{fig:bench_stats}. Our natural language specifications contain a median of 5k words across 23 sections, 679 reference test functions $T^*$, and 67 mutants $M$. The reference implementation length varies significantly from 538 to 1.6M LoC with a median of 7k.

\subsection{Evaluation Metrics}
\label{subsec:metrics}

We measure the quality of a test suite $\hat{T}$ with two core metrics: First, its \emph{Mutation Score (\ms)} is the fraction of mutants $M$ that are killed by $\hat{T}$.
Second, its \emph{Validity Rate (\vr)} is the fraction of generated tests $\hat{t} \in \hat{T}$ that do not kill the reference executable $E^*$.
Importantly, since test functions that kill all executables, would achieve artificially high $\mathcal{M}$, we filter out all tests $t$ that fail on the reference executable $E^*$ when calculating $\mathcal{M}$.
To handle flaky tests that kill non-deterministically, we run test suites 3 times on the reference executable and remove tests for which behavior changes across runs.

For implementations $\hat{I}$, we measure correctness using the \emph{Reference Pass Rate (\pr)}, the fraction of reference tests $t \in T^*$ that do \emph{not} kill $\hat{I}$.

To prevent cheating, agents are run inside Docker containers without networking and CLI search tools disabled but all standard libraries including the dependencies of the reference implementation pre-installed.

%% file: src/figure/construction.tex
\begin{figure*}[t]
    \centering
    \resizebox{0.95\linewidth}{!}{\input{src/figure/construction.tikz}}
    \par\vspace{1pt}
    \resizebox{0.95\linewidth}{!}{\input{src/figure/conceptual.tikz}}
    \vspace{-4mm}
    \caption{
         (Top) To construct \benchmark{},
        we first obtain a reference test suites $T^*$, implementation $I^*$ and its executable $E^*$ from a GitHub repository $R$.
        We then create a draft specification $S'$, and mutant candidates $M$ that violate individual requirements, before refining the mutants to reduce unrelated failures against $T^*$.
        We then refine the specification $S'$ by iteratively generating implementations and test suites from it, analysing resulting errors, and addressing the ones attributed to the specification. (Bottom) In the space of implementations, we first define the set of admissible implementations using $T$, and align it with the set according to the specification $S$.
    }
    \label{fig:construction}
    \label{fig:conceptual-progression}
\end{figure*}
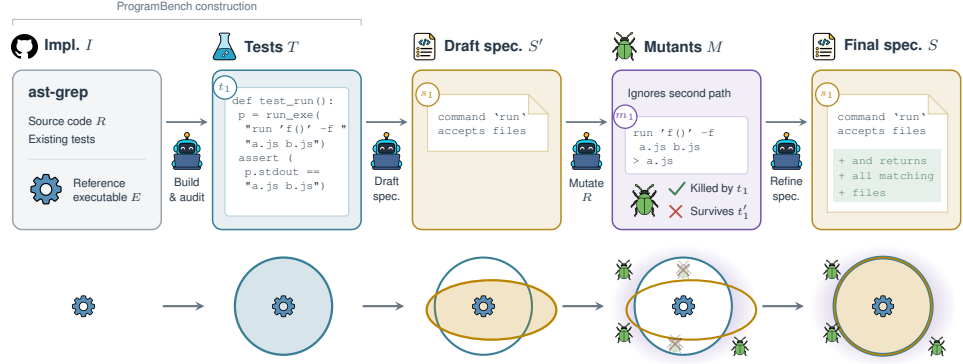

%% file: src/figure/construction.tikz
\begingroup
\definecolor{tbInk}{HTML}{263747}
\definecolor{tbMuted}{HTML}{637382}
\definecolor{tbGold}{HTML}{B78A20}
\definecolor{tbBlue}{HTML}{397D96}
\definecolor{tbPurple}{HTML}{7756AC}
\definecolor{tbGreen}{HTML}{43815B}
\definecolor{tbRed}{HTML}{BA554F}
\begin{tikzpicture}[
    x=1cm, y=1cm, text=tbInk,
    font=\fontfamily{phv}\fontsize{6}{7.5}\selectfont,
    every node/.style={inner sep=0pt, outer sep=0pt},
    title/.style={font=\fontfamily{phv}\bfseries\fontsize{8}{10}\selectfont},
    code/.style={font=\fontfamily{pcr}\fontsize{6}{7.5}\selectfont},
    flow/.style={->, >=stealth, draw=tbMuted, line width=.9pt},
    card/.style={rounded corners=5pt, line width=.8pt}
]
\path[use as bounding box] (-.15,2.35) rectangle (17.72,6.75);

\node at (.22,5.84) {\includegraphics[height=.46cm]{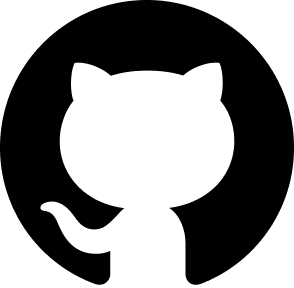}};
\node[title, anchor=west] at (.59,5.84) {Impl. $I$};
\node at (3.92,5.84) {\includegraphics[height=.56cm]{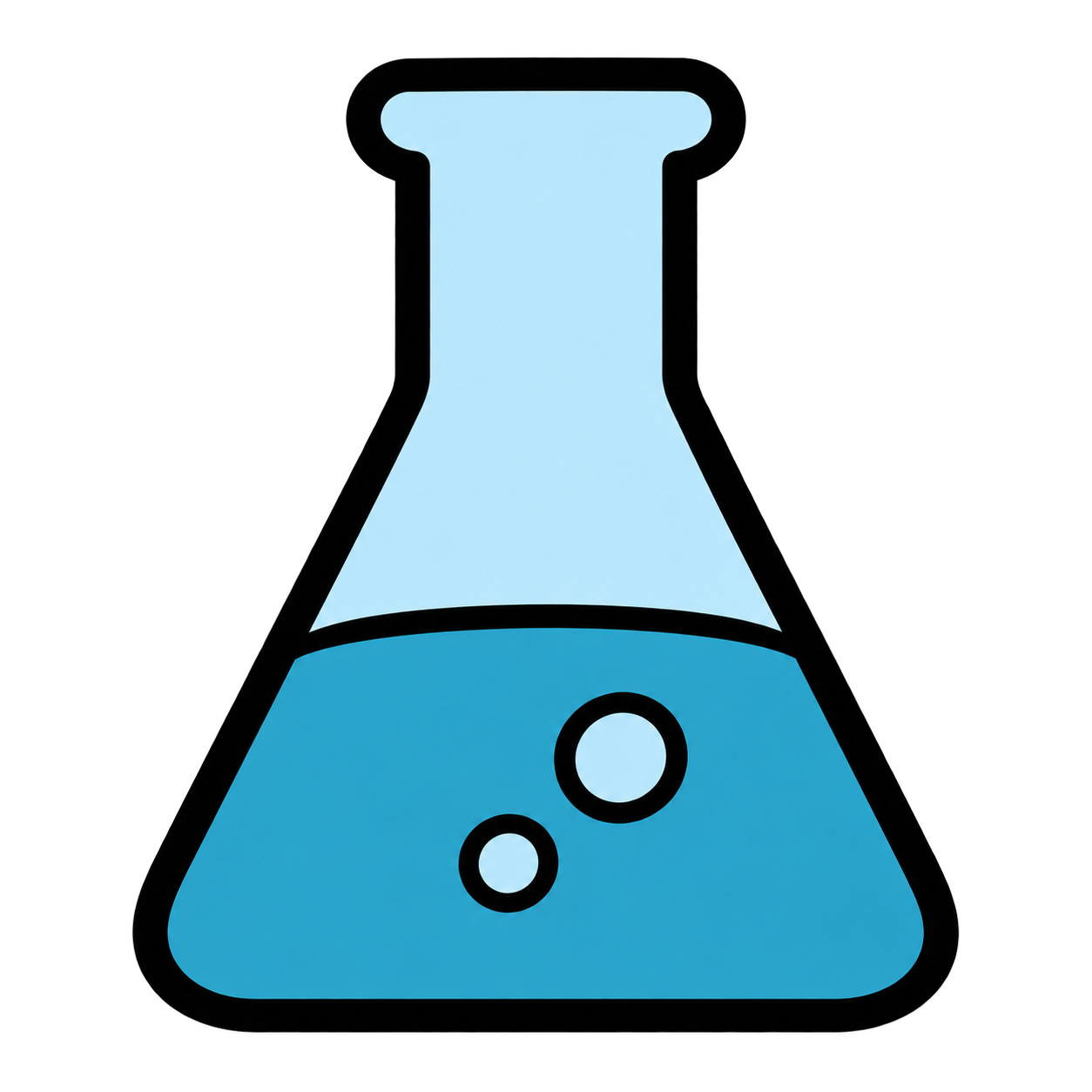}};
\node[title, anchor=west] at (4.29,5.84) {Tests $T$};
\node at (7.62,5.84) {\includegraphics[height=.56cm]{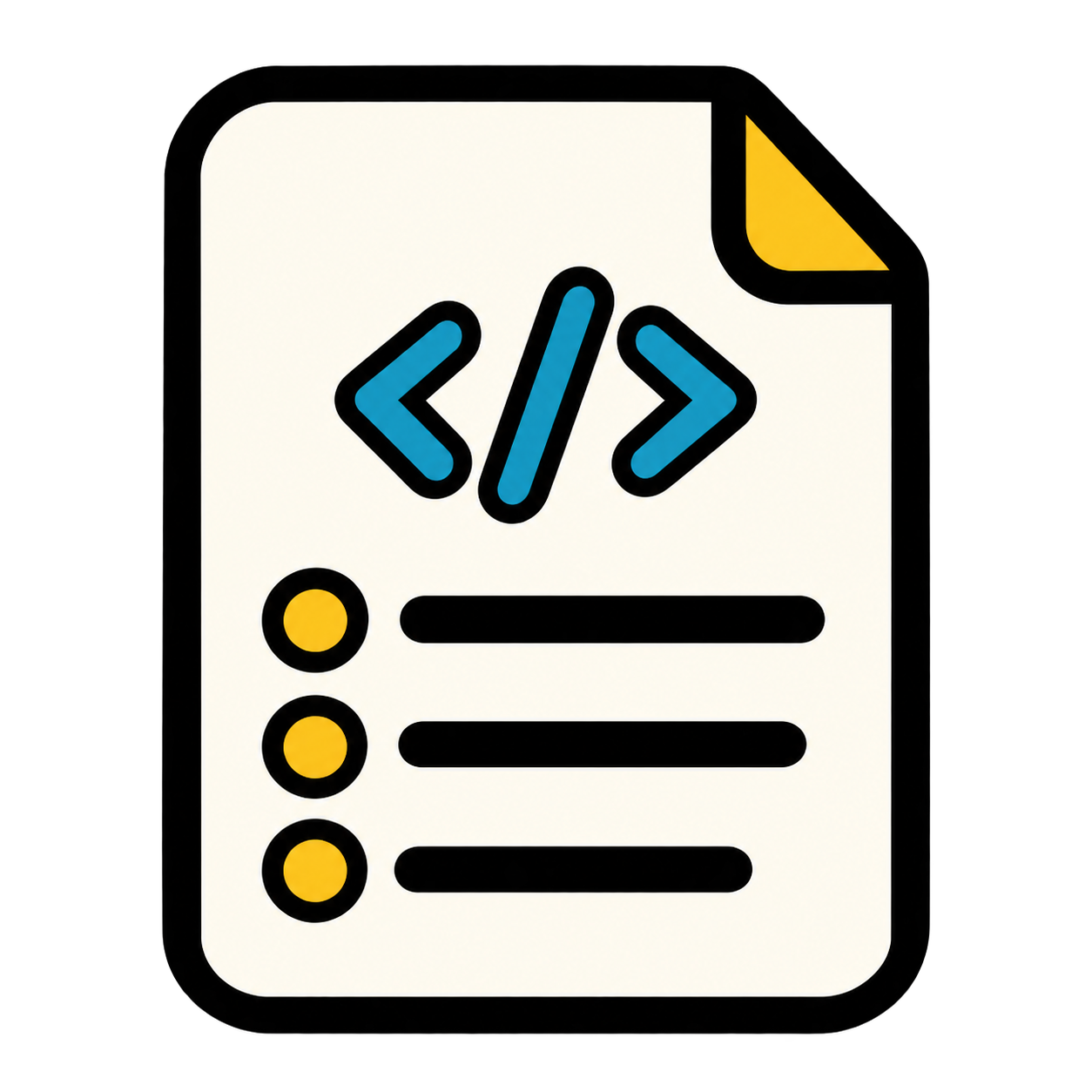}};
\node[title, anchor=west] at (7.99,5.84) {Draft spec. $S'$};
\node at (11.32,5.84) {\includegraphics[height=.56cm]{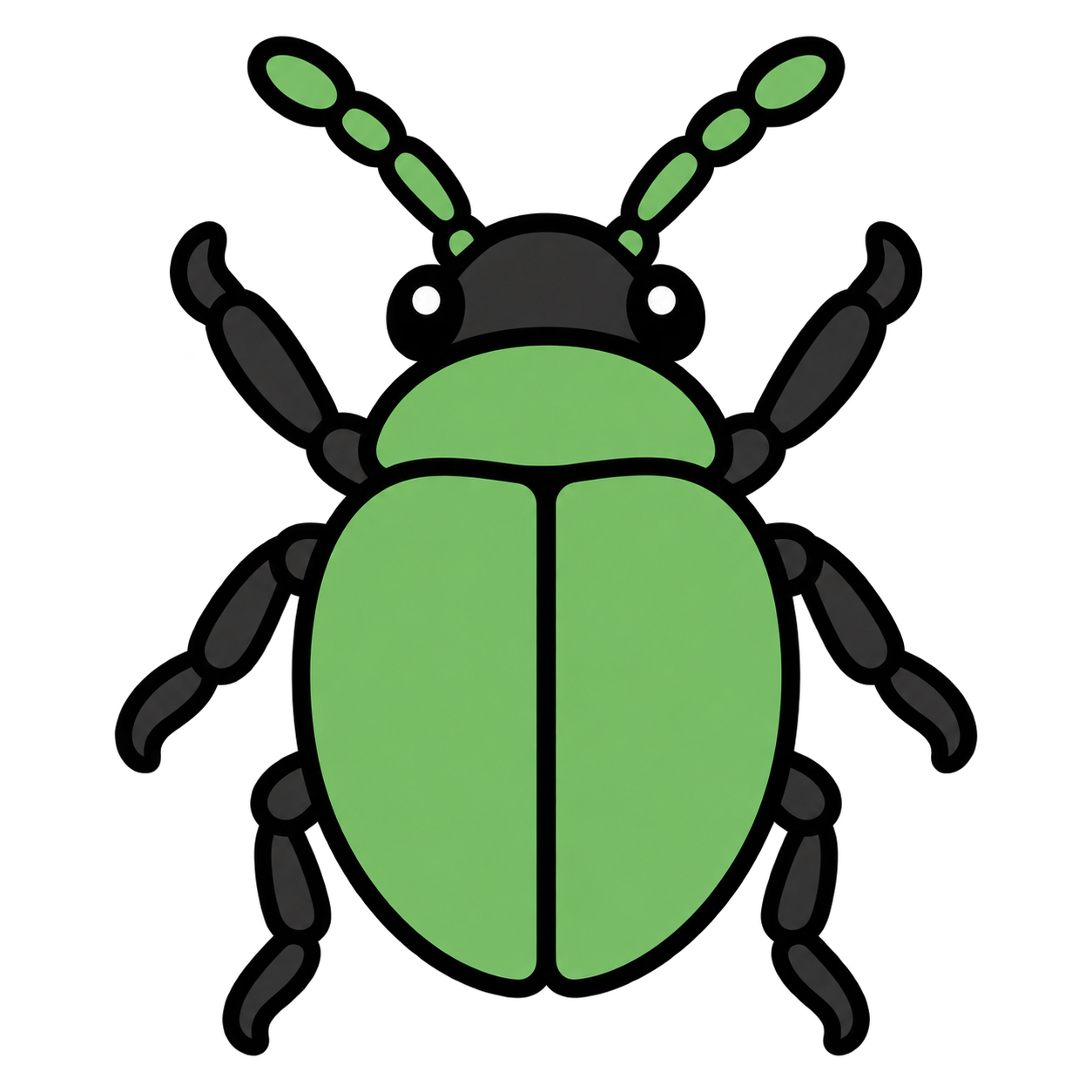}};
\node[title, anchor=west] at (11.69,5.84) {Mutants $M$};
\node at (15.02,5.84) {\includegraphics[height=.56cm]{src/figure/assets/document.png}};
\node[title, anchor=west] at (15.39,5.84) {Final spec. $S$};

\draw[card, draw=tbMuted!65, fill=tbMuted!7] (0,2.45) rectangle (2.75,5.40);
\node[title, anchor=west] at (.29,4.96) {ast-grep};
\node[anchor=west] at (.29,4.47) {Source code $R$};
\node[anchor=west] at (.29,4.10) {Existing tests};
\draw[draw=tbMuted!25] (.29,3.75) -- (2.46,3.75);
\node at (.62,3.19) {\includegraphics[width=.62cm]{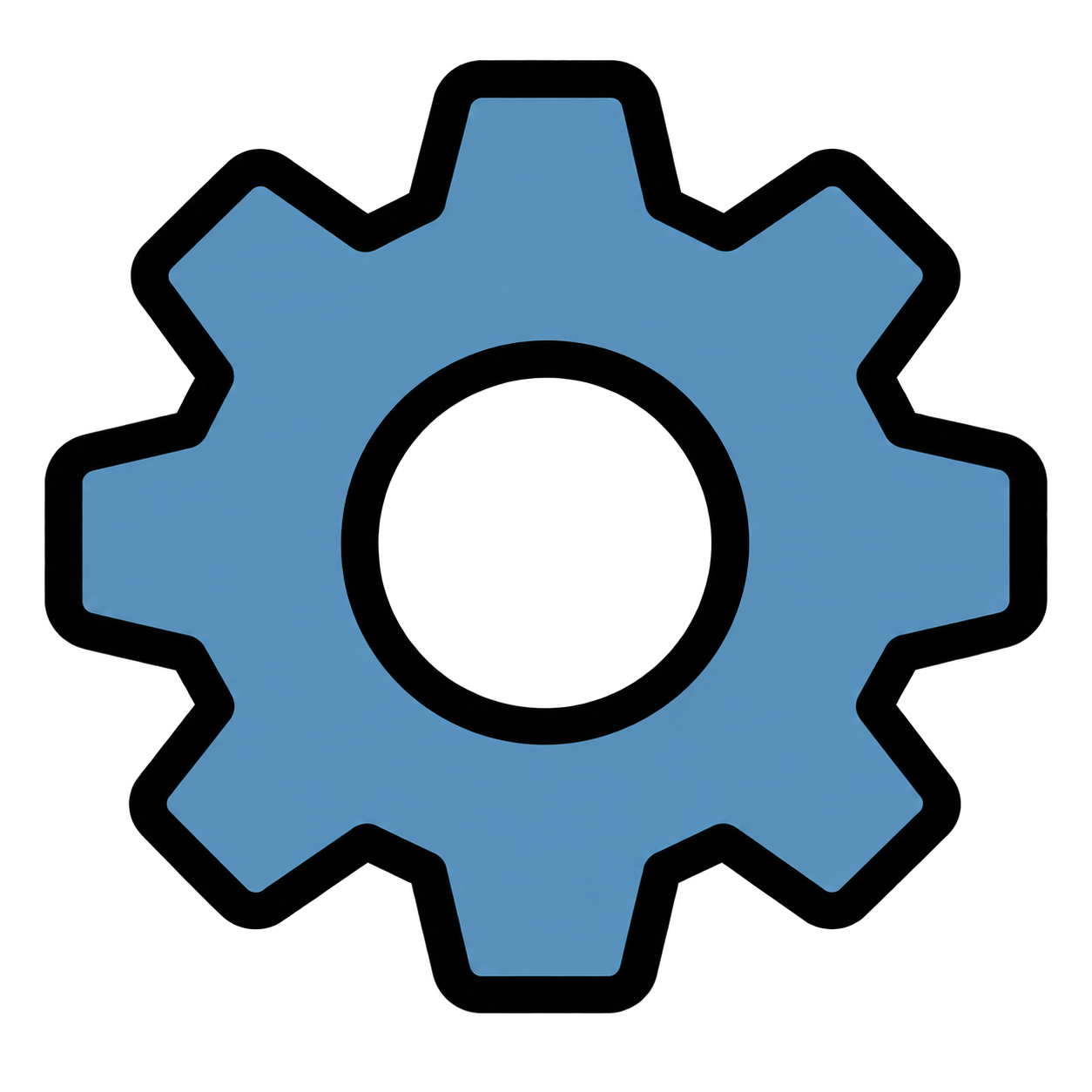}};
\node[anchor=west, align=left] at (1.12,3.19) {Reference\\executable $E$};

\draw[card, draw=tbBlue, fill=tbBlue!12] (3.7,2.45) rectangle (6.45,5.40);
\draw[draw=tbBlue!35, fill=white, rounded corners=2pt, line width=.6pt]
    (3.93,2.70) rectangle (6.22,5.08);
\node[anchor=north west, code, align=left] at (4.06,4.91) {%
    def test\_run():\\
    \hspace*{.12cm}p = run\_exe(\\
    \hspace*{.24cm}"run 'f()' -f "\\
    \hspace*{.24cm}"a.js b.js")\\
    \hspace*{.12cm}assert (\\
    \hspace*{.24cm}p.stdout ==\\
    \hspace*{.24cm}"a.js b.js")};
\node[circle, draw=tbBlue, fill=white, line width=.7pt,
    minimum size=.43cm, text=tbBlue] at (3.93,5.08) {$t_1$};

\draw[card, draw=tbGold, fill=tbGold!13] (7.4,2.45) rectangle (10.15,5.40);
\draw[draw=tbGold!40, fill=white, line width=.6pt]
    (7.68,3.90) -- (9.91,3.90) -- (9.91,4.59) -- (9.55,4.95)
    -- (7.68,4.95) -- cycle;
\draw[draw=tbGold!40, fill=tbGold!10, line width=.6pt]
    (9.55,4.95) -- (9.55,4.59) -- (9.91,4.59) -- cycle;
\node[anchor=west, code, align=left] at (7.87,4.40) {command `run`\\accepts files};
\node[circle, draw=tbGold, fill=white, line width=.7pt,
    minimum size=.43cm, text=tbGold] at (7.68,4.95) {$s_1$};

\draw[card, draw=tbPurple, fill=tbPurple!10] (11.1,2.45) rectangle (13.85,5.40);
\node[anchor=west] at (11.39,4.96) {Ignores second path};
\draw[draw=tbPurple!30, fill=white, rounded corners=2pt, line width=.6pt]
    (11.34,3.48) rectangle (13.61,4.54);
\node[anchor=north west, code, align=left] at (11.48,4.32) {%
    run 'f()' -f\\
    \hspace*{.12cm}a.js b.js\\
    > a.js};
\node[circle, draw=tbPurple, fill=white, line width=.7pt,
    minimum size=.43cm, text=tbPurple] at (11.34,4.54) {$m_1$};
\node at (11.70,2.99) {\includegraphics[height=.62cm]{src/figure/assets/mutant.png}};
\draw[draw=tbGreen, line width=1.1pt, line cap=round, line join=round]
    (12.16,3.20) -- (12.26,3.10) -- (12.44,3.32);
\node[anchor=west] at (12.55,3.20) {Killed by $t_1$};
\draw[draw=tbRed, line width=1pt, line cap=round]
    (12.17,2.68) -- (12.38,2.89) (12.17,2.89) -- (12.38,2.68);
\node[anchor=west] at (12.55,2.78) {Survives $t_1'$};

\draw[card, draw=tbGold, fill=tbGold!13] (14.8,2.45) rectangle (17.55,5.40);
\draw[draw=tbGold!40, fill=white, line width=.6pt]
    (15.08,2.77) -- (17.31,2.77) -- (17.31,4.59) -- (16.95,4.95)
    -- (15.08,4.95) -- cycle;
\draw[draw=tbGold!40, fill=tbGold!10, line width=.6pt]
    (16.95,4.95) -- (16.95,4.59) -- (17.31,4.59) -- cycle;
\node[anchor=west, code, align=left] at (15.27,4.40) {command `run`\\accepts files};
\fill[tbGreen!13] (15.20,2.97) rectangle (17.19,3.95);
\node[anchor=west, code, text=tbGreen] at (15.29,3.73) {+ and returns};
\node[anchor=west, code, text=tbGreen] at (15.29,3.43) {+ all matching};
\node[anchor=west, code, text=tbGreen] at (15.29,3.13) {+ files};
\node[circle, draw=tbGold, fill=white, line width=.7pt,
    minimum size=.43cm, text=tbGold] at (15.08,4.95) {$s_1$};

\foreach \x/\start/\end in {3.225/2.84/3.61,6.925/6.54/7.31,10.625/10.24/11.01,14.325/13.94/14.71} {
    \draw[flow] (\start,4.44) -- (\end,4.44);
    \node at (\x,3.89) {\includegraphics[height=.77cm]{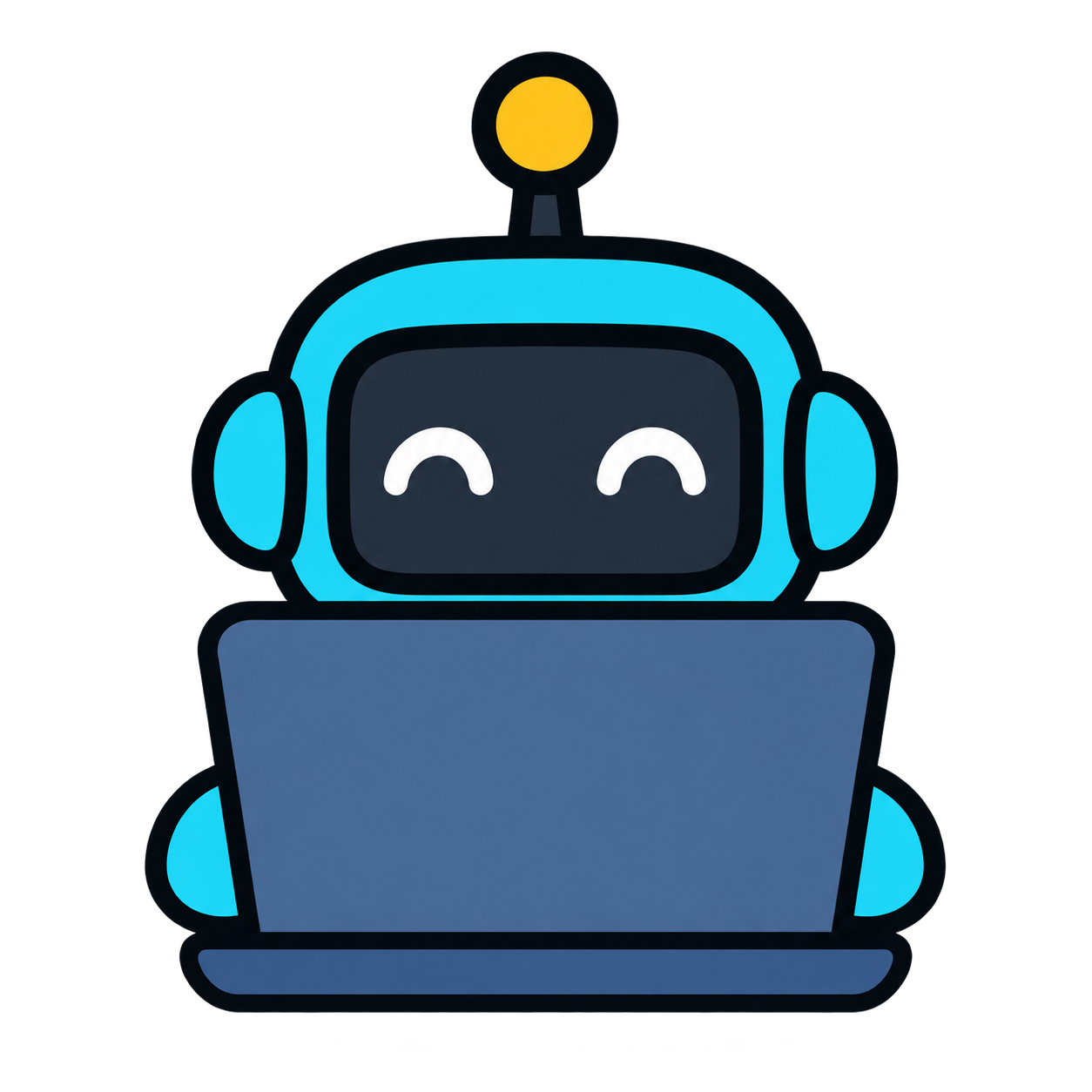}};
}
\node[align=center] at (3.225,3.19) {Build\\\& audit};
\node[align=center] at (6.925,3.19) {Draft\\spec.};
\node[align=center] at (10.625,3.19) {Mutate\\$R$};
\node[align=center] at (14.325,3.19) {Refine\\spec.};

\node[text=tbMuted] at (3.225,6.57) {ProgramBench construction};
\draw[draw=tbMuted!45, line width=.6pt] (0,6.23) -- (0,6.34) -- (6.45,6.34) -- (6.45,6.23);

\end{tikzpicture}
\endgroup

%% file: src/figure/conceptual.tikz
\begingroup
\definecolor{tbInk}{HTML}{263747}
\definecolor{tbMuted}{HTML}{637382}
\definecolor{tbGold}{HTML}{C48B00}
\definecolor{tbBlue}{HTML}{397D96}
\definecolor{tbPurple}{HTML}{7756AC}
\definecolor{tbGreen}{HTML}{2D7D46}
\definecolor{tbRed}{HTML}{BA554F}
\begin{tikzpicture}[
    x=1cm, y=1cm, text=tbInk,
    font=\fontsize{7}{9}\selectfont,
    every node/.style={inner sep=0pt, outer sep=0pt},
    transarrow/.style={->, >=stealth, draw=tbMuted, line width=1.1pt}
]
\path[use as bounding box] (-.15, -0.08) rectangle (17.72, 2.37);

\begin{scope}[xshift=-.225cm]

\node at (1.60, 1.15) {\includegraphics[height=.44cm]{src/figure/assets/gear.png}};
\end{scope}

\draw[transarrow] (2.84, 1.15) -- (3.61, 1.15);

\begin{scope}[xshift=-.125cm]

\fill[tbBlue, fill opacity=0.20] (5.2, 1.15) circle (0.90cm);
\draw[draw=tbBlue, line width=1.2pt] (5.2, 1.15) circle (0.90cm);

\node at (5.2, 1.15) {\includegraphics[height=.44cm]{src/figure/assets/gear.png}};
\end{scope}

\draw[transarrow] (6.54, 1.15) -- (7.31, 1.15);

\begin{scope}[xshift=-.025cm]

\draw[draw=tbBlue, line width=1.1pt] (8.8, 1.15) circle (0.90cm);

\fill[tbGold, fill opacity=0.22] (8.95, 1.08) ellipse (1.20cm and 0.55cm);
\draw[draw=tbGold!95!black, line width=1.2pt] (8.95, 1.08) ellipse (1.20cm and 0.55cm);
\node at (8.8, 1.15) {\includegraphics[height=.44cm]{src/figure/assets/gear.png}};
\end{scope}

\draw[transarrow] (10.24, 1.15) -- (11.01, 1.15);

\shade[inner color=tbPurple!65, outer color=white]
    (12.475, 1.15) ellipse (1.36cm and 1.14cm);
\begin{scope}[xshift=.075cm]

\fill[white] (12.4, 1.15) circle (0.90cm);
\draw[draw=tbBlue, line width=1.1pt] (12.4, 1.15) circle (0.90cm);

\draw[draw=tbGold!95!black, line width=1.1pt] (12.55, 1.08) ellipse (1.20cm and 0.55cm);

\node at (12.4, 1.15) {\includegraphics[height=.44cm]{src/figure/assets/gear.png}};

\node[opacity=0.30] at (12.40, 1.82) {\includegraphics[height=.34cm]{src/figure/assets/mutant.png}};
\draw[draw=tbRed!80!black, line width=.8pt, opacity=0.55]
    (12.29, 1.71) -- (12.51, 1.93) (12.29, 1.93) -- (12.51, 1.71);

\node[opacity=0.30] at (12.25, 0.45) {\includegraphics[height=.34cm]{src/figure/assets/mutant.png}};
\draw[draw=tbRed!80!black, line width=.8pt, opacity=0.55]
    (12.14, 0.34) -- (12.36, 0.56) (12.14, 0.56) -- (12.36, 0.34);

\node at (11.30, 1.82) {\includegraphics[height=.38cm]{src/figure/assets/mutant.png}};
\node at (11.30, 0.65) {\includegraphics[height=.38cm]{src/figure/assets/mutant.png}};
\node at (13.45, 0.40) {\includegraphics[height=.38cm]{src/figure/assets/mutant.png}};
\end{scope}

\draw[transarrow] (13.94, 1.15) -- (14.71, 1.15);

\shade[inner color=tbPurple!65, outer color=white]
    (16.175, 1.15) ellipse (1.36cm and 1.14cm);
\begin{scope}[xshift=.175cm]

\fill[white] (16.0, 1.15) circle (0.90cm);
\fill[tbGold, fill opacity=0.22] (16.0, 1.15) circle (0.90cm);
\draw[draw=tbBlue, line width=2pt] (16.0, 1.15) circle (0.90cm);
\draw[draw=tbGold!95!black, line width=.9pt] (16.0, 1.15) circle (0.90cm);
\node at (16.0, 1.15) {\includegraphics[height=.44cm]{src/figure/assets/gear.png}};

\node at (15.05, 1.82) {\includegraphics[height=.38cm]{src/figure/assets/mutant.png}};
\node at (15.05, 0.65) {\includegraphics[height=.38cm]{src/figure/assets/mutant.png}};
\node at (17.00, 0.40) {\includegraphics[height=.38cm]{src/figure/assets/mutant.png}};
\end{scope}

\end{tikzpicture}
\endgroup

%% file: src/figure/method_spec_polish/spec_polish.tex
\begin{wrapfigure}[23]{r}{0.45\linewidth}
    \centering
    \begin{tabular}{@{}c@{}}
        \includegraphics[width=0.85\linewidth]{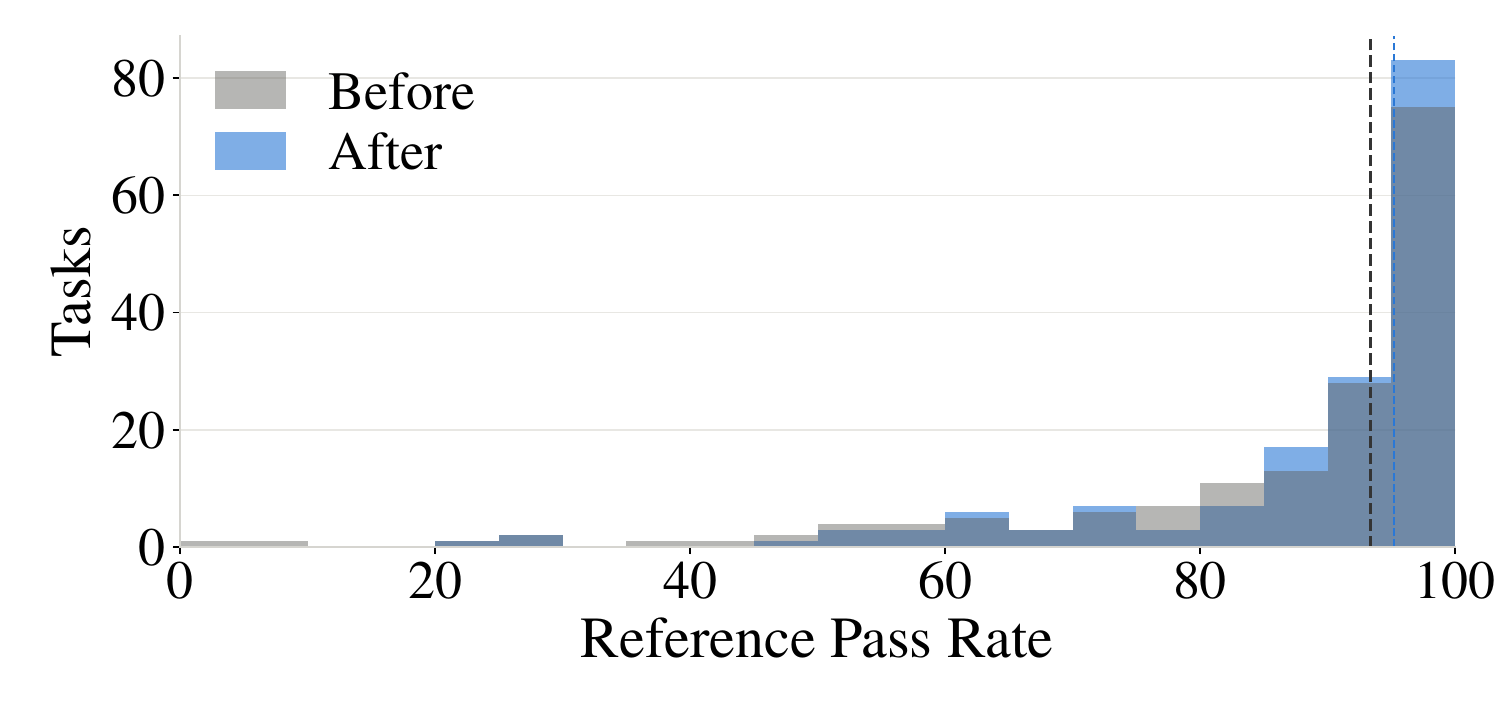} \\
        \includegraphics[width=0.85\linewidth]{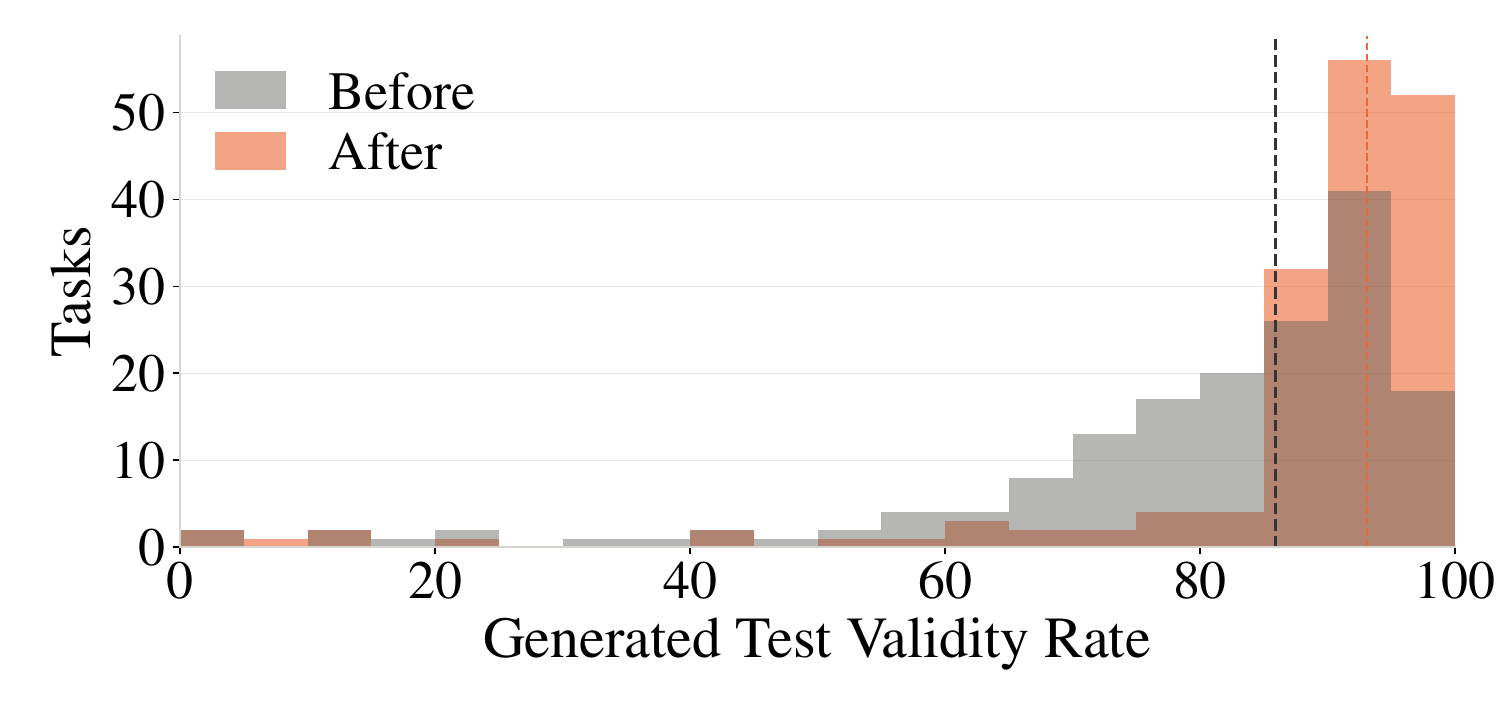} \\
        \includegraphics[width=0.85\linewidth]{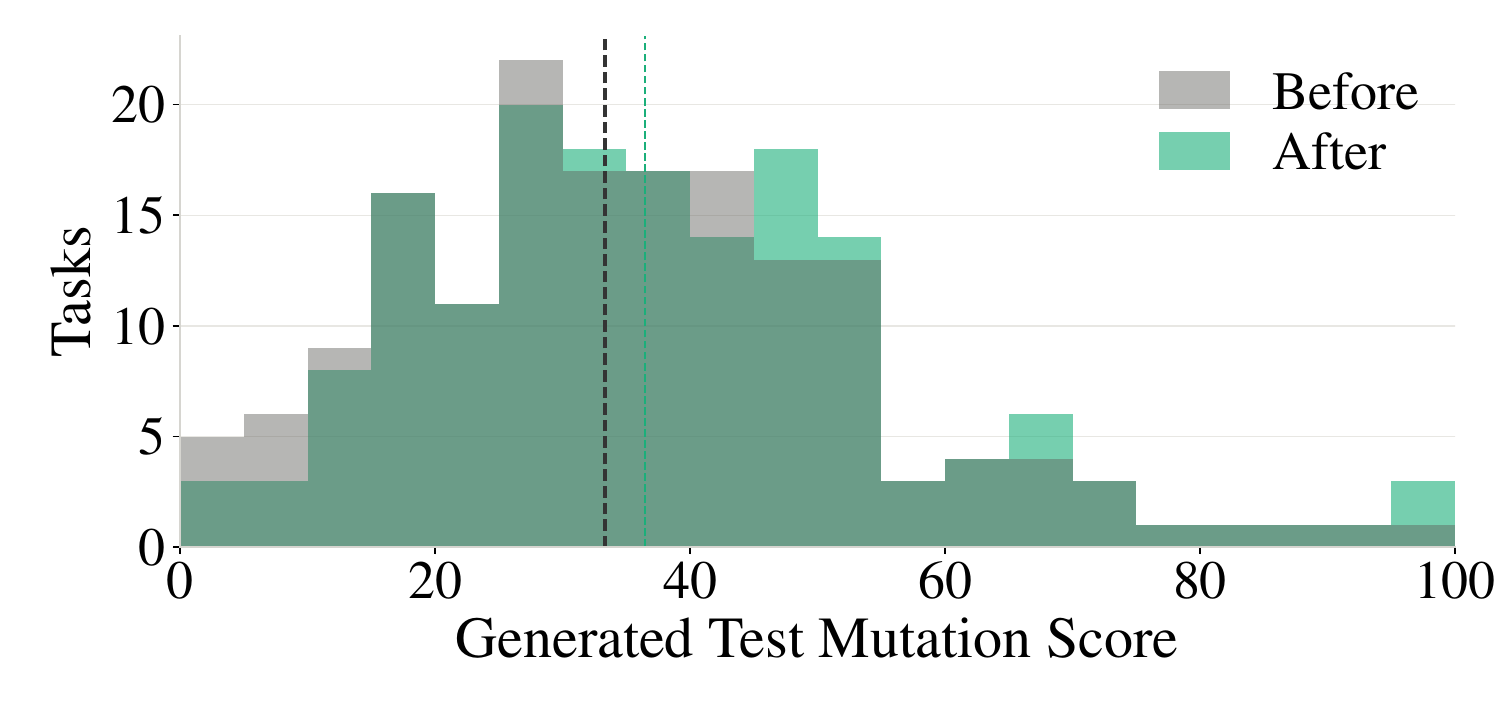} \\
    \end{tabular}
    \vspace{-3mm}
    \caption{
        The iterative specification refinement increases the implementation's reference pass rate and test's validity rate, while mutation score remains low.
    }
    \label{fig:spec-polish}
\end{wrapfigure}

%% file: src/table/stats.tex
\begin{wraptable}[24]{r}{0.35\linewidth}
    \vspace{-5mm}
    \centering
    \caption{\benchmark Statistics across benchmark instances.}
    \label{tab:stats}
    \setlength{\tabcolsep}{3pt}
    \resizebox{\linewidth}{!}{
    \begin{tabular}{@{}llrrr@{}}
        \toprule
        & & Min & Max & Median \\
        \midrule
        $S$     & \# Words    & 2,065     & 27,406    & 5,635 \\
                & \# Lines    & 158       & 3,258     & 576   \\
                & \# Sections    & 10        & 50        & 23  \\
                & \# Files & 1         & 46        & 1     \\
        \midrule
        $I$    & \# LOC    & 538       & 1,613,380 & 7,008 \\
        \midrule
        $T$    & \# Func   & 85        & 13,298    & 679   \\
        \midrule
        $M$    & \# Amount & 27        & 148       & 67    \\
        \bottomrule
    \end{tabular}
    }
    \input{src/figure/bench_stats/bench_stats.tex}
\end{wraptable}

%% file: src/figure/bench_stats/bench_stats.tex
\vspace{2mm}
\centering
\includegraphics[width=\linewidth]{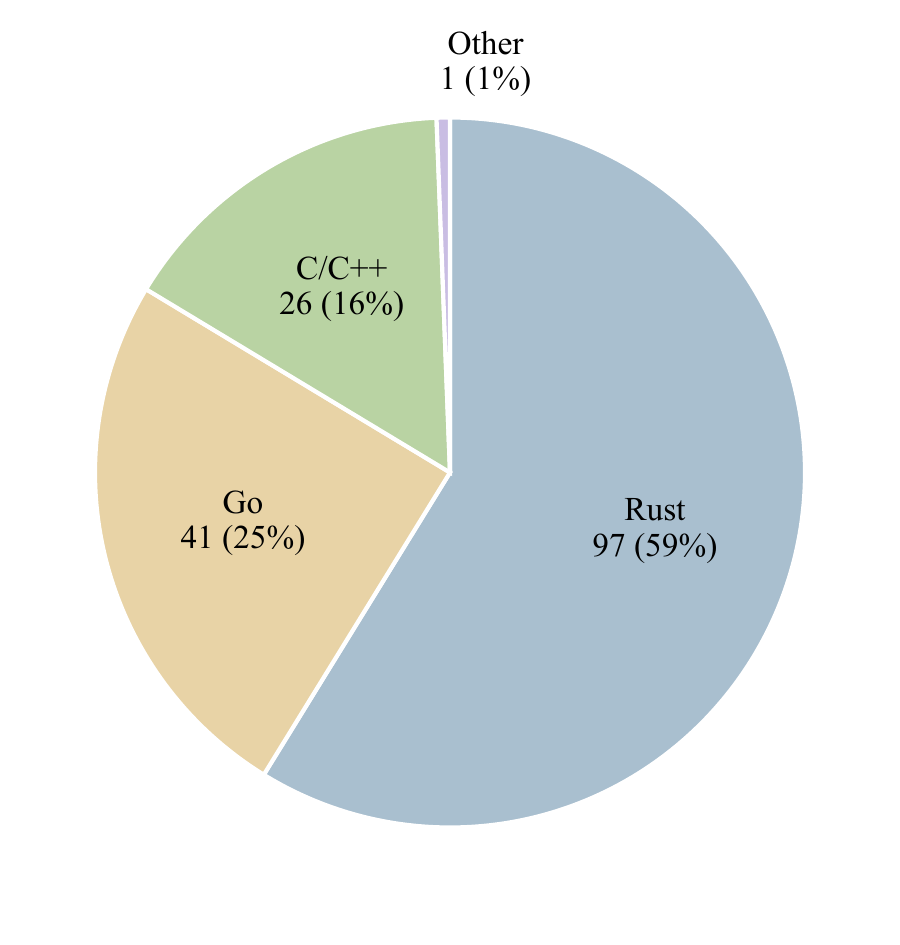}
\vspace{-3em}
\captionof{figure}{Implementation languages of \benchmark instances.}
\label{fig:bench_stats}

%% file: src/experiment.tex
\section{Experimental Evaluation}
\label{sec:experiment}

\paragraph{Models and Setup}
We consider GPT-5.6 Sol and Luna \citep{gpt56}, GPT-6 Astra \citep{gpt6}, Claude Opus 5 \citep{opus5}, Gemini-3.8 Flash \citep{gemini38}, DeepSeek V4 Flash \citep{deepseek}.
Unless otherwise indicated, we use the default reasoning effort and vendor-recommended agent harnesses, enforcing a 6h timeout on agents.
We follow the inference and evaluation setup outlined in \cref{sec:overview} and provide additional details in \cref{appsec:setup}.

\subsection{Main Results}
\input{src/table/main.tex}
\paragraph{Agents Struggle to Generate Strong Tests.}
We compare different coding agents across test and program generation in \cref{tab:main-results} and observe that they all struggle to generate strong test suites, as measured by their mutation score $\mathcal{M}$. Claude Opus 5 achieves the highest score at $55.5\%$, at a cost of $\$10.56$ per instance and taking on average \SI{37}{\minute}. It is followed by GPT 6 Astra which achieves $50.1\%$, at only $\$2.18$ and \SI{12.4}{\minute}. All other agents achieve less than $40\%$.

\paragraph{Agents Generate Valid Tests and Implementations}
Interestingly, all models but DeepSeek V4 Flash and GPT 5.6 Luna generally produce valid tests and implementations with validity scores \vr between 88\%-96\% and  reference test pass rates \pr between 86\%-93\%.
Both imply that the specification is clear and precise enough to provide a detailed understanding of the expected behaviour, validating our construction process. In particular the high \pr implies that the specification covers all the requirements in sufficient detail to pass the reference test suite. This makes it all the more surprising that all models still fail at generating strong test suites and points to a gap in their capabilities.

\paragraph{Valid but Weak Tests}
The large gap between validity scores \vr (88\%-96\%) and mutation scores \ms (38\%-56\%), in particular, has a concerning implication for practitioners. While the models are capable of writing valid tests, giving the impression of well tested software, they fail to write discriminative tests that would actually surface specification violations. This directly limits the capability of current agents for autonomous software engineering and reflects commonly observed patterns of code agents writing large test suites yet not detecting failures in their implementations.

\paragraph{Cost-Strength Tradeoff}
Overall, we observe a clear logarithmic relationship between cost and mutation score $\mathcal{M}$ with more recent frontier models achieving the best performance at the highest prices. All evaluated models except for Gemini 3.8 Flash lie on the pareto frontier.

\paragraph{Reasoning Effort}
Ablating over reasoning effort for GPT 5.6 Sol, we see the same trend continue. A higher reasoning effort increases both cost and mutation score \ms from 31.3\% at \$0.45 per instance on low reasoning to 46.5\% at \$2.4 per instance on xhigh reasoning.

\subsection{Additional Analysis}
\label{subsec:analysis}

\paragraph{Model Test Suites are (Too) Small}
We investigate the observed low mutation scores comparing the number of generated tests with those in the reference test suites in  \cref{fig:correlation_numtest}.
While all models, in expectation generate smaller test suites than the reference, the gap ($\Delta = -8.9\%$) is smallest for Opus with Sol ($\Delta = -79.9\%$) producing roughly 5x smaller test suites than the reference. This correlation is unsurprising as a high mutation score \ms hinges on exhaustively covering all requirements, which generally requires a large number of tests.
We observe a general trend that (i) instances with fewer reference tests (and thus fewer requirements to cover), and (ii) a larger number of generated tests lead to a better $\mathcal{M}$.
We provide a more comprehensive analysis in \cref{appsubsec:full_correlation_numtest}.

\paragraph{Agents Struggle with Large Specifications}
In \cref{fig:correlation_speclength} (full results in \cref{appsubsec:full_correlation_speclength}), we visualize the relationship between the number of specification sections and test generation performance ($\mathcal{M}$).
Across all models, we observe a negative correlation (Opus: $\rho = -0.16$, Sol: $\rho = -0.42$), indicating that models struggle to generate strong test suites for large requirement sets.

\input{src/figure/correlation/correlation.tex}

\paragraph{Executable/Mutant Creation during Test Generation}
\input{src/table/self_mutation.tex}

We have observed some agents to create implementations and mutants to help test suite generation, without them being instructed to do so.
Here we investigate whether such an approach improves the resulting test suites.
We design a rule-based parser to detect whether the test generation trajectory contains code generation attempts (\eg write \texttt{executable}), or mutation attempts (\eg shell action mentioning mutation).

As shown in \cref{tab:self_mutation}, Opus creates both executables and mutants most frequently, agreeing well with its strong mutation score $\mathcal{M}$.
Yet, even Opus creates mutants in only 33 of 165 instances, indicating potential for mutant based test suite generation methods.
Interestingly, however, mutant and executable generation do not seem to be necessary to generate strong test suites; Sol creates by far the fewest executables and no mutants, yet still achieves better $\mathcal{M}$ than DeepSeek.
This allows Sol to create complete test suites in much fewer steps than any other model (\cref{fig:distribution_cumulative}).

\paragraph{Error Analysis}

\input{src/figure/error_analysis/ast_grep_opus_only/multipath_mutant_main.tex}
To provide a more in depth understanding of common failure modes, we show a small case study in \cref{fig:error_analysis_main}. We consider a mutant from the \texttt{ast-grep} instance, where the specification mentions that \texttt{run} (the pattern-search command) accepts multiple files.
The mutant breaks this contract, ignoring the second item, \texttt{b.js}.
While the Opus-generated test suite captures this requirement, generating a test that provides two files and successfully identifies the mutation, the Sol-generated test suite tests \texttt{run} only with a single input file and therefore fails to kill the mutant.
We observe this to be a common pattern where core functionality is tested but input combinations and edge cases are omitted from test suites.
We provide additional examples in \cref{appsubsec:full_error_analysis}.

\paragraph{Providing Reference Executables for Test Generation}
\input{src/table/code_as_input.tex}
We assess whether providing the reference executable in addition to the specification and thus disambiguating any potential remaining uncertainties improves performance using GPT 5.6 Sol and showing results in \cref{tab:code_as_input}.
As expected, the validity rate $\mathcal{V}$ increases even further from $94.4\%$ to $98.4\%$ as the model can directly inspect the reference executable when constructing its tests.
In contrast, however, the mutation score barely moves, from $37.7\%$ to $38.1\%$.
This indicates that a lack of clarity in the specification that could make the model hesitate to add invalid tests does not explain the low mutation scores observed across this analysis.

\paragraph{Comparison with Developer-Written Documentation}
\input{src/table/developer_spec.tex}
In our benchmark, instead of relying on developer-written documentation that is often brief and incomplete, we generate rich specifications.
To quantify this difference, we subsample 30 instances that provide extensive developer documentation beyond their README.
Replacing our specifications with this documentation, Sol's mutation score drops from $25.5\%$ to $4.0\%$ (\cref{tab:developer_spec}), as it writes $40\%$ fewer tests ($2{,}274$ vs.\ $3{,}789$), of which fewer are valid ($75.4\%$ vs.\ $86.7\%$).
Classifying the failing tests as in \cref{subsec:iterative_spec}, we observe the original documentation induces three times as many underspecification failures per test.

%% file: src/table/main.tex
\begin{table}[t]
    \centering
    \begin{minipage}[t]{0.54\linewidth}
        \centering
        \caption{
            Frontier agents struggle to achieve $\mathcal{M}$ beyond $56\%$, even when $\mathcal{V}$ and $\mathcal{R}$ are almost $100\%$.
        }
        \vspace{-2mm}
        \label{tab:main-results}
        \setlength{\tabcolsep}{3pt}
        \resizebox{0.95\linewidth}{!}{%
        \input{src/table/main-rows.tex}        }
    \end{minipage}\hfill
    \begin{minipage}[t]{0.43\linewidth}
        \centering
        \label{fig:pareto}
        \includegraphics[width=0.99\linewidth]{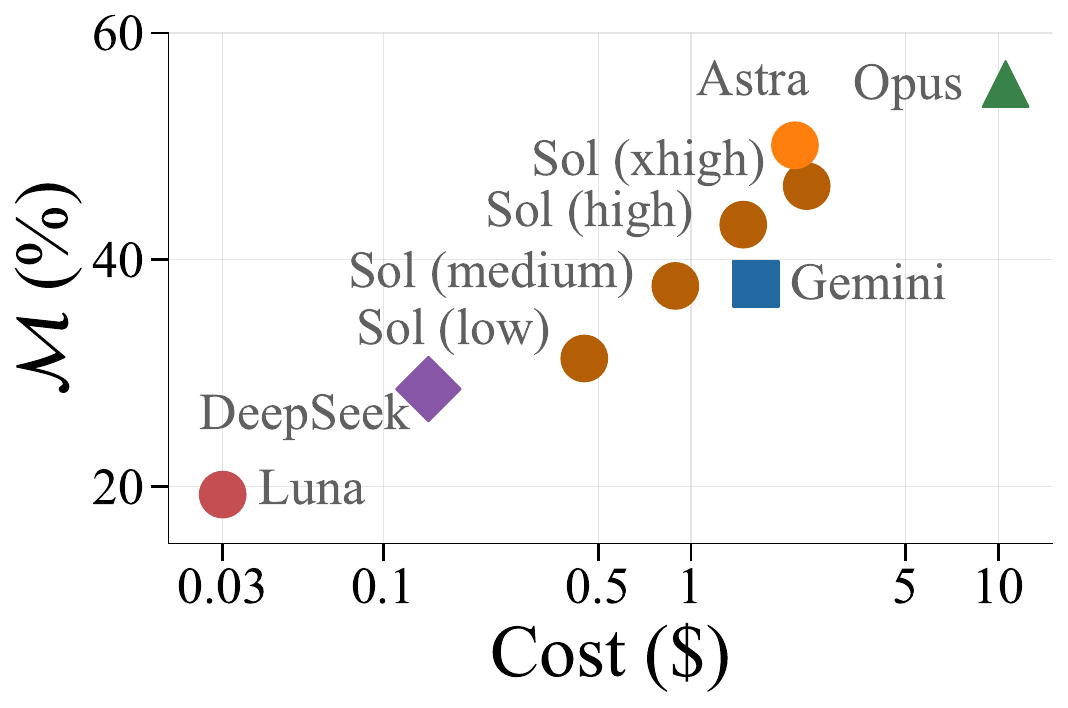}
        \vspace{-7mm}
        \captionof{figure}{
            Increased cost provides marginal gains for $\mathcal{M}$.
        }
    \end{minipage}
\end{table}

%% file: src/table/main-rows.tex
\begin{tabular}{@{}lccccc@{}}
\toprule
& \multicolumn{4}{c}{Test Generation} & \\
\cmidrule(lr){2-5}
Model & {\boldmath$\mathcal{M}$} & $\mathcal{V}$ & \$ & $t$ (min) & $\mathcal{R}$ \\
\midrule
Claude Opus 5 & \textbf{55.5} & \textcolor{black!55}{\textbf{96.1}} & \textcolor{black!55}{10.56} & \textcolor{black!55}{37.0} & \textcolor{black!55}{91.8} \\
GPT 6 Astra & 50.1 & \textcolor{black!55}{94.3} & \textcolor{black!55}{2.18} & \textcolor{black!55}{12.4} & \textcolor{black!55}{\textbf{92.7}} \\
Gemini 3.8 Flash & 37.9 & \textcolor{black!55}{88.2} & \textcolor{black!55}{1.63} & \textcolor{black!55}{9.3} & \textcolor{black!55}{86.2} \\
GPT 5.6 Sol & 37.7 & \textcolor{black!55}{94.4} & \textcolor{black!55}{0.89} & \textcolor{black!55}{8.2} & \textcolor{black!55}{86.9} \\
DeepSeek V4 Flash & 28.6 & \textcolor{black!55}{78.1} & \textcolor{black!55}{0.14} & \textcolor{black!55}{19.0} & \textcolor{black!55}{66.8} \\
GPT 5.6 Luna & 19.3 & \textcolor{black!55}{74.1} & \textcolor{black!55}{\textbf{0.03}} & \textcolor{black!55}{\textbf{4.0}} & \textcolor{black!55}{65.4} \\
\bottomrule
\end{tabular}

%% file: src/figure/correlation/correlation.tex
\begin{figure}[t]
    \centering
    \begin{subfigure}[t]{0.48\linewidth}
        \centering
        \includegraphics[width=\linewidth]{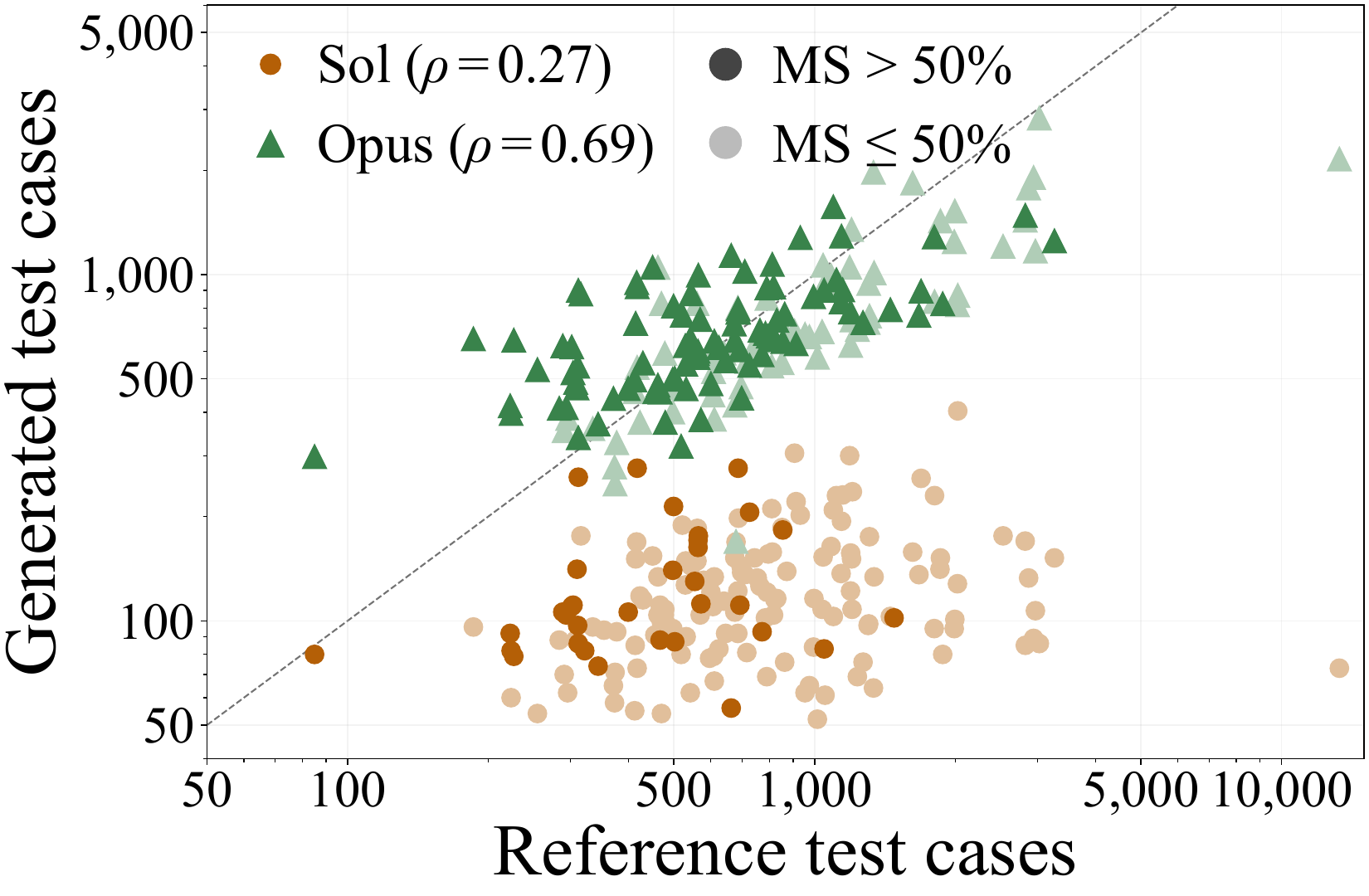}
        \caption{Reference and generated test counts.}
        \label{fig:correlation_numtest}
    \end{subfigure}\hfill
    \begin{subfigure}[t]{0.48\linewidth}
        \centering
        \includegraphics[width=\linewidth]{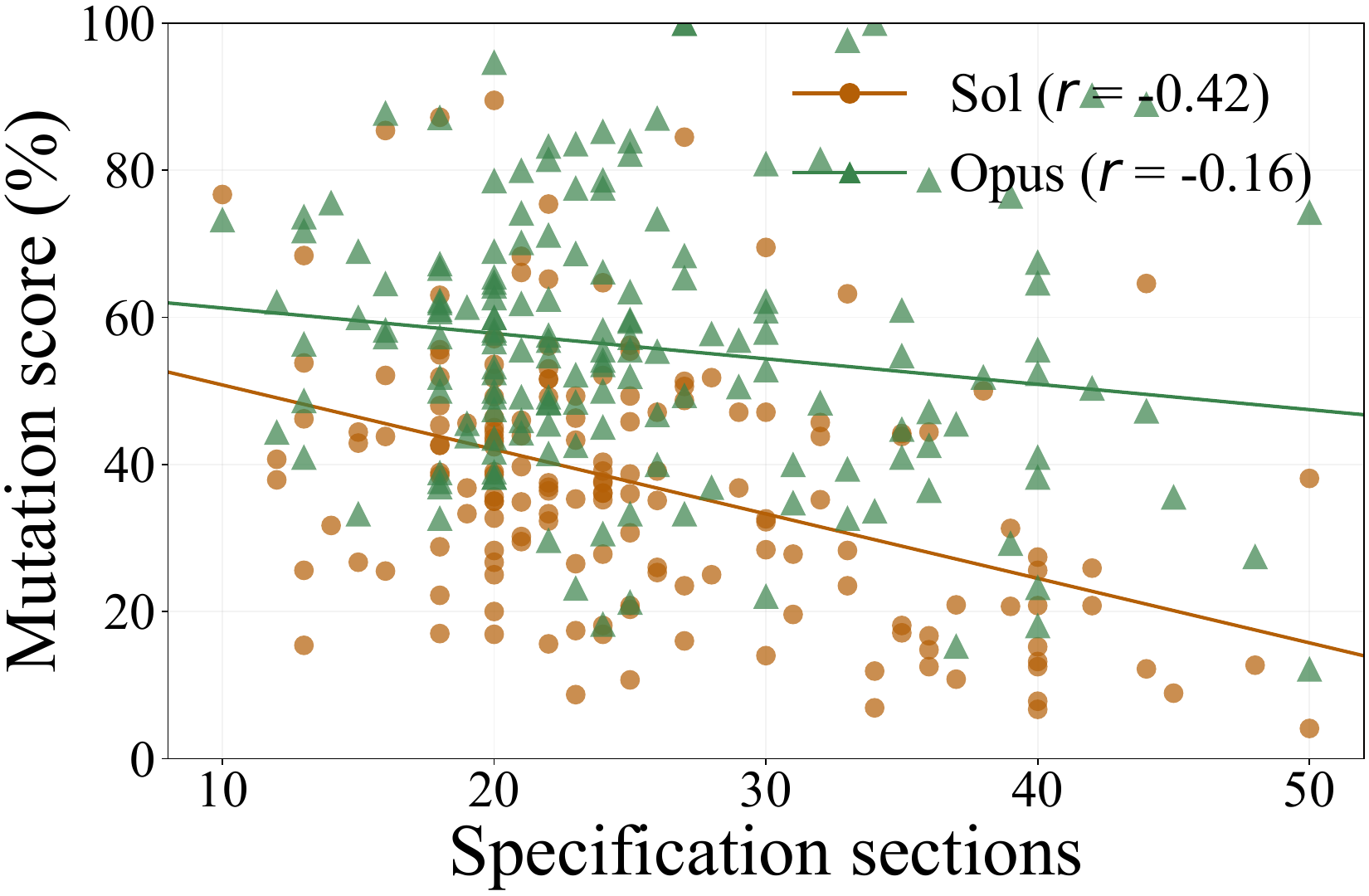}
        \caption{Specification length and mutation score.}
        \label{fig:correlation_speclength}
    \end{subfigure}
    \caption{
        \textbf{Left:} Opus generates more tests than Sol with stronger correlation to the reference test count.
        \textbf{Right:} Both models struggle to generate strong test suites for larger specifications.
    }
    \label{fig:correlation}
\end{figure}

%% file: src/table/self_mutation.tex
\begin{wraptable}[12]{r}{0.4\linewidth}
    \centering
    \vspace{-1em}
    \caption{
        Executable and mutant creation attempt detected in test generation trajectories.
        Opus has the highest signal, which could explain its highest $\mathcal{M}$.
    }
    \label{tab:self_mutation}
    \setlength{\tabcolsep}{4pt}
    \resizebox{0.95\linewidth}{!}{
    \begin{tabular}{@{}lrrr@{}}
        \toprule
        Model & Exec. & Mutants & $\mathcal{M}$ \\
        \midrule
        Opus & 149 & 33 & \textcolor{black!55}{55.5} \\
        Gemini & 102 & 12 & \textcolor{black!55}{37.9} \\
        Sol & 4 & 0 & \textcolor{black!55}{37.7} \\
        DeepSeek & 58 & 4 & \textcolor{black!55}{28.6} \\
        \bottomrule
    \end{tabular}
    }
\end{wraptable}

%% file: src/figure/error_analysis/ast_grep_opus_only/multipath_mutant_main.tex
\begin{figure*}[t]
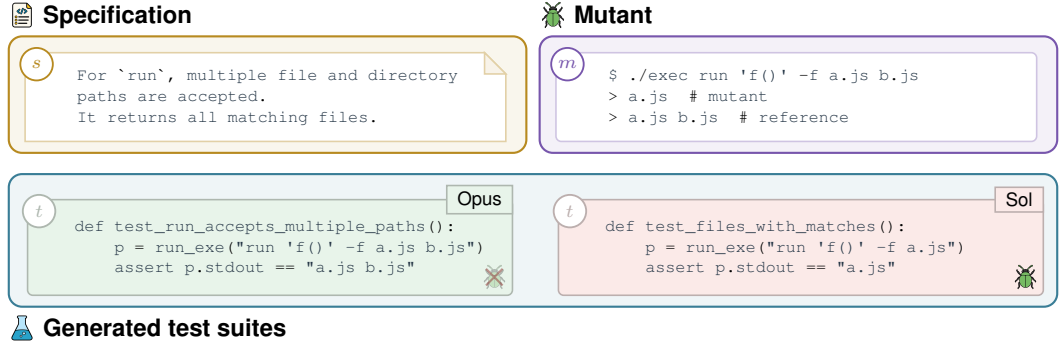

    \centering
    \begingroup
    \definecolor{tbInk}{HTML}{263747}
    \definecolor{tbMuted}{HTML}{637382}
    \definecolor{tbGold}{HTML}{B78A20}
    \definecolor{tbBlue}{HTML}{397D96}
    \definecolor{tbPurple}{HTML}{7756AC}
    \definecolor{tbGreen}{HTML}{43815B}
    \definecolor{tbRed}{HTML}{BA554F}
    \definecolor{tbOpusTest}{HTML}{E8F4EA}
    \definecolor{tbSolTest}{HTML}{FBEAE8}

    \lstset{
        basicstyle=\fontfamily{pcr}\fontsize{6.3}{7.8}\selectfont,
        columns=fullflexible,keepspaces=true,showstringspaces=false,
        mathescape=false,escapechar=,numbers=none,
        keywordstyle=\color{tbInk},stringstyle=\color{tbInk},
        commentstyle=\color{tbMuted},identifierstyle=\color{tbInk},
        breaklines=true,breakatwhitespace=false,
        frame=none,aboveskip=0pt,belowskip=0pt
    }
    \newsavebox{\specbox}
    \newsavebox{\mutantbox}
    \newsavebox{\opustestbox}
    \newsavebox{\soltestbox}
    \begin{lrbox}{\specbox}
    \begin{minipage}{.42\textwidth}
\begin{lstlisting}[breakindent=0pt,breakautoindent=false]
For `run`, multiple file and directory paths are accepted.
It returns all matching files.
\end{lstlisting}
    \end{minipage}
    \end{lrbox}
    \begin{lrbox}{\mutantbox}
    \begin{minipage}{.42\textwidth}
\begin{lstlisting}
$ ./exec run 'f()' -f a.js b.js
> a.js  # mutant
> a.js b.js  # reference
\end{lstlisting}
    \end{minipage}
    \end{lrbox}
    \begin{lrbox}{\opustestbox}
    \begin{minipage}{.434\textwidth}
\begin{lstlisting}[language=Python]
def test_run_accepts_multiple_paths():
    p = run_exe("run 'f()' -f a.js b.js")
    assert p.stdout == "a.js b.js"
\end{lstlisting}
    \end{minipage}
    \end{lrbox}
    \begin{lrbox}{\soltestbox}
    \begin{minipage}{.434\textwidth}
\begin{lstlisting}[language=Python]
def test_files_with_matches():
    p = run_exe("run 'f()' -f a.js")
    assert p.stdout == "a.js"
\end{lstlisting}
    \end{minipage}
    \end{lrbox}
    \tikzset{
        example snippet/.style={inner xsep=7pt,inner ysep=5pt,outer sep=0pt,
            anchor=north west,line width=.6pt},
        example badge/.style={circle,fill=white,line width=.7pt,
            minimum size=.43cm,inner sep=0pt,outer sep=0pt,
            font=\fontsize{7}{8}\selectfont}
    }
    \newcommand{\examplesnippet}[3]{%
        \begin{tikzpicture}
            \node[example snippet,draw=#1!35,fill=white,rounded corners=2pt]
                (snippet) {\usebox{#2}};
            \node[example badge,draw=#1,text=#1] at (snippet.north west) {$#3$};
        \end{tikzpicture}%
    }
    \newlength{\examplecolumnwidth}
    \setlength{\examplecolumnwidth}{.47\textwidth}
    \newlength{\examplefullwidth}
    \setlength{\examplefullwidth}{\dimexpr .94\textwidth+13pt\relax}
    \newcommand{\exampletest}[4]{%
        \begin{tikzpicture}[baseline=(snippet.north)]
            \node[example snippet,inner xsep=5pt,minimum height=36pt,draw=#4!75!black,fill=#4,rounded corners=2pt]
                (snippet) {\usebox{#2}};
            \node[anchor=east,inner xsep=4pt,inner ysep=2pt,text=black,align=center,
                draw=#4!75!black,fill=#4,sharp corners,line width=.6pt,
                font=\fontfamily{phv}\fontsize{7}{8}\mdseries\selectfont]
                at (snippet.north east) {#1};
            \node[example badge,draw=#4!75!black,text=#4!75!black] at (snippet.north west) {$t$};
            \node[anchor=south east,inner sep=2pt,opacity=#3] (bug)
                at (snippet.south east) {\includegraphics[height=.34cm]{src/figure/assets/mutant.png}};
            \ifdim #3pt<1pt
                \draw[sharp corners,draw=tbRed!80!black,line width=.8pt,opacity=.55]
                    ([xshift=-.11cm,yshift=-.11cm]bug.center) -- ([xshift=.11cm,yshift=.11cm]bug.center)
                    ([xshift=-.11cm,yshift=.11cm]bug.center) -- ([xshift=.11cm,yshift=-.11cm]bug.center);
            \fi
        \end{tikzpicture}%
    }
    \begin{tikzpicture}[
        every node/.style={inner sep=4pt,outer sep=0pt},
        card/.style={rounded corners=5pt,line width=.8pt,anchor=north west},
        font=\fontfamily{phv}\footnotesize\selectfont,text=tbInk
    ]
        \node[card,draw=tbGold,fill=tbGold!8] (spec) at (0,0) {%
            \begin{minipage}{\examplecolumnwidth}
                \begin{tikzpicture}
                    \node[example snippet,inner ysep=5pt] (snippet) {\usebox{\specbox}};
                    \draw[sharp corners,draw=tbGold!40,fill=white,line width=.6pt]
                        (snippet.south west) -- (snippet.south east)
                        -- ([yshift=-8pt]snippet.north east)
                        -- ([xshift=-8pt]snippet.north east)
                        -- (snippet.north west) -- cycle;
                    \draw[sharp corners,draw=tbGold!40,fill=tbGold!10,line width=.6pt]
                        ([xshift=-8pt]snippet.north east)
                        -- ([xshift=-8pt,yshift=-8pt]snippet.north east)
                        -- ([yshift=-8pt]snippet.north east) -- cycle;
                    \node[example snippet,inner ysep=5pt] at (snippet.north west) {\usebox{\specbox}};
                    \node[example badge,draw=tbGold,text=tbGold] at (snippet.north west) {$s$};
                \end{tikzpicture}
            \end{minipage}
        };
        \node[card,draw=tbPurple,fill=tbPurple!8] (mutant)
            at ([xshift=5pt]spec.north east) {%
            \begin{minipage}{\examplecolumnwidth}
                \examplesnippet{tbPurple}{\mutantbox}{m}
            \end{minipage}
        };
        \node[card,draw=tbBlue,fill=tbBlue!8] (tests)
            at ([yshift=-8pt]spec.south west) {%
            \begin{minipage}{\examplefullwidth}
                \begin{minipage}[t]{\examplecolumnwidth}
                    \centering\exampletest{Opus}{\opustestbox}{.30}{tbOpusTest}
                \end{minipage}\hfill%
                \begin{minipage}[t]{\examplecolumnwidth}
                    \centering\exampletest{Sol}{\soltestbox}{1}{tbSolTest}
                \end{minipage}
            \end{minipage}
        };
        \foreach \card/\asset/\heading in {
            spec/document/Specification,
            mutant/mutant/Mutant} {
            \node[anchor=south west,inner sep=0pt,text=black,font=\fontfamily{phv}\footnotesize\bfseries\selectfont]
                at ([yshift=3pt] \card.north west) {%
                    \raisebox{-.2\height}{\includegraphics[height=.35cm]{src/figure/assets/\asset.png}}\hspace{3pt}\heading};
        }
        \node[anchor=north west,inner sep=0pt,text=black,font=\fontfamily{phv}\footnotesize\bfseries\selectfont]
            at ([yshift=-3pt]tests.south west) {%
                \raisebox{-.2\height}{\includegraphics[height=.35cm]{src/figure/assets/flask.png}}\hspace{3pt}Generated test suites};
    \end{tikzpicture}
    \endgroup
    \caption{
        The specification requires \texttt{run} to accept multiple paths.
        The mutant ignores the second path, \texttt{b.js}.
        The Opus test checks both paths and kills the mutant.
        In contrast, the Sol test only checks one path and misses the mutation.
        Complete version of error analysis is provided in \cref{appsubsec:full_error_analysis}.
    }
    \label{fig:error_analysis_main}
\end{figure*}

%% file: src/table/code_as_input.tex
\begin{wraptable}[11]{r}{0.35\linewidth}
    \vspace{-4mm}
    \centering
    \caption{
        Input ablation of test generation (Sol).
        Even with the access to the reference executable, mutation score remains low.
    }
    \label{tab:code_as_input}
    \vspace{-2mm}
    \setlength{\tabcolsep}{3pt}
    \resizebox{0.9\linewidth}{!}{
    \begin{tabular}{@{}lrr@{}}
        \toprule
        Setting & $\mathcal{V}$ & $\mathcal{M}$ \\
        \midrule
        Spec only & 94.4\% & 37.7\% \\
        + Executable    & 98.4\% & 38.1\% \\
        \bottomrule
    \end{tabular}
    }
\end{wraptable}

%% file: src/table/developer_spec.tex
\begin{wraptable}[10]{r}{0.35\linewidth}
    \vspace{-4mm}
    \centering
    \small
    \caption{
        Performance with the developer specification (Sol).
        30 instances with the official documentation are subsampled.
    }
    \vspace{-2mm}
    \label{tab:developer_spec}
    \setlength{\tabcolsep}{3pt}
    \resizebox{0.85\linewidth}{!}{
    \begin{tabular}{@{}lrr@{}}
        \toprule
        Setting & $\mathcal{V}$ & $\mathcal{M}$ \\
        \midrule
        Official Docs & 75.4\% & 4.0\% \\
        Our Spec & 86.7\% & 25.5\% \\
        \bottomrule
    \end{tabular}
    }
\end{wraptable}

%% file: src/related_work.tex
\section{Related Work}
\label{sec:related}

\paragraph{Repository-Level Coding Benchmarks}
Recently multiple coding benchmarks task agents with implementing full projects from scratch.
NL2Repo-Bench \citep{ding2025nl2repo} provides agents with an extremely detailed specification and ProgramBench \citep{yang2026programbench} with reference executable. Both task the agent with (re)implementing the program.
In contrast, we measure the ability to generate a test suite given only the specification.

\input{src/table/related_work.tex}

\paragraph{Test Generation Benchmarks}
SWT-Bench \citep{mundler2024swtbench} measures whether agents can add regression tests reproducing user-reported bugs.
SWE-Mutation \citep{sun2026swemutation} evaluates issue-directed test generation and program repair using both bug reproduction and mutation testing.
TestGenEval \citep{jain2025testgeneval} evaluates tests generation for individual source files, measuring test validity, coverage, and mutation scores.
MultiFileTest \citep{wang2026multifiletest} evaluates test generation from source code for 59 released projects from 21 repositories across Python, Java, and JavaScript, primarily through direct prompting and error-fixing loops; it also includes a four-project agent case study.
RESTestBench \citep{kogler2026restestbench} evaluates individual test generation for specified scenarios against three REST services using manually constructed mutants.

We compare our work against these benchmarks in \cref{tab:related-work} and highlight two crucial differences. i) prior work generally provides access to an implementation or executable and ii) prior work restricts test generation by focusing on specific behaviors, individual files, or concrete issues. As a result, the golden reference test suite size per task ($|T|/N$) and thus the expected size of generated test suites is orders of magnitudes smaller than ours.

%% file: src/table/related_work.tex
\begin{table}[t]
    \centering
    \caption{Comparison to prior work. Inputs are implementations $I$, executables $E$, and specifications $S$. We distinguish whether the evaluation employs mutation score $\mathcal{M}$ and validity rate $\mathcal{V}$, and whether the evaluation target is mainly coding agents or LLMs.
    }
    \label{tab:related-work}
    \vspace{-3mm}
    \small
    \setlength{\tabcolsep}{4pt}
    \renewcommand{\arraystretch}{1.1}
    \begin{tabularx}{\textwidth}{@{}>{\raggedright\arraybackslash}Xccccccc@{}}
        \toprule
        Benchmark  & \# Repos. & Scope & Inputs & $|T|/N$ &  $\mathcal{M}$ &$ \mathcal{V}$ & Agents \\
        \midrule
        SWT-Bench \citep{mundler2024swtbench}
        & 12 & Issue & I & 3 & \xmark & \cmark & \cmark \\
        TestGenEval \citep{jain2025testgeneval}
        & 11 & File & I & 57 & \cmark & \cmark & \xmark \\
        MultiFileTest \citep{wang2026multifiletest}
        & 21 & Few-File & I & -- & \xmark & \cmark & \xmark \\
        SWE-Mutation \citep{sun2026swemutation}
        & 53 & Issue & I & 62 & \cmark & \cmark & \cmark \\
        RESTestBench \citep{kogler2026restestbench}
        & 3 & Scenario & S, E & 1 & \cmark & \cmark & \xmark \\
        \midrule
        \benchmark (ours)
        & 165 & Repository & S & 944 & \cmark & \cmark & \cmark \\
        \bottomrule
    \end{tabularx}
\end{table}

%% file: src/discussion.tex
\section{Limitations and Future Work}
\label{sec:limitations}

\paragraph{Mutants as a Proxy for Test Strength}
To keep construction and evaluation tractable, our mutants target a sample of requirements, selected uniformly across the specification (\cref{subsec:mutant_construction}).
Gaps in untargeted requirements thus go unpenalized, but since agents cannot know which requirements are targeted, only suites covering the whole specification achieve a high expected score.
Further, selecting mutants that are killed by a single reference test risks favoring mutants triggered only by that test's exact inputs.
We mitigate this by instructing the mutation agent to mutate the logic implementing a requirement and removing mutants that hard-code behavior for specific inputs in a manual and agentic audit (\cref{subsec:manual_check}).
Finally, each mutant violates a single requirement by design and need not reflect the distribution of real-world faults.

\paragraph{Specification Realism}
We derive specifications from reference test suites and refine them until they suffice for test and code generation, making them far more detailed than typical developer documentation, with which mutation scores increase from $4.0\%$ to $25.5\%$ (\cref{tab:developer_spec}).
However, specifications sufficient to build software from scratch, such as product requirement documents, must be similarly detailed, making ours a reasonable proxy for this setting.
Importantly, having access to such detailed specifications provides a potentially more favorable condition than those encountered in practice, suggesting that our evaluation may still overestimate agents' test-generation performance.

\paragraph{Reference Behavior as Ground Truth}
Our construction process aligns specifications, reference tests, and mutants with the reference implementation, so all of them inherit its bugs and incidental behavior.
We manually removed obvious cases, such as crashes expected at specific source lines (\cref{appsubsec:construction_manual_check}), but subtler ones may persist.
We accept this trade-off, as grounding all artifacts in an existing implementation is what makes our construction scale.

\paragraph{Scope}
Building on \programbench, \benchmark is limited to command-line applications in compiled languages, as \programbench provides agents the reference executable, which would expose the source code for interpreted languages.
In contrast, \benchmark provides only the specification.
Therefore, our pipeline can be extended to interpreted languages such as Python and TypeScript in future work.

%% file: src/conclusion.tex
\section{Conclusion}
\label{sec:conclusion}

We introduced \benchmark, a benchmark of 165 open-source applications that measures whether coding agents can turn a natural-language specification into a strong test suite, using mutants that each violate a single requirement.
We find that frontier agents can build software they cannot verify: their tests are largely valid and their implementations pass most reference tests, yet even the strongest agent's suites kill just over half of the mutants.
Giving agents the reference executable leaves mutation scores unchanged, pointing to test design rather than specification quality as the bottleneck: generated suites are often too small and miss the input combinations needed to expose violated requirements.
Closing this gap is a prerequisite for fully autonomous development of new software, and \benchmark makes progress toward it measurable.

\clearpage
\section*{Reproducibility Statement}
We describe our benchmark construction process in detail in \cref{sec:construction} and our evaluation methodology in \cref{sec:experiment}. We also provide the full prompts used for construction and evaluation in \cref{appsec:prompt}. The full benchmark data and the code required to reproduce it will be released upon publication.

\section*{AI Usage Statement}
In this work, we used generative AI tools to generate synthetic data and to clean datasets: coding agents (Codex with GPT-5.6 Sol) construct and refine the reference test suites, specifications, and mutants of \benchmark, as described in \cref{sec:construction}.
We have not used generative AI tools to develop conceptual frameworks, propose or refine hypotheses, design or provide feedback on research methodology or experiments, implement methods, or interpret results, and the remaining required disclosure tasks (formulating mathematical claims, proofs, translation, and qualitative or thematic data analysis) are not applicable to this work.
Additionally, we used generative AI tools to create and modify figures, to draft parts of the paper, and to polish the writing for readability.
We have reviewed all AI-assisted work: agent-generated benchmark artifacts are validated by the automated checks of our construction pipeline and a manual quality check (\cref{subsec:manual_check}), and all AI-assisted text and figures were reviewed and edited by the authors.
We take responsibility for the final content of this work, including text, claims, and artifacts produced with the aid of generative AI.

%% file: appendix.tex
\input{src/appendix/more_results.tex}
\input{src/appendix/construction_detail.tex}
\input{src/appendix/eval_detail.tex}
\input{src/appendix/prompt.tex}

\input{src/appendix/task_list.tex}

%% file: src/appendix/more_results.tex
\section{Additional Experimental Results}
\label{appsec:more_results}

\subsection{Correlation between Reference and Generated Test Suite}
\label{appsubsec:full_correlation_numtest}

\input{src/figure/correlation/correlation_test_counts_full.tex}

In \cref{fig:correlation_numtest_full}, we provide a full version of \cref{fig:correlation_numtest}, \ie correlation analysis between the number of reference tests and the number of generated tests.
We observe a general trend that the number of generated tests is generally lower than the reference test suite, with an exception of Opus generating similar number of tests as the reference test suite, and that the $\mathcal{M}$ is generally higher when the reference suite is smaller, and generated tests is closer to the reference test suite.

\subsection{Correlation between Spec Length and Mutation Score}
\label{appsubsec:full_correlation_speclength}

\input{src/figure/correlation/correlation_speclength_full.tex}

In \cref{fig:correlation_speclength_full}, we provide a full version of \cref{fig:correlation_speclength}, \ie correlation analysis between the spec length and the mutation score.
Across all agents, we observe a weak negative correlation between the spec length and the mutation score, indicating that longer specifications with more specified requirements makes it more challenging for agents to generate strong test suites.

\subsection{Score and Cost Distribution}
\label{appsubsec:full_cost_score_distribution}

\input{src/figure/distribution/density.tex}

\cref{fig:distribution_cumulative} shows the distribution of Reference Pass Rate ($\mathcal{R}$) for code generation, Validity Rate ($\mathcal{V}$) and Mutation Score ($\mathcal{M}$) for test generation, along with the cost, time, and number of steps for each setting.

\subsection{Tool Use Over Trajectories}
\label{appsubsec:full_tool_use_trajectories}

\input{src/figure/tool_trajectory/tool_trajectory.tex}

In \cref{fig:tool_use_distribution}, we show the distribution of tool use over the trajectory of each model, by classifying the actions into \texttt{read} (reading file contents, such as \texttt{cat}), \texttt{search} (finding files or text, such as \texttt{ls, grep}),\texttt{write} (create or edit files), \texttt{execute} (run tests or other shell commands), and \texttt{other}.

We observe a general trend that first few steps are dominated by \texttt{read} and \texttt{search}, and majority of the later sections are dominated by \texttt{write} and \texttt{execute}.
This is indicates that the model quickly digests the provided specification (usually a single file) within a few steps, and  they start writing while executing it to check the correctness of their written code.
Overall, \texttt{write} dominates the trajectory, with Opus having 60\% of its actions being \texttt{write}.

\subsection{More Ablation}
\label{appsubsec:full_ablation}

\input{src/figure/mutant_ben_diagram/mutant_ben_diagram.tex}

\paragraph{Overlap of Killed Mutants}
\label{appsubsec:full_overlap_killed_mutants}
We consider three representative models (Opus, Sol, Gemini) and analyze the overlap in the mutants they kill (\cref{fig:killed_mutant_overlap}).
Out of 12{,}130 mutants, only 3{,}080 ($25\%$) are killed by all three models, and the two weaker models each kill over 200 mutants that neither other model kills.
Combining the three suites increases the number of killed mutants from 6{,}584 ($54\%$) for Opus alone to 7{,}134 ($59\%$), leaving 4{,}996 mutants ($41\%$) that no model kills.
Combining agents thus yields only modest gains, and we leave more effective ways to combine them to future work.

\paragraph{Realism of the Mutation}
We aim to create mutants that breaks a specific requirement, and, as a verifiable proxy for that, we aim to create mutants that are killed only by a single test in the reference suite.
As we discussed in \cref{sec:limitations}, this approach risks creating mutants that are unrealistically niche (\eg mutants that only fail under a specific input string that are used for \texttt{pytest.mark.parametrize}).
We mitigate by instructing the mutation agent to mutate the logic.
Here, we ablate the effect of this mitigation by comparing the mutants created with and without such instruction.

\input{src/figure/ablation_mutation_prompt/literal_copy_examples.tex}

\Cref{fig:literal_copy} shows two pairs of mutants generated for the same target test, one with and one without the instruction.
All four mutants are killed only by their target test, so they are indistinguishable with respect to the single-test criterion.
Without the instruction, however, the agent copies concrete values from the target test into the mutant.
In \cref{fig:literal_copy_hostctl}, the mutant fails only when the three domains are exactly \texttt{site1.test}, \texttt{site2.test}, and \texttt{site3.test}; in \cref{fig:literal_copy_skeema}, it exits early only when the configuration contains \texttt{host=localhost:3307} and \texttt{port=3308}.
Renaming the domains or changing the port numbers exercises the same requirement, yet this mutant then behaves like the reference.
With the instruction, the mutants instead break the behavior for such inputs as well: \texttt{hostctl} drops all but the first of three domains whatever their names, and \texttt{Skeema} lets the separate port override any conflicting port in the host name.

\subsection{More Error Analysis}
\label{appsubsec:full_error_analysis}

\input{src/figure/error_analysis/full_error_analysis.tex}

\Cref{fig:error_analysis_full} extends the error analysis in \cref{subsec:analysis} with four mutants from \texttt{ast-grep} and \texttt{cmatrix}.

\cref{fig:error_analysis_astgrep_1} is the full version of \cref{fig:error_analysis_main}.
\Cref{fig:error_analysis_cmatrix_all_bold} is another mutant killed only by Opus.
The specification requires \texttt{-B} to render all rain characters in bold, but the mutant aborts after parsing whenever all-bold mode is selected.
Opus starts the renderer with \texttt{-B} and detects the failure. Sol's comparable live rendering test uses \texttt{-a -b -r}, selecting occasional bold (not all-bold) characters.

The other two mutants survive both suites.
In \cref{fig:error_analysis_astgrep_await}, the mutant ignores JavaScript \texttt{await} expressions when \texttt{run} is given no path and thus searches the current directory.
Opus tests \texttt{await} only in Python and tests nested JavaScript only with a different pattern, each on an explicitly given file.
Sol never tests \texttt{await}; its closest test searches deeply nested JavaScript read from standard input.
In \cref{fig:error_analysis_astgrep_elvish}, the mutant generates a Zsh completion script when Elvish is requested.
Opus requests Elvish completions, but checks only that the script is long and mentions commands; the mutated version (\ie a Zsh script) still passes those checks. Sol tests completion help and failed shell inference without validating a generated Elvish script.

%% file: src/figure/correlation/correlation_test_counts_full.tex
\begin{figure}[t]
    \centering

    \includegraphics[width=\linewidth]{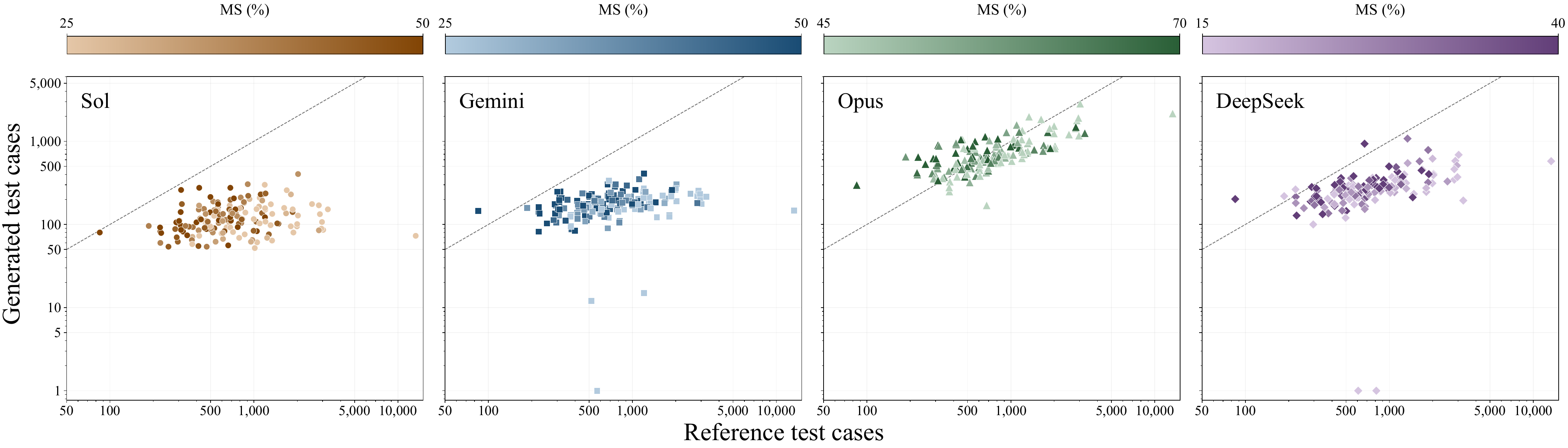}
    \caption{Correlation between number of reference and generated test functions.}
    \label{fig:correlation_numtest_full}
\end{figure}

%% file: src/figure/correlation/correlation_speclength_full.tex
\begin{figure}[t]
    \centering

    \includegraphics[width=\linewidth]{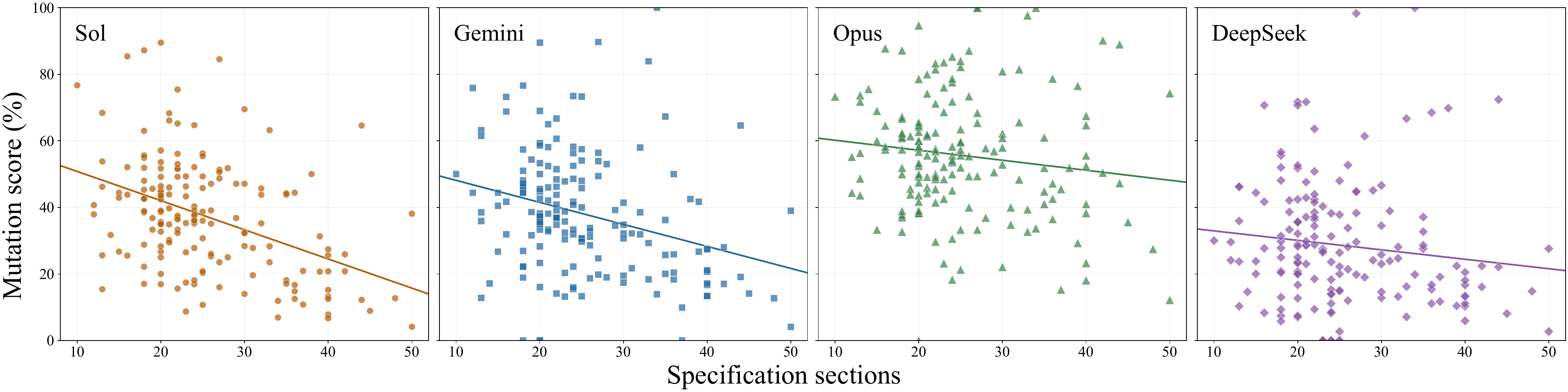}
    \caption{Correlation between the number of spec section and mutation score.}
    \label{fig:correlation_speclength_full}
\end{figure}

%% file: src/figure/distribution/density.tex
\begin{figure}[t]
    \centering
    \includegraphics[width=\linewidth]{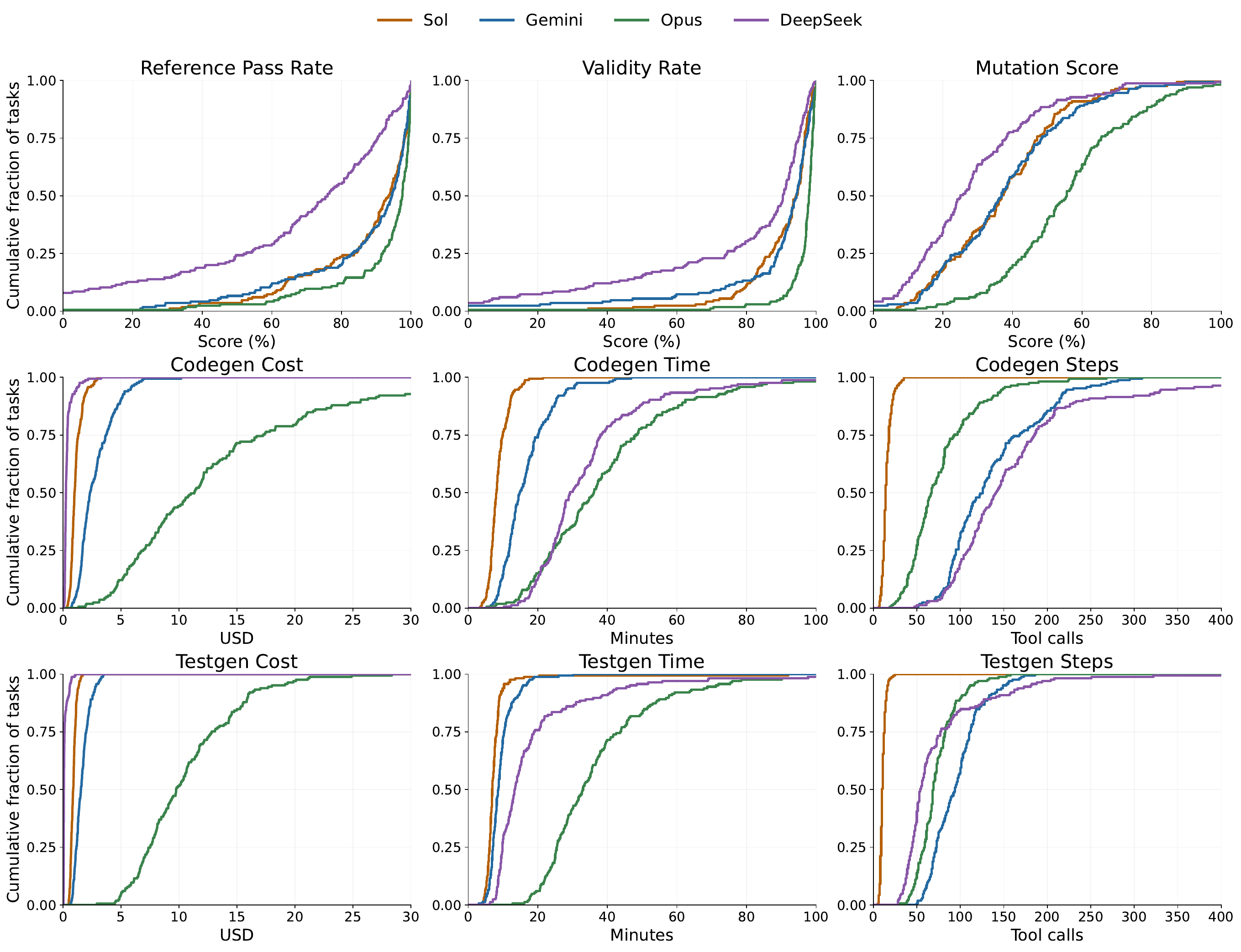}
    \caption{Cumulative distribution of scores and costs across models.}
    \label{fig:distribution_cumulative}
    \vspace{4em}
\end{figure}

%% file: src/figure/tool_trajectory/tool_trajectory.tex
\begin{figure}[t]
    \centering
    \includegraphics[width=\linewidth]{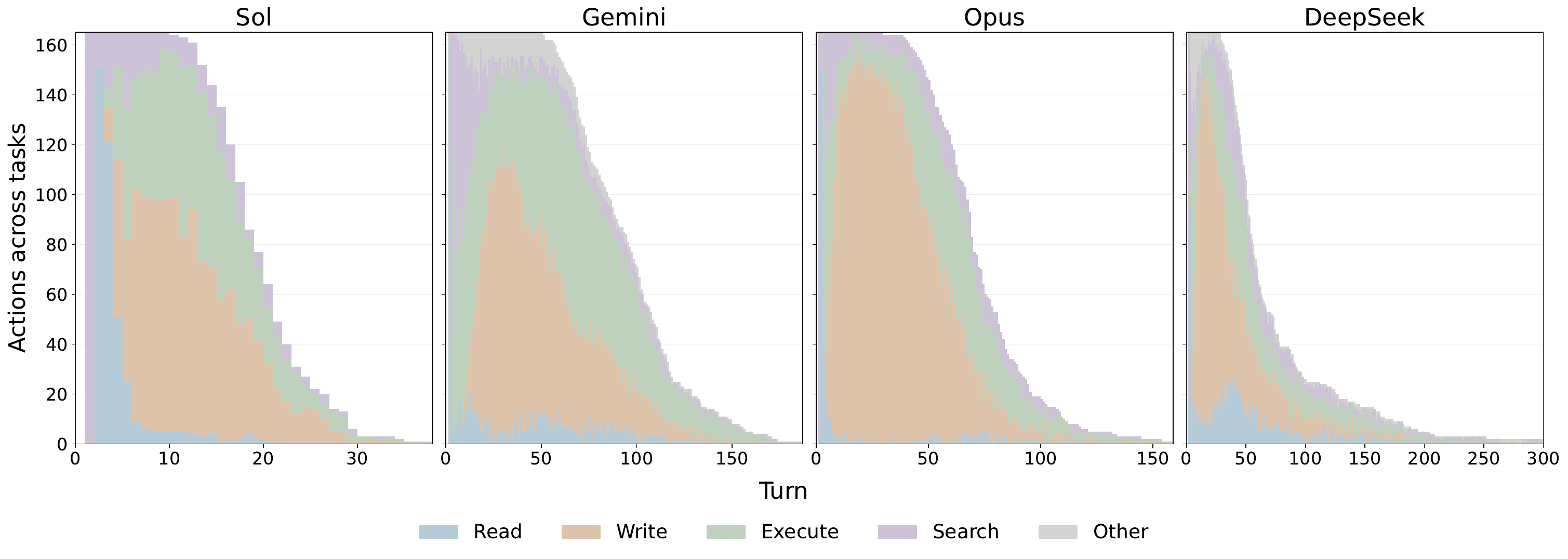}
    \caption{Distribution of tool use across agent trajectories.}
    \label{fig:tool_use_distribution}
    \vspace{3em}
\end{figure}

%% file: src/figure/mutant_ben_diagram/mutant_ben_diagram.tex
\begin{figure}[t]
    \centering
    \includegraphics[width=0.3\linewidth]{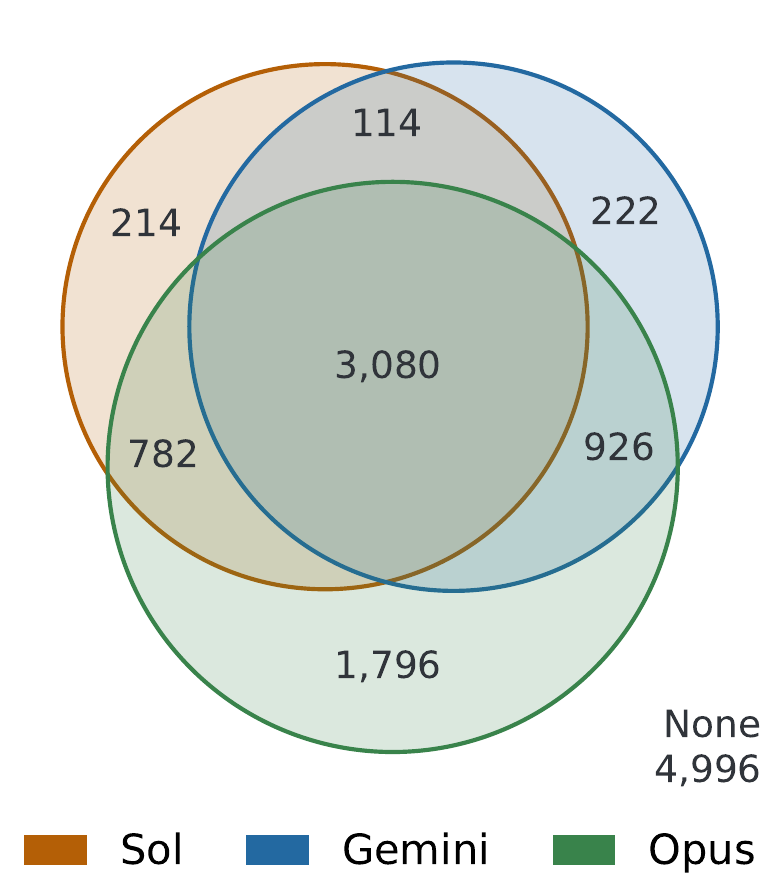}
    \caption{Overlap in mutants killed by test suites generated by different models.}
    \label{fig:killed_mutant_overlap}
\end{figure}

%% file: src/figure/ablation_mutation_prompt/literal_copy_examples.tex
\begin{figure*}[t]
    \centering
    \begingroup
    \definecolor{tbInk}{HTML}{263747}
    \definecolor{tbMuted}{HTML}{637382}
    \definecolor{tbBlue}{HTML}{397D96}
    \definecolor{tbPurple}{HTML}{7756AC}
    \definecolor{tbRed}{HTML}{BA554F}
    \definecolor{tbLogicMut}{HTML}{E8F4EA}
    \definecolor{tbLiteralMut}{HTML}{FBEAE8}

    \lstset{
        basicstyle=\fontfamily{pcr}\fontsize{5.6}{6.9}\selectfont,
        columns=fullflexible,keepspaces=true,showstringspaces=false,
        mathescape=false,escapechar=|,numbers=none,
        keywordstyle=\color{tbInk},stringstyle=\color{tbInk},
        commentstyle=\color{tbMuted},identifierstyle=\color{tbInk},
        breaklines=true,breakatwhitespace=false,
        frame=none,aboveskip=0pt,belowskip=0pt
    }
    \newcommand{\lit}[1]{\textcolor{tbRed}{\bfseries #1}}
    \newsavebox{\litTestBox}
    \newsavebox{\litLogicBox}
    \newsavebox{\litLiteralBox}
    \newlength{\litMutHeight}
    \newlength{\litColumnWidth}
    \setlength{\litColumnWidth}{.47\textwidth}
    \newlength{\litFullWidth}
    \setlength{\litFullWidth}{\dimexpr .94\textwidth+13pt\relax}
    \newcommand{\litmatchheight}{%
        \setlength{\litMutHeight}{\dimexpr\ht\litLogicBox+\dp\litLogicBox\relax}%
        \ifdim\dimexpr\ht\litLiteralBox+\dp\litLiteralBox\relax>\litMutHeight
            \setlength{\litMutHeight}{\dimexpr\ht\litLiteralBox+\dp\litLiteralBox\relax}%
        \fi
    }
    \tikzset{
        example snippet/.style={inner xsep=7pt,inner ysep=4pt,outer sep=0pt,
            anchor=north west,line width=.6pt},
        example badge/.style={circle,fill=white,line width=.7pt,
            minimum size=.4cm,inner sep=0pt,outer sep=0pt,
            font=\fontsize{7}{8}\selectfont}
    }
    \newcommand{\litmutant}[3]{%
        \begin{tikzpicture}[baseline=(snippet.north)]
            \node[example snippet,inner xsep=5pt,draw=#3!75!black,fill=#3,rounded corners=2pt]
                (snippet) {\parbox[c][\litMutHeight][t]{\wd#2}{\usebox{#2}}};
            \node[anchor=east,inner xsep=4pt,inner ysep=2pt,text=black,align=center,
                draw=#3!75!black,fill=#3,sharp corners,line width=.6pt,
                font=\fontfamily{phv}\fontsize{7}{8}\mdseries\selectfont]
                at ([yshift=3pt]snippet.north east) {#1};
            \node[example badge,draw=#3!75!black,text=#3!75!black] at (snippet.north west) {$m$};
        \end{tikzpicture}%
    }
    \newcommand{\literalpanel}[1]{%
        \litmatchheight
        \begin{tikzpicture}[
            every node/.style={inner sep=3pt,outer sep=0pt},
            card/.style={rounded corners=5pt,line width=.8pt,anchor=north west},
            font=\fontfamily{phv}\footnotesize\selectfont,text=tbInk
        ]
            \node[card,draw=tbBlue,fill=tbBlue!8] (test) at (0,0) {%
                \begin{minipage}{\litFullWidth}
                    \begin{tikzpicture}
                        \node[example snippet,draw=tbBlue!35,fill=white,rounded corners=2pt]
                            (snippet) {\usebox{\litTestBox}};
                        \node[example badge,draw=tbBlue,text=tbBlue] at (snippet.north west) {$t$};
                    \end{tikzpicture}
                \end{minipage}
            };
            \node[card,draw=tbPurple,fill=tbPurple!8] (mutants)
                at ([yshift=-5pt]test.south west) {%
                \begin{minipage}{\litFullWidth}
                    \begin{minipage}[t]{\litColumnWidth}
                        \centering\litmutant{w/ logic instruction}{\litLogicBox}{tbLogicMut}
                    \end{minipage}\hfill%
                    \begin{minipage}[t]{\litColumnWidth}
                        \centering\litmutant{w/o logic instruction}{\litLiteralBox}{tbLiteralMut}
                    \end{minipage}
                \end{minipage}
            };
            \node[anchor=south west,inner sep=0pt,text=black,font=\fontfamily{phv}\footnotesize\bfseries\selectfont]
                at ([yshift=4pt]test.north west) {#1};
        \end{tikzpicture}%
    }

    \begin{lrbox}{\litTestBox}
    \begin{minipage}{.92\textwidth}
\begin{lstlisting}
def test_add_domains_multiple(temp_hosts_file):
    result = run("add", "domains", "multi", "|\lit{site1.test}|", "|\lit{site2.test}|", "|\lit{site3.test}|",
                 "--host-file", temp_hosts_file)
    ...
\end{lstlisting}
    \end{minipage}
    \end{lrbox}
    \begin{lrbox}{\litLogicBox}
    \begin{minipage}{.434\textwidth}
\begin{lstlisting}
$ ./exec add domains multi \
    |\lit{site1.test}| |\lit{site2.test}| |\lit{site3.test}|
> adds site1.test only  # mutant
$ ./exec add domains multi \
    a.test b.test c.test
> adds a.test only  # mutant
\end{lstlisting}
    \end{minipage}
    \end{lrbox}
    \begin{lrbox}{\litLiteralBox}
    \begin{minipage}{.434\textwidth}
\begin{lstlisting}
$ ./exec add domains multi \
    |\lit{site1.test}| |\lit{site2.test}| |\lit{site3.test}|
> exit 1, adds nothing  # mutant
$ ./exec add domains multi \
    a.test b.test c.test
> adds all three  # same as reference
\end{lstlisting}
    \end{minipage}
    \end{lrbox}
    \phantomsubcaption\label{fig:literal_copy_hostctl}
    \literalpanel{(a) \texttt{hostctl}: adding multiple domains}

    \vspace{4pt}
    \begin{lrbox}{\litTestBox}
    \begin{minipage}{.92\textwidth}
\begin{lstlisting}
def test_port_in_hostname_overrides_port_option(...):
    (test_dir / ".skeema").write_text("schema=testdb\n|\lit{host=localhost:3307}|\n|\lit{port=3308}|\nuser=root\n")
    result = run_skeema(["diff", "."], cwd=str(test_dir))
    ...
\end{lstlisting}
    \end{minipage}
    \end{lrbox}
    \begin{lrbox}{\litLogicBox}
    \begin{minipage}{.434\textwidth}
\begin{lstlisting}
# .skeema: |\lit{host=localhost:3307}| |\lit{port=3308}|
$ ./exec diff .
> uses port 3308  # mutant
# .skeema: host=localhost:4307 port=4308
$ ./exec diff .
> uses port 4308  # mutant
\end{lstlisting}
    \end{minipage}
    \end{lrbox}
    \begin{lrbox}{\litLiteralBox}
    \begin{minipage}{.434\textwidth}
\begin{lstlisting}
# .skeema: |\lit{host=localhost:3307}| |\lit{port=3308}|
$ ./exec diff .
> exit 0 before connecting  # mutant
# .skeema: host=localhost:4307 port=4308
$ ./exec diff .
> uses port 4307  # same as reference
\end{lstlisting}
    \end{minipage}
    \end{lrbox}
    \phantomsubcaption\label{fig:literal_copy_skeema}
    \literalpanel{(b) \texttt{Skeema}: port embedded in the host name}
    \endgroup
    \caption{
        \textbf{Mutants generated with and without the logic-level instruction.}
        Each panel shows the target reference test ($t$) and the behavior of the two mutants ($m$) generated for it, on the test's input and on the same input with its literals changed.
        In both panels, both mutants are killed by the target test and by no other reference test.
        Literals from the test that the mutant without the instruction hard-codes are highlighted in red; changing them makes this mutant behave like the reference.
    }
    \label{fig:literal_copy}
\end{figure*}

%% file: src/figure/error_analysis/full_error_analysis.tex
\begin{figure*}[p]
    \centering
    \begingroup
    \definecolor{tbInk}{HTML}{263747}
    \definecolor{tbMuted}{HTML}{637382}
    \definecolor{tbGold}{HTML}{B78A20}
    \definecolor{tbBlue}{HTML}{397D96}
    \definecolor{tbPurple}{HTML}{7756AC}
    \definecolor{tbGreen}{HTML}{43815B}
    \definecolor{tbRed}{HTML}{BA554F}
    \definecolor{tbOpusTest}{HTML}{E8F4EA}
    \definecolor{tbSolTest}{HTML}{FBEAE8}

    \lstset{
        basicstyle=\fontfamily{pcr}\fontsize{5.6}{6.9}\selectfont,
        columns=fullflexible,keepspaces=true,showstringspaces=false,
        mathescape=false,escapechar=,numbers=none,
        keywordstyle=\color{tbInk},stringstyle=\color{tbInk},
        commentstyle=\color{tbMuted},identifierstyle=\color{tbInk},
        breaklines=true,breakatwhitespace=false,
        frame=none,aboveskip=0pt,belowskip=0pt
    }
    \newsavebox{\errSpecBox}
    \newsavebox{\errMutantBox}
    \newsavebox{\errOpusTestBox}
    \newsavebox{\errSolTestBox}
    \newlength{\errTopHeight}
    \newlength{\errTestHeight}
    \newlength{\errColumnWidth}
    \setlength{\errColumnWidth}{.47\textwidth}
    \newlength{\errFullWidth}
    \setlength{\errFullWidth}{\dimexpr .94\textwidth+13pt\relax}
    \newcommand{\errmatchheight}[3]{%
        \setlength{#1}{\dimexpr\ht#2+\dp#2\relax}%
        \ifdim\dimexpr\ht#3+\dp#3\relax>#1
            \setlength{#1}{\dimexpr\ht#3+\dp#3\relax}%
        \fi
    }
    \newcommand{\topsnippet}[1]{\parbox[c][\errTopHeight][t]{\wd#1}{\usebox{#1}}}
    \newcommand{\testsnippet}[1]{\parbox[c][\errTestHeight][t]{\wd#1}{\usebox{#1}}}
    \tikzset{
        example snippet/.style={inner xsep=7pt,inner ysep=4pt,outer sep=0pt,
            anchor=north west,line width=.6pt},
        example badge/.style={circle,fill=white,line width=.7pt,
            minimum size=.4cm,inner sep=0pt,outer sep=0pt,
            font=\fontsize{7}{8}\selectfont}
    }
    \newcommand{\examplesnippet}[3]{%
        \begin{tikzpicture}
            \node[example snippet,draw=#1!35,fill=white,rounded corners=2pt]
                (snippet) {\topsnippet{#2}};
            \node[example badge,draw=#1,text=#1] at (snippet.north west) {$#3$};
        \end{tikzpicture}%
    }
    \newcommand{\exampletest}[4]{%
        \begin{tikzpicture}[baseline=(snippet.north)]
            \node[example snippet,inner xsep=5pt,draw=#4!75!black,fill=#4,rounded corners=2pt]
                (snippet) {\testsnippet{#2}};
            \node[anchor=east,inner xsep=4pt,inner ysep=2pt,text=black,align=center,
                draw=#4!75!black,fill=#4,sharp corners,line width=.6pt,
                font=\fontfamily{phv}\fontsize{7}{8}\mdseries\selectfont]
                at (snippet.north east) {#1};
            \node[example badge,draw=#4!75!black,text=#4!75!black] at (snippet.north west) {$t$};
            \node[anchor=south east,inner sep=2pt,opacity=#3] (bug)
                at (snippet.south east) {\includegraphics[height=.32cm]{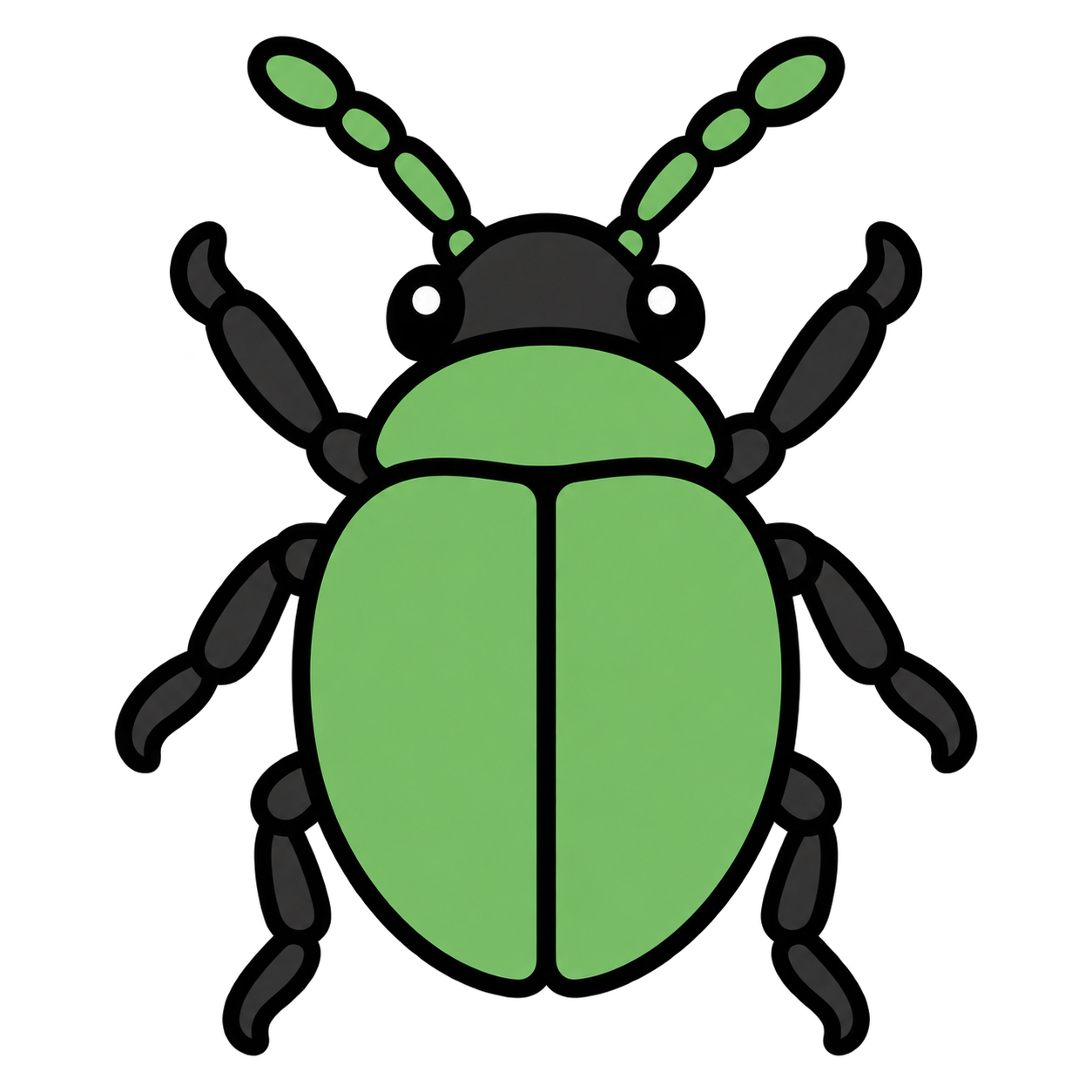}};
            \ifdim #3pt<1pt
                \draw[sharp corners,draw=tbRed!80!black,line width=.8pt,opacity=.55]
                    ([xshift=-.11cm,yshift=-.11cm]bug.center) -- ([xshift=.11cm,yshift=.11cm]bug.center)
                    ([xshift=-.11cm,yshift=.11cm]bug.center) -- ([xshift=.11cm,yshift=-.11cm]bug.center);
            \fi
        \end{tikzpicture}%
    }
    \newcommand{\errorpanel}[7]{%
        \errmatchheight{\errTopHeight}{\errSpecBox}{\errMutantBox}%
        \errmatchheight{\errTestHeight}{\errOpusTestBox}{\errSolTestBox}%
        \begin{tikzpicture}[
            every node/.style={inner sep=3pt,outer sep=0pt},
            card/.style={rounded corners=5pt,line width=.8pt,anchor=north west},
            font=\fontfamily{phv}\footnotesize\selectfont,text=tbInk
        ]
            \node[card,draw=tbGold,fill=tbGold!8] (spec) at (0,0) {%
                \begin{minipage}{\errColumnWidth}
                    \begin{tikzpicture}
                        \node[example snippet] (snippet) {\topsnippet{\errSpecBox}};
                        \draw[sharp corners,draw=tbGold!40,fill=white,line width=.6pt]
                            (snippet.south west) -- (snippet.south east)
                            -- ([yshift=-8pt]snippet.north east)
                            -- ([xshift=-8pt]snippet.north east)
                            -- (snippet.north west) -- cycle;
                        \draw[sharp corners,draw=tbGold!40,fill=tbGold!10,line width=.6pt]
                            ([xshift=-8pt]snippet.north east)
                            -- ([xshift=-8pt,yshift=-8pt]snippet.north east)
                            -- ([yshift=-8pt]snippet.north east) -- cycle;
                        \node[example snippet] at (snippet.north west) {\topsnippet{\errSpecBox}};
                        \node[example badge,draw=tbGold,text=tbGold] at (snippet.north west) {$s$};
                    \end{tikzpicture}
                \end{minipage}
            };
            \node[card,draw=tbPurple,fill=tbPurple!8] (mutant)
                at ([xshift=5pt]spec.north east) {%
                \begin{minipage}{\errColumnWidth}
                    \examplesnippet{tbPurple}{\errMutantBox}{m}
                \end{minipage}
            };
            \node[card,draw=tbBlue,fill=tbBlue!8] (tests)
                at ([yshift=-5pt]spec.south west) {%
                \begin{minipage}{\errFullWidth}
                    \begin{minipage}[t]{\errColumnWidth}
                        \centering\exampletest{#2}{\errOpusTestBox}{#4}{#6}
                    \end{minipage}\hfill%
                    \begin{minipage}[t]{\errColumnWidth}
                        \centering\exampletest{#3}{\errSolTestBox}{#5}{#7}
                    \end{minipage}
                \end{minipage}
            };
            \node[anchor=south west,inner sep=0pt,text=black,font=\fontfamily{phv}\footnotesize\bfseries\selectfont]
                at ([yshift=4pt]spec.north west) {#1};
        \end{tikzpicture}%
    }

    \begin{lrbox}{\errSpecBox}
    \begin{minipage}{.42\textwidth}
\begin{lstlisting}[breakindent=0pt,breakautoindent=false]
For `run` and `scan`, omitted paths mean
the current directory, and multiple file
and directory paths are accepted. [...]
\end{lstlisting}
    \end{minipage}
    \end{lrbox}
    \begin{lrbox}{\errMutantBox}
    \begin{minipage}{.42\textwidth}
\begin{lstlisting}
$ ./exec run -p 'hit()' -l js \
    --files-with-matches a.js b.js
> a.js  # mutant
> a.js b.js  # reference
\end{lstlisting}
    \end{minipage}
    \end{lrbox}
    \begin{lrbox}{\errOpusTestBox}
    \begin{minipage}{.434\textwidth}
\begin{lstlisting}[language=Python]
def test_run_accepts_multiple_paths(work):
    r = run_exe(
        ["run", "-p", "console.log($A)",
         "-l", "js", "--json=compact",
         "a.js", "b.js"], work)
    files = sorted(m["file"] for m in r.json())
    assert files == ["a.js", "b.js"], r
\end{lstlisting}
    \end{minipage}
    \end{lrbox}
    \begin{lrbox}{\errSolTestBox}
    \begin{minipage}{.434\textwidth}
\begin{lstlisting}[language=Python]
def test_files_with_matches(run, path):
    write(path / "a.js", "hit(); hit();")
    p = run(
        "run", "-p", "hit()", "-l", "js",
        "--files-with-matches", "a.js")
    assert p.stdout == "a.js\n"
\end{lstlisting}
    \end{minipage}
    \end{lrbox}
    \phantomsubcaption\label{fig:error_analysis_astgrep_1}
    \errorpanel{(a) \texttt{ast-grep}: multiple paths}{Opus}{Sol}{.30}{1}{tbOpusTest}{tbSolTest}

    \vspace{4pt}
    \begin{lrbox}{\errSpecBox}
    \begin{minipage}{.42\textwidth}
\begin{lstlisting}[breakindent=0pt,breakautoindent=false]
-B makes all rain characters bold.
Combining or repeating intensity options
must remain valid. [...]
\end{lstlisting}
    \end{minipage}
    \end{lrbox}
    \begin{lrbox}{\errMutantBox}
    \begin{minipage}{.42\textwidth}
\begin{lstlisting}
$ ./exec -B
> error: unsupported all-bold
  rendering mode; exit 1  # mutant
> (live animation)  # reference
\end{lstlisting}
    \end{minipage}
    \end{lrbox}
    \begin{lrbox}{\errOpusTestBox}
    \begin{minipage}{.434\textwidth}
\begin{lstlisting}[language=Python]
def test_all_bold_mode_emits_bold_attribute():
    with H.running_render(["-B"]) as p:
        ok = p.wait_for(lambda q:
            _has_bold(q.output()), 5)
        assert ok, ...
\end{lstlisting}
    \end{minipage}
    \end{lrbox}
    \begin{lrbox}{\errSolTestBox}
    \begin{minipage}{.434\textwidth}
\begin{lstlisting}[language=Python]
# async-bold-rainbow case
flags = ("-a", "-b", "-r")
first, second, errors, returncode = \
    live_capture(flags, first=0.14, second=0.10)
assert len(first + second) > 200
assert first != second
\end{lstlisting}
    \end{minipage}
    \end{lrbox}
    \phantomsubcaption\label{fig:error_analysis_cmatrix_all_bold}
    \errorpanel{(b) \texttt{cmatrix}: all-bold startup mode}{Opus}{Sol}{.30}{1}{tbOpusTest}{tbSolTest}

    \vspace{4pt}
    \begin{lrbox}{\errSpecBox}
    \begin{minipage}{.42\textwidth}
\begin{lstlisting}[breakindent=0pt,breakautoindent=false]
JavaScript is supported. [...]
Each AST node in each selected source is
considered, including nodes at arbitrary
nesting depth. [...]
\end{lstlisting}
    \end{minipage}
    \end{lrbox}
    \begin{lrbox}{\errMutantBox}
    \begin{minipage}{.42\textwidth}
\begin{lstlisting}
$ cat a.js
async function f() { await g(); }
$ ./exec run -p 'await $X' -l js \
    --json=compact
> []  # mutant
> [{"text":"await g()"}]  # reference
\end{lstlisting}
    \end{minipage}
    \end{lrbox}
    \begin{lrbox}{\errOpusTestBox}
    \begin{minipage}{.434\textwidth}
\begin{lstlisting}[language=Python]
def test_arbitrary_nesting_depth(work):
    write(work / "a.js", NESTED_JS)
    r = run_exe(
        ["run", "-p", "console.log($A)",
         "-l", "js", "--json=compact",
         "a.js"], work)
    assert len(r.json()) == 1, r

def test_python_await_syntax(work):
    write(work / "a.py", AWAIT_PY)
    r = run_exe(
        ["run", "-p", "await $X",
         "-l", "py", "--json=compact",
         "a.py"], work)
    assert r.json()[0]["text"] == "await g()"
\end{lstlisting}
    \end{minipage}
    \end{lrbox}
    \begin{lrbox}{\errSolTestBox}
    \begin{minipage}{.434\textwidth}
\begin{lstlisting}[language=Python]
def test_deep_tree_has_no_shallow_limit(run):
    source = "(" * 180 + "target" + ")" * 180
    p = run(
        "run", "-p", "target", "-l", "js",
        "--stdin", "--json=compact",
        stdin=source)
    assert compact_json(p)[0]["text"] \
        == "target"
\end{lstlisting}
    \end{minipage}
    \end{lrbox}
    \phantomsubcaption\label{fig:error_analysis_astgrep_await}
    \errorpanel{(c) \texttt{ast-grep}: JavaScript \texttt{await}}{Opus}{Sol}{1}{1}{tbSolTest}{tbSolTest}

    \vspace{4pt}
    \begin{lrbox}{\errSpecBox}
    \begin{minipage}{.42\textwidth}
\begin{lstlisting}[breakindent=0pt,breakautoindent=false]
completions SHELL emits a native
completion script for bash, elvish,
fish, powershell, or zsh. [...]
\end{lstlisting}
    \end{minipage}
    \end{lrbox}
    \begin{lrbox}{\errMutantBox}
    \begin{minipage}{.42\textwidth}
\begin{lstlisting}
$ ./exec completions elvish
> Zsh completion script  # mutant
> Elvish script  # reference
\end{lstlisting}
    \end{minipage}
    \end{lrbox}
    \begin{lrbox}{\errOpusTestBox}
    \begin{minipage}{.434\textwidth}
\begin{lstlisting}[language=Python]
SHELLS = ["bash", "elvish", "fish",
          "powershell", "zsh"]
@pytest.mark.parametrize("shell", SHELLS)
def test_completions_for_shell(work, shell):
    r = run_exe(["completions", shell], work)
    assert r.code == 0
    assert len(r.out) > 100
\end{lstlisting}
    \end{minipage}
    \end{lrbox}
    \begin{lrbox}{\errSolTestBox}
    \begin{minipage}{.434\textwidth}
\begin{lstlisting}[language=Python]
# shell inference failure case
p = run("completions",
    env={"SHELL": "/bin/not-a-shell"})
assert p.returncode == 10
assert p.stdout == ""
\end{lstlisting}
    \end{minipage}
    \end{lrbox}
    \phantomsubcaption\label{fig:error_analysis_astgrep_elvish}
    \errorpanel{(d) \texttt{ast-grep}: Elvish completions}{Opus}{Sol}{1}{1}{tbSolTest}{tbSolTest}
    \endgroup
    \caption{
        \textbf{Error analysis on \texttt{ast-grep} and \texttt{cmatrix}.}
        Each panel shows the specification ($s$), the mutant ($m$), and the relevant Opus and Sol tests ($t$).
        A crossed-out mutant icon marks a test that kills the mutant.
        Opus kills the mutants in (a) and (b); both suites miss those in (c) and (d).
    }
    \label{fig:error_analysis_full}
\end{figure*}

%% file: src/appendix/construction_detail.tex
\section{Benchmark Construction Detail}
\label{appsec:construction_detail}

\subsection{Details on Filtering}
\label{appsec:test_filtering}
As briefly described in \cref{subsec:initial_test_spec}, we filter out some functions from the \programbench reference suite based on the following criteria:
\begin{enumerate}
    \item Excessive Resources. We limit the execution resources for a test environment run to 2 CPU cores, \SI{60}{GB} RAM and \SI{1}{\hour}. We evaluate reference executable $E$ against $T$ and filter all tests that fail under these constraints.
    \item Internet Access. Reliance on internet access tends to make tests flaky, we therefore exclude repositories whose reference test suites require internet access. To detect test cases that require internet, we execute the reference test suite $T$ against $E$ without network access.
    \item Flaky tests. Whenever a test fails on $I$ in any step of the pipeline, we exclude it.
\end{enumerate}

\subsection{Optimal Selection of Mutant Targets}
\label{appsubsec:minflow}

Here, we detail the process of selecting the mutation target (\cref{subsec:mutant_construction}). Each mutant targets a pair $(s,t)$, where $s\in S$ is a specification
section and $t\in T$ is a reference-test function that evaluates behavior
described by $s$. We add an edge $(s,t)$ whenever this relationship holds.

We augment the resulting bipartite graph with a source $\alpha$ and a sink
$\omega$. The edge capacities are
\[
u(\alpha,s)=6,\qquad u(s,t)=1,\qquad u(t,\omega)=1.
\]
Thus, each specification section receives at most six target tests, while
each test can be assigned to at most one mutant. Every unit-carrying path
$\alpha\to s\to t\to\omega$ represents one selected target pair.

To choose among maximum-cardinality assignments, we assign cost
\[
w(s,t)=d(t),
\]
where $d(t)$ is the number of specification sections linked to $t$.
We then solve a min-cost max-flow problem.
Consequently, the solver favors section-specific tests over tests shared by many sections and uses the test identifier to break remaining ties.

\subsection{Findings in Manual Quality Check}
\label{appsubsec:construction_manual_check}

After the spec iteration (\cref{subsec:iterative_spec}), we found 20 instances where the spec contained a specific location of an error (\eg \texttt{stdout should include `panicked at src/main.rs:24:40`}).
All mutants could be trivially killed by a single test function as the error location differs in each mutant.
In such a case, we replace it with an obfuscated \texttt{<error location>} and remove the test function from the reference suite if there are functions that check such particular property.
Further, we found 14 instances where the spec mentions behaviors that are related to tests flagged to be skipped (\eg requiring internet). We drop these behaviors from the spec in such cases.

%% file: src/appendix/eval_detail.tex
\section{Experimental Setup}
\label{appsec:setup}
\paragraph{Models and Agent Harnesses}
We consider GPT-5.6 Sol \citep{gpt56}, GPT-5.6 Luna \citep{gpt56} GPT-6 Astra \citep{gpt6}, Claude Opus 5 \citep{opus5}, Gemini-3.8 Flash \citep{gemini38}, DeepSeek V4 Flash \citep{deepseek}.
Unless otherwise indicated, we use default reasoning effort and vendor-recommended agent harnesses

We use default reasoning effort for our main experiment.
We use vendor-default harnesses (Codex, Claude Code, Gemini CLI, DeepSeek Harness) unless noted otherwise.
Each model runs inside a container with 2 CPUs and 256 GB RAM, with a 6-hour limit per instance (each for code and test generation).

\paragraph{Solution Generation}
For each task, we use the same container for code generation and test generation.
The container is preinstalled with libraries used by the reference $I, T$, and is the same across instances except for a few instances that have unique libraries which are \eg incompatible with the global library version setup.
The internet is disabled and the only accessible point is the API provider's server endpoint (\eg \texttt{openai.com}).

For code generation, following \programbench \citep{yang2026programbench}, we ask the agent to generate \texttt{build.sh} that writes \texttt{executable}.
The test suite then checks behaviors of the executable instead of the code, allowing unified grading of language-agnostic implementation by the agent.

The test suite generation task asks agents to write a suite that generates its results in JUnit XML format, which is supported by various libraries in various languages, including pytest.
While this design conceptually allows cross-language test generation, we in practice find that models struggle to generate the XML themselves. We therefore gave a hint that \texttt{pytest --junitxml} can generate the result in the expected format; all observed submissions then used pytest.

\paragraph{Evaluation}
We use the same container for evaluation as the solution generation container.
The generated code is evaluated against the reference suite, and the generated test is evaluated against reference code and mutants.
Each code-test pair is evaluated in a separate container (each with 2 CPUs, 60 GB RAM), with internet being disabled.

As we described in \cref{subsec:initial_test_spec}, our primary focus is  \textbf{Mutation Score ($\mathcal{M}$)} (fraction of mutants killed by the generated suite) of the test generation setup.
Along with this, we calculate \textbf{Validity Rate ($\mathcal{V}$)} (fraction of generated test suite that passed reference implementation) for test generation, and \textbf{Reference Pass Rate ($\mathcal{R}$)} (fraction of reference test suite that passed the implementation) for code generation.

To avoid the trivial test generation described in \cref{subsec:initial_test_spec}, we run the test suite on reference $I$ three times, and only use valid tests when calculating $\mathcal{M}$.
This evaluation design is explicitly mentioned in the prompt, saying that the test should be "break them down into smaller, focused tests rather than writing one huge test function" so that gradable part will not be massively skipped for $\mathcal{M}$ calculation because only part of it fails.

%% file: src/appendix/prompt.tex
\section{Prompt}
\label{appsec:prompt}

\subsection{Construction Steps}

\input{src/prompt/spec_init.tex}
\input{src/prompt/mutant_init.tex}
\input{src/prompt/mutant_polish.tex}

\input{src/prompt/classify_acc.tex}
\input{src/prompt/classify_precision.tex}
\input{src/prompt/classify_recall.tex}

\subsection{Evaluation}
Below are the prompts used for evaluating the models.
These are also used for the \textbf{Mock Run} step during the spec improvement iteration (\cref{subsec:iterative_spec}).

\input{src/prompt/codegen.tex}
\input{src/prompt/testgen.tex}

%% file: src/prompt/spec_init.tex
\begin{promptbox}[Specification-initialization prompt]
You are given two directories and must clarify a task specification.

\vspace{1mm}
\textbf{Workspace.}
\begin{itemize}
    \item \texttt{specification/}: contains the current task specification. A program and its compiled executable will be created based solely on this specification.
    \item \texttt{test-branches/}: contains the test suite and, when provided, \texttt{skipped\_tests.json}. Treat every listed function as absent: it is excluded from evaluation and must not influence the specification.
\end{itemize}

\vspace{1mm}
\textbf{Task.}

Update the contents of \texttt{specification/} so that it provides enough information for someone to implement an executable that passes all tests in \texttt{test-branches/}, without relying on any information outside the specification.

\vspace{1mm}
\textbf{Workflow.}
\begin{enumerate}
    \item Prepare a new file \texttt{spec.md}.
    \item For each tested behavior, create a section titled \texttt{\#\# Spec \{i\}}, and thoroughly explain the required behavior to the point where an executable can be created based on the document that can pass the test.
\end{enumerate}

\vspace{1mm}
\textbf{Rules.}
\begin{itemize}
    \item Make sure each section is titled \texttt{\#\# Spec \{i\}}.
    \item The spec may have at least 10 sections, and at most 50 sections. Effectively aggregate similar requirements into a single section.
    \item The spec should be an abstracted description of the expected behaviors, i.e., what will be tested. Do not simply enumerate over all test functions.
    \item The required artifact that will be made based on the spec is exactly \texttt{executable}. If \texttt{eval/run.sh} of the test builds another command from repository source, that command has to be harness-provided. Do not assign its behavior to executable, and write the spec as if they are simply provided and available.
    \item Write a self-contained specification that does not rely on information outside \texttt{specification/}.
\end{itemize}

\vspace{1mm}
\textbf{Output.}

Leave the clarified specification in \texttt{specification/spec.md}.
\end{promptbox}

%% file: src/prompt/mutant_init.tex
\begin{promptbox}[Mutant-suite initialization prompt]
You are creating a mutant suite.

\vspace{1mm}
\textbf{Workspace.}

\begin{itemize}
    \item \texttt{specification/}: contains the task specification. We aim to perform mutation testing to detect executables that violate the specification.
    \item \texttt{repo/}: contains the ground-truth implementation and \texttt{build.sh}, which creates an executable binary.
    \item \texttt{reference-tests/}: contains the immutable active reference-test sources used to construct the plan.
    \item \texttt{mutant\_plan.json}: the authoritative list of \texttt{(target\_spec, target\_test)} assignments.
    \item \texttt{mutants.json}: a manifest pre-seeded with exactly the planned mutant IDs and targets.
\end{itemize}

\vspace{1mm}
\textbf{Task.}

Create exactly the mutants in \texttt{mutant\_plan.json}. Each mutant must violate its assigned specification rule in the narrowest way that is expected to make its assigned test fail while every other active reference-test function continues to pass.

\vspace{1mm}
\textbf{Workflow.}

\begin{enumerate}
    \item Read every assignment in \texttt{mutant\_plan.json} and inspect its exact \texttt{target\_spec} and \texttt{target\_test} source.
    \item For every assignment, make exactly one copy of \texttt{repo/} whose directory name equals \texttt{mutant\_id}.
    \item Modify source code or build inputs so the assigned specification rule is violated specifically at the behavior asserted by \texttt{target\_test}. Leave a \texttt{[MUTATED]} comment identifying that scope.
    \item Keep each copied repository \texttt{build.sh} compatible with the original repository build contract.
    \item After creating all mutants and updating \texttt{mutants.json}, stop. The script will compile the repositories after your turn and will resume this same session with a build report if repairs are needed.
\end{enumerate}

\vspace{1mm}
\textbf{Rules.}

\begin{itemize}
    \item The plan is immutable: do not add, remove, rename, merge, or retarget assignments.
    \item Create every planned repository exactly once and create no unplanned \texttt{repo\_*} repositories.
    \item A valid mutant is intended to have the singleton kill set \texttt{\{target\_test\}}. Avoid changes that affect other reference tests, unrelated crashes, global timeouts, or broad output corruption.
    \item Do not detect the test environment, test name, harness process, or fixture inputs merely to force the assigned test to fail.
    \item Mutate the logic that produces the behavior: do not be based on literal argument, filename, or payload strings copied from \texttt{target\_test}, and do not post-process a correct result back into a wrong one. A mutant must be a rule a re-implementer could plausibly get wrong across all inputs in its class.
    \item Include every generated \texttt{repo\_*} repository exactly once in \texttt{/workspace/mutants.json}.
    \item Avoid exact full error-message assertions unless they are important; exit codes and functional effects are more important.
    \item Do not create wrapper/proxy executables that delegate to \texttt{/workspace/executable}, \texttt{./executable}, or another reference executable. Mutants must be self-contained outputs of their repository build, because evaluation installs each candidate as \texttt{/workspace/executable}.
    \item Do not replace \texttt{build.sh} with a script whose main purpose is to write a shell wrapper instead of building the mutated implementation.
    \item Do not compile or run the mutants yourself: do not invoke \texttt{build.sh}, \texttt{make}, or a parallel-build helper. The script owns compilation so resource limits and logs remain deterministic.
\end{itemize}

\vspace{1mm}
\textbf{Output.}

\begin{itemize}
    \item Leave every generated mutant in a numbered \texttt{repo\_*} repository.
    \item After defining all repositories, edit \texttt{/workspace/mutants.json} so its top-level \texttt{mutants} array contains one entry per generated mutant.
\end{itemize}

Here are the entries you have to fill; the rest are pre-seeded:

\begin{itemize}
    \item \texttt{mutation}: required. The behavior summary and the mutation details -- which specification rule is violated, and how the source change violates it.
\end{itemize}

Leave every other field -- \texttt{id}, \texttt{assignment\_id}, \texttt{candidate\_index}, \texttt{target\_spec}, \texttt{target\_test}, and \texttt{target\_test\_description} -- exactly as seeded. Do not add, remove, or reorder entries.
\end{promptbox}

%% file: src/prompt/mutant_polish.tex
\begin{promptbox}[Mutant-polishing prompt]
You are repairing ProgramBench mutants for \placeholder{task} after fixed-reference-suite mutation evaluation (attempt \placeholder{attempt}).

\vspace{1mm}
\textbf{Workspace.}

\begin{itemize}
    \item \texttt{specification/}: the current task specification.
    \item \texttt{reference-tests/}: the immutable extracted test sources.
    \item \texttt{skipped\_tests.json}: functions excluded from the fixed reference suite; do not use them as repair evidence.
    \item \texttt{repair\_required.json}: the repair evidence.
    \item \texttt{mutants/}: only the affected repositories: \placeholder{list of ids}.
\end{itemize}

\vspace{1mm}
\textbf{Task.}

Try your best to make every listed mutant fail its assigned \texttt{target\_test} while minimizing failures of other active tests and retaining its initial scope. An exact singleton kill set is desirable, but it is not required.

\vspace{1mm}
\textbf{Rules.}

\begin{itemize}
    \item Modify only repositories already present under \texttt{mutants/}.
    \item Keep every mutant ID and its original \texttt{target\_spec} and \texttt{target\_test} unchanged. Do not add, remove, rename, or merge mutants.
    \item Preserve the behavioral topic of each original mutation. Strengthen or simplify how that same behavior is violated; do not switch to a different spec behavior merely because another test is easy to fail.
    \item For every edited mutant, the assigned \texttt{target\_test} must fail. Aim for every other active reference-test function to pass.
    \item Prefer the smallest source-level change that creates a focused, deterministic violation in the original scope.
    \item Inspect only relevant reference tests. Never modify tests, the specification, or \texttt{repair\_required.json}; host-side copyback excludes them.
    \item Rebuild each edited repository with its existing \texttt{build.sh} and leave a self-contained readable \texttt{executable}.
    \item Do not create wrapper/proxy executables, delegate to a reference executable, detect the test environment, key behavior on test names, or introduce unrelated crashes/timeouts.
    \item If the original change is invalid, replace it with a valid focused violation of the same \texttt{target\_spec}.
    \item Some immutable tests may exercise indistinguishable behavior. If further focus would require detecting the test harness or violating these rules, leave that mutant completely unchanged and explain why in \texttt{repair\_outcome.json}.
\end{itemize}

\vspace{1mm}
\textbf{Output.}

\begin{itemize}
    \item Leave each repaired mutant in its existing repository under \texttt{mutants/}.
    \item Map each mutant ID to the strengthened violation in its retained scope.
    \item Write \texttt{repair\_outcome.json} with two objects: \texttt{changed\_mutants}, mapping each edited mutant ID to a concise summary, and \texttt{kept\_as\_is}, mapping each unedited mutant ID to the reason further focus is impractical. Every listed mutant ID must occur in exactly one object. This file is authoritative.
\end{itemize}
\end{promptbox}

%% file: src/prompt/classify_acc.tex
\begin{promptbox}[Accuracy-classification prompt]
You are attributing candidate-implementation failures for a task.

\vspace{1mm}
\textbf{Workspace.}

\begin{itemize}
    \item \texttt{/workspace/spec/}: the current specification for the task.
    \item \texttt{/workspace/failures.json}: tests of the fixed reference suite that the candidate implementation failed, one record each. Every record is a candidate failure: the same suite passes on the reference implementation.
    \item \texttt{/workspace/reference-tests/}: the fixed reference suite itself, read-only: test sources, conftest fixtures and golden files. Read the failing test there when the traceback is truncated or the expectation lives in a fixture. Traceback paths are the eval container's: a file quoted under \texttt{/workspace/eval/} is staged here at \texttt{/workspace/reference-tests/<branch>/eval/}.
    \item \texttt{/workspace/executable}: the reference implementation. Run it to check what the real behavior is.
\end{itemize}

Each record carries the test \texttt{name}, \texttt{branch}, \texttt{status}, the assertion \texttt{message}, the pytest \texttt{traceback}, and \texttt{relevant\_specs} (the specification sections the test was linked to, when known). Any output quoted inside a traceback is output of the candidate implementation, which the test invoked as \texttt{/workspace/executable}. The fixed suite passes on the reference implementation, so what a test asserts is the real behavior, and a record always shows the candidate deviating from it.

\vspace{1mm}
\textbf{Task.}

There are \placeholder{failure\_count} failures in \texttt{failures.json}, all from test branch \placeholder{branch}. For each failure, determine whether the specification or candidate executable is at fault.

\vspace{1mm}
\textbf{Classification criteria.}

\begin{itemize}
    \item \texttt{executable}: the specification legitimately defines the behavior and the candidate implementation is wrong. Recommended action: \texttt{ignore}.
    \item \texttt{spec\_underspecification}: the specification is missing or ambiguous with respect to the failed behavior. Recommended action: \texttt{polish\_spec}.
    \item \texttt{spec\_contradiction}: the specification is misleading or contradictory with respect to the failed behavior. Recommended action: \texttt{polish\_spec}.
    \item \texttt{error}: the failure is an artifact of the environment rather than a judgement about the candidate. Recommended action: \texttt{ignore}.
    \item \texttt{unknown}: the failure record and specification do not provide enough information. Recommended action: \texttt{manual\_review}.
\end{itemize}

Choose \texttt{spec\_underspecification} or \texttt{spec\_contradiction} only when the failure is clearly due to the specification. \texttt{spec\_contradiction} requires a conflict you can show, not one you infer: you must be able to quote the specification sentence that states something different from what the test asserts. If the specification is merely silent, vague, or less precise than the test expectation, that is \texttt{spec\_underspecification}. A test expectation the specification simply fails to mention is never a contradiction. As long as the most natural interpretation should lead to the tested behavior, it is not the spec's fault, but it is the executable's fault, even if the expected behavior is not stated explicitly.

\vspace{1mm}
\textbf{Rules.}

\begin{itemize}
    \item Decide from the record, the specification, and what \texttt{/workspace/executable} does when you run it. If they leave unclear what the test expected, use \texttt{unknown} rather than guessing.
    \item Judge every record from its own traceback. Tests in one file routinely fail for different reasons, so never carry a verdict from one record to the next. For each record copy the failing assert line into \texttt{failing\_assertion} and the candidate output that made it fail into \texttt{candidate\_observed}, both verbatim from that record's traceback. A verdict whose copied evidence is not found in its own record is rejected and the record is re-judged.
    \item In a pytest diff, \texttt{-} lines are the assert's right operand and \texttt{+} lines its left one. A \texttt{CompletedProcess(...)} repr is the candidate's output. These rules can be different if the test is written in other frameworks. Read the whole failing assertion, not only its first clause, and check which flag or argument the test actually passed.
    \item Before emitting \texttt{spec\_contradiction}, reconstruct the failing invocation from the traceback, run it on \texttt{/workspace/executable}, and state in the rationale the quoted specification claim, the test expectation, and the reference output that confirms the test. Copy the sentence from \texttt{/workspace/spec/} as it appears. If no such sentence can be quoted, or the invocation cannot be run here, then the fault is not a contradiction: use \texttt{spec\_underspecification} when the specification is merely silent or vague, and \texttt{executable} otherwise. Put that sentence in \texttt{spec\_quote}; a contradiction whose \texttt{spec\_quote} is not a literal substring of the specification is rejected and the verdict is discarded, so never paraphrase, trim or reflow it.
    \item The executable was evaluated with networking disabled; a failure whose proximate cause is the missing network is `error`, never evidence about the specification or the executable.
    \item Preserve each failure's \texttt{name} and \texttt{branch} exactly so the result can be matched to its input.
\end{itemize}

\vspace{1mm}
\textbf{Output.}

Write \texttt{classifier\_result.json} containing exactly one JSON object with this shape:
\begin{minipage}{\linewidth}
\ttfamily\scriptsize\raggedright
\{\\
\ \ "failures": [\\
\ \ \ \ \{\\
\ \ \ \ \ \ "name": "same failure name from failures.json",\\
\ \ \ \ \ \ "branch": "same branch from failures.json",\\
\ \ \ \ \ \ "fault": "executable|spec\_underspecification|spec\_contradiction|error|unknown",\\
\ \ \ \ \ \ "subreason": "short\_snake\_case\_reason providing detail beyond fault",\\
\ \ \ \ \ \ "confidence": 0.0,\\
\ \ \ \ \ \ "rationale": "a few sentences that explicitly name every relevant section as Spec N and cover (i) which specification number provides relevant information, (ii) what the test evaluates in that regard, (iii) why the executable failed, and (iv) why it is classified as executable/spec\_underspecification/spec\_contradiction/error/unknown",\\
\ \ \ \ \ \ "failing\_assertion": "the failing assert line, copied verbatim from this record's traceback",\\
\ \ \ \ \ \ "candidate\_observed": "the candidate output or status that made it fail, copied verbatim from this record's traceback",\\
\ \ \ \ \ \ "spec\_quote": "for spec\_contradiction only: the conflicting sentence copied verbatim from /workspace/spec/, otherwise empty",\\
\ \ \ \ \ \ "recommended\_action": "ignore|polish\_spec|manual\_review"\\
\ \ \ \ \}\\
\ \ \ ]\\
\}
\end{minipage}

\vspace{1mm}
\textbf{Checkpointing.}

Persist classifications incrementally to \texttt{classifier\_result.json} in the workspace. After each compact group, atomically rewrite that valid JSON object with every completed entry so far. If time or output budget runs out, leave as many completed entries as possible.

\vspace{1mm}
\textbf{Final response.}

Do not return failures inline. When finished, return only:
\texttt{\{"output\_path": "classifier\_result.json"\}}
Do not add markdown or other text. The checkpoint file is authoritative.
\end{promptbox}

%% file: src/prompt/classify_precision.tex
\begin{promptbox}[Precision-classification prompt]
You are classifying generated tests that failed on the ground-truth executable.

\vspace{1mm}
\textbf{Workspace.}

\begin{itemize}
    \item \texttt{/workspace/spec/}: the current specification used to generate the test suite.
    \item \texttt{/workspace/test/}: the Codex-generated test suite produced from \texttt{/workspace/spec/}.
    \item \texttt{/workspace/generated\_test\_failures.json}: generated tests that failed on the ground-truth executable.
    \item \texttt{/workspace/executable}: the ground-truth executable itself. Run it to see what the real behavior is.
\end{itemize}

\vspace{1mm}
\textbf{Task.}

There are \placeholder{failure\_count} generated-test failures in \texttt{generated\_test\_failures.json}. For each generated-test failure, decide whose fault it is.

\vspace{1mm}
\textbf{Classification criteria.}

\begin{itemize}
    \item \texttt{test}: the generated test exercised the executable but asserted something the specification does not entail: it misread a sentence, demanded an exact incidental value, or depends on timing. Recommended action: ignore for spec/mutant polishing.
    \item \texttt{spec\_underspecification}: the specification is missing or ambiguous about the behavior, causing the generated test to make an unsupported assumption. Recommended action: polish the specification.
    \item \texttt{spec\_contradiction}: the specification is misleading or contradicts the reference behavior, causing the generated test to encode the wrong behavior. Recommended action: polish the specification.
    \item \texttt{error}: the failure says nothing about the specification or the behavior: the test never reached the executable, its own fixture crashed or was written wrongly (an unexecutable fake tool, a missing input file), or the sandbox interfered. The executable was evaluated with networking disabled; a failure whose proximate cause is the missing network is \texttt{error}, never evidence about the specification or the executable. A package install that could not reach an index also belongs here. Recommended action: ignore for spec/mutant polishing.
    \item \texttt{unknown}: there is not enough information to decide.
\end{itemize}

Choose \texttt{spec\_underspecification} or \texttt{spec\_contradiction} only when the failure is clearly due to the specification. If the most natural interpretation should not result in this error, it is the test's fault, even if the expected behavior is not stated explicitly. Never classify an environment failure as a specification fault.

\vspace{1mm}
\textbf{Rules.}

\begin{itemize}
    \item Explore only the relevant specification and generated tests. Do not do broad exploration.
    \item Run \texttt{/workspace/executable} on the failing invocation before deciding. A failure that never reached the executable is \texttt{error}; one whose expectation the executable in fact satisfies is \texttt{test}.
    \item Before emitting \texttt{spec\_contradiction}, copy the misleading specification sentence into \texttt{spec\_quote} exactly as it appears in \texttt{/workspace/spec/}, and give the observed reference output that conflicts with it in the rationale. A \texttt{spec\_quote} that is not a literal substring of the specification is rejected and the verdict is discarded, so never paraphrase, trim or reflow it.
    \item Before emitting \texttt{test}, copy into \texttt{spec\_quote} the specification sentence that shows the test's expectation is not entailed: the sentence the test misread, or the one that states the behavior the reference actually has. The same literal-substring check applies. If no sentence addresses the behavior the test assumed, the specification is silent about it: that is \texttt{spec\_underspecification}, not \texttt{test}, unless the test's own record shows it never exercised the executable.
    \item Each record carries \texttt{timing\_sensitive} (the test sleeps, polls or waits with a fixed deadline) and \texttt{flaky\_siblings\_in\_file} (tests in the same file that passed and failed across reference trials). When both are nonzero and the specification describes the behavior as asynchronous, consolidated or best-effort, the failure is a race in the test, not a contradiction: classify it as \texttt{test} unless the reference fails the invocation deterministically when you replay it.
    \item You may repair a broken fixture (for example a fake tool with a malformed shebang) to replay the test and learn what the reference does. When your verdict rests on such a replay, set \texttt{replayed\_with\_fix} to true and describe the repair in \texttt{fix\_description}; the recorded failure itself is then also an \texttt{error}, and the spec finding is what the polish step consumes.
    \item Prefer a specification fault only when clarifying the specification would plausibly prevent this kind of generated test failure. Use \texttt{spec\_underspecification} for missing or ambiguous guidance and \texttt{spec\_contradiction} for misleading or contradictory guidance. Prefer \texttt{test} when the test asserts behavior not entailed by the spec, exact incidental values, timing-sensitive behavior, or behavior contradicted by the ground truth.
\end{itemize}

\vspace{1mm}
\textbf{Output.}

Return exactly one JSON object with this shape:\\
\begin{minipage}{\linewidth}
\ttfamily\scriptsize\raggedright
\{\\
"test\_failures": [\\
\ \ \ \ \{\\
\ \ \ \ \ \ "name": "same name from generated\_test\_failures.json",\\
\ \ \ \ \ \ "fault": "spec\_underspecification|spec\_contradiction|test|error|unknown",\\
\ \ \ \ \ \ "subreason": "short\_snake\_case\_reason",\\
\ \ \ \ \ \ "confidence": 0.0,\\
\ \ \ \ \ \ "rationale": "a few sentences that explicitly name every relevant section as Spec N and cover (i) which specification number provides relevant information, (ii) what the test evaluates in that regard, (iii) why the executable fails, and (iv) why it is classified as spec\_underspecification/spec\_contradiction/test/error/unknown",\\
\ \ \ \ \ \ "spec\_quote": "for spec\_contradiction and test: the sentence copied verbatim from /workspace/spec/, otherwise empty",\\
\ \ \ \ \ \ "replayed\_with\_fix": false,\\
\ \ \ \ \ \ "fix\_description": "when replayed\_with\_fix is true: what you repaired to replay the test, otherwise empty",\\
\ \ \ \ \ \ "recommended\_action": "polish\_spec|ignore|manual\_review"\\
\ \ \ \ \}\\
\ \ \ ]\\
\}
\end{minipage}

Return one entry for every case in \texttt{generated\_test\_failures.json}, in the same order.

\vspace{1mm}
\textbf{Checkpointing.}

Write the object to \texttt{/workspace/classifier\_result.json}. Persist incrementally: after each compact group, atomically rewrite that valid JSON object with every completed entry so far. If time or output budget runs out, leave as many completed entries as possible. Every entry needs its own \texttt{subreason} and rationale; never reuse one verdict across a group.

\vspace{1mm}
\textbf{Final response.}

Do not return the entries inline. When finished, return only:
\texttt{\{"output\_path": "classifier\_result.json"\}}
Do not add markdown or other text. The checkpoint file is authoritative.
\end{promptbox}

%% file: src/prompt/classify_recall.tex
\begin{promptbox}[Recall-classification prompt]
You are classifying ProgramBench specification clarity and generated-test adequacy using survived immutable mutants.

\vspace{1mm}
\textbf{Workspace.}

\begin{itemize}
    \item \texttt{/workspace/spec/}: the current specification for the task.
    \item \texttt{/workspace/test/}: an agent-generated test suite produced from \texttt{/workspace/spec/}.
    \item \texttt{/workspace/mutants/}: the relevant mutant repositories or executables.
    \item \texttt{/workspace/mutant\_cases.json}: survived mutants to classify.
    \item \texttt{/workspace/executable}: the ground-truth executable. Run it to see what the real behavior is.
\end{itemize}

\vspace{1mm}
\textbf{Task.}

There are \placeholder{mutant\_count} mutant cases in \texttt{mutant\_cases.json}. For each mutant case, decide whether the specification needs clarification.

\vspace{1mm}
\textbf{Evidence fields.}

\begin{itemize}
    \item \texttt{n\_generated\_tests\_observed}: all generated tests observed in the evaluation matrix.
    \item \texttt{n\_generated\_tests\_passing\_ground\_truth}: generated tests accepted as valid because they pass on the ground-truth executable.
    \item \texttt{n\_generated\_tests\_passing\_both\_ground\_truth\_and\_mutant}: valid generated tests that also pass on this mutant.
    \item \texttt{generated\_tests\_passing\_ground\_truth\_but\_failing\_mutant}: valid generated tests that kill this mutant.
    \item \texttt{generated\_tests\_passing\_ground\_truth}: the valid generated-test baseline; this is not a list of tests that pass on the mutant.
\end{itemize}

\vspace{1mm}
\textbf{Classification criteria.}

\begin{itemize}
    \item \texttt{spec\_underspecification}: a survived mutant exposes behavior that the specification leaves missing or ambiguous, so future generated suites need clearer instructions. Recommended action: \texttt{polish\_spec}.
    \item \texttt{spec\_contradiction}: a survived mutant exposes behavior for which the specification is misleading or contradicts the reference behavior. Recommended action: \texttt{polish\_spec}.
    \item \texttt{test}: a survived mutant violates behavior already stated clearly in the specification, so this generated suite is weak or wrong. Recommended action: \texttt{ignore}.
    \item \texttt{unknown}: there is not enough information to decide. Recommended action: \texttt{manual\_review}.
\end{itemize}

Choose \texttt{spec\_underspecification} or \texttt{spec\_contradiction} only when the failure is clearly due to the specification. If the most natural interpretation should not result in this error, it is the test's fault, even if the expected behavior is not stated explicitly.

\vspace{1mm}
\textbf{Rules.}

\begin{itemize}
    \item Explore only the relevant specification, generated tests, and immutable mutant files. Do not do broad exploration.
    \item Never recommend changing, rebuilding, removing, strengthening, or narrowing a mutant.
\end{itemize}

\vspace{1mm}
\textbf{Output.}

Return exactly one JSON object with this shape:\\
\begin{minipage}{\linewidth}
\ttfamily\scriptsize\raggedright
\{\\
\ \ "mutants": [\\
\ \ \ \ \{\\
\ \ \ \ \ \ "mutant\_id": "same mutant\_id from mutant\_cases.json",\\
\ \ \ \ \ \ "fault": "test|spec\_underspecification|spec\_contradiction|unknown",\\
\ \ \ \ \ \ "subreason": "short\_snake\_case\_reason",\\
\ \ \ \ \ \ "confidence": 0.0,\\
\ \ \ \ \ \ "rationale": "a few sentences that explicitly name the mutant's target\_spec as Spec N, name any other relevant sections as Spec N, and cover (i) which specification number provides relevant information, (ii) what the test evaluates in that regard, (iii) what the scope of the mutant is, (iv) why the mutant survived, and (v) why it is classified as test|spec\_underspecification|spec\_contradiction|unknown",\\
\ \ \ \ \ \ "recommended\_action": "ignore|polish\_spec|manual\_review"\\
\ \ \ \ \}\\
\ \ \ ]\\
\}
\end{minipage}

Return one entry for every case in \texttt{mutant\_cases.json}, in the same order.

\vspace{1mm}
\textbf{Checkpointing.}

Write the object to \texttt{/workspace/classifier\_result.json}. Persist incrementally: after each compact group, atomically rewrite that valid JSON object with every completed entry so far. If time or output budget runs out, leave as many completed entries as possible. Every entry needs its own \texttt{subreason} and rationale; never reuse one verdict across a group.

\vspace{1mm}
\textbf{Final response.}

Do not return the entries inline. When finished, return only:
\texttt{\{"output\_path": "classifier\_result.json"\}}
Do not add markdown or other text. The checkpoint file is authoritative.
\end{promptbox}

%% file: src/prompt/codegen.tex
\begin{promptbox}[Code-generation prompt]
You are reimplementing a program from its specification.

\vspace{1mm}
\textbf{Workspace.}

The workspace contains the program specification.

\vspace{1mm}
\textbf{Task.}

Reimplement a new, original codebase from scratch that satisfies the specification and produces the required executable.

\vspace{1mm}
\textbf{Rules.}

\begin{itemize}
    \item You may read bundled documentation and the specification.
    \item Do not search the internet, clone repositories, or download the original project from package registries.
    \item Do not run package-manager updates or installs, curl/wget, networked git commands, or any other command that fetches task dependencies.
    \item Use only tools and dependencies already present when the task starts. Installing something in the live container does not make it available on a fresh grading checkout.
\end{itemize}

\vspace{1mm}
\textbf{Output.}

\begin{itemize}
    \item Create \texttt{./compile.sh}.
    \item \texttt{./compile.sh} must be executable and must build \texttt{./executable} in the workspace root on a fresh checkout.
    \item Only \texttt{./executable} is retained when tests run; the rest of the workspace is removed. It must be self-contained and must not depend on sibling files.
\end{itemize}
\end{promptbox}

%% file: src/prompt/testgen.tex
\begin{promptbox}[Test-generation prompt]
You are creating a test suite.

\vspace{1mm}
\textbf{Workspace.}

The workspace contains a program specification.

\vspace{1mm}
\textbf{Task.}

Create a comprehensive test suite under \texttt{/workspace/eval}.

\vspace{1mm}
\textbf{Evaluation design.}

\begin{itemize}
    \item The test suite will go through mutation testing. It will be evaluated against fresh workspaces containing the program under test at \texttt{/workspace/executable}.
    \item The specification is available only while authoring the suite.
    \item Tests must invoke \texttt{/workspace/executable}, or compute it robustly relative to \texttt{eval}.
    \item Do not hard-code \texttt{./executable} inside tests that run after \texttt{run.sh} has changed into \texttt{/workspace/eval}.
\end{itemize}

\vspace{1mm}
\textbf{Rules.}

\begin{itemize}
    \item Cover documented functional behavior thoroughly. The test suite must kill any mutants that go against the spec, while it has to pass any implementation as long as it follows the spec.
    \item Pay particular attention to explicitly numbered or sectioned requirements in the specification. Cover each documented behavior with at least one meaningful test.
    \item When creating tests, please break them down into smaller, focused tests rather than writing one huge test function; tests that kill the ground truth are not used when calculating the mutation score.
\end{itemize}

\vspace{1mm}
\textbf{Output.}

\begin{itemize}
    \item \texttt{/workspace/eval/run.sh} must exist and be executable.
    \item \texttt{/workspace/eval/run.sh} must always write valid JUnit XML to \texttt{/workspace/eval/results.xml}.
    \item \texttt{/workspace/eval/run.sh} must run the whole generated suite, even after individual test failures.
    \item The report must have a \texttt{<testsuites>} root containing \texttt{<testsuite>} elements (or a single \texttt{<testsuite>} root) and one named \texttt{<testcase>} per test. Represent non-passing outcomes with a nested \texttt{<failure>}, \texttt{<error>}, or \texttt{<skipped>} element.
    \item For pytest suites, this is achieved simply by running \texttt{pytest --junitxml=results.xml} from \texttt{/workspace/eval}.
    \item JUnit XML must escape XML metacharacters and must not contain raw control characters or binary stdout/stderr; encode such output with \texttt{repr}, base64, or another XML-safe textual representation.
    \item When performing parametrized testing, always provide short, descriptive \texttt{ids=} values. Never allow large strings, bytes, or object representations to become test IDs.
    \item Put all helper files, fixtures, scripts, and generated test assets under \texttt{/workspace/eval}.
\end{itemize}
\end{promptbox}

%% file: src/appendix/task_list.tex
\section{List of Instances}
\label{appsec:task_list}
We list all \benchmark instances in \cref{tab:task_list}.

\input{src/table/task_list.tex}

%% file: src/table/task_list.tex
{\scriptsize
\setlength{\tabcolsep}{3pt}
\begin{longtable}{@{}lrrr@{\hspace{12pt}}lrrr@{}}
\caption{All 165 \benchmark instances with the number of specification sections $|S|$, mutants $|M|$, and reference test functions $|T|$.}
\label{tab:task_list} \\
\toprule
Repository & $|S|$ & $|M|$ & $|T|$ & Repository & $|S|$ & $|M|$ & $|T|$ \\
\midrule
\endfirsthead
\toprule
Repository & $|S|$ & $|M|$ & $|T|$ & Repository & $|S|$ & $|M|$ & $|T|$ \\
\midrule
\endhead
\bottomrule
\endfoot
\texttt{abishekvashok/cmatrix} & 16 & 48 & 502 & \texttt{agourlay/zip-password-finder} & 20 & 60 & 679 \\
\texttt{ajeetdsouza/zoxide} & 13 & 39 & 531 & \texttt{alecthomas/chroma} & 20 & 57 & 186 \\
\texttt{alexpovel/srgn} & 33 & 99 & 1,807 & \texttt{altdesktop/i3-style} & 20 & 59 & 518 \\
\texttt{anordal/shellharden} & 15 & 45 & 1,085 & \texttt{antonmedv/fx} & 32 & 91 & 2,026 \\
\texttt{antonmedv/walk} & 26 & 70 & 412 & \texttt{ariga/atlas} & 31 & 92 & 1,302 \\
\texttt{arq5x/bedtools2} & 44 & 131 & 1,053 & \texttt{arthursonzogni/json-tui} & 21 & 63 & 754 \\
\texttt{ast-grep/ast-grep} & 23 & 69 & 863 & \texttt{astro/deadnix} & 20 & 60 & 600 \\
\texttt{bellard/quickjs} & 48 & 142 & 3,029 & \texttt{bensadeh/tailspin} & 21 & 63 & 610 \\
\texttt{blacknon/hwatch} & 23 & 68 & 791 & \texttt{blake3-team/blake3} & 16 & 47 & 645 \\
\texttt{bootandy/dust} & 24 & 72 & 521 & \texttt{boyter/scc} & 40 & 116 & 462 \\
\texttt{brocode/fblog} & 22 & 66 & 975 & \texttt{burntsushi/ripgrep} & 40 & 117 & 1,991 \\
\texttt{burntsushi/xsv} & 27 & 78 & 1,048 & \texttt{byron/dua-cli} & 24 & 72 & 675 \\
\texttt{canop/broot} & 14 & 41 & 530 & \texttt{canop/rhit} & 24 & 72 & 817 \\
\texttt{cheat/cheat} & 25 & 73 & 294 & \texttt{chirlu/sox} & 40 & 105 & 1,097 \\
\texttt{chmln/handlr} & 18 & 50 & 698 & \texttt{chmln/sd} & 22 & 65 & 810 \\
\texttt{clog-tool/clog-cli} & 20 & 60 & 569 & \texttt{cmatsuoka/figlet} & 27 & 81 & 872 \\
\texttt{codesnap-rs/codesnap} & 22 & 66 & 563 & \texttt{cordx56/rustowl} & 20 & 57 & 554 \\
\texttt{crowdagger/crowbook} & 42 & 125 & 790 & \texttt{cslarsen/jp2a} & 20 & 59 & 624 \\
\texttt{cweill/gotests} & 22 & 64 & 596 & \texttt{dalance/amber} & 25 & 74 & 563 \\
\texttt{dandavison/delta} & 40 & 117 & 932 & \texttt{direnv/direnv} & 23 & 67 & 812 \\
\texttt{doxygen/doxygen} & 16 & 41 & 85 & \texttt{drew-alleman/datasurgeon} & 21 & 62 & 499 \\
\texttt{ducaale/xh} & 18 & 52 & 1,149 & \texttt{duckdb/duckdb} & 13 & 39 & 2,825 \\
\texttt{dundee/gdu} & 23 & 67 & 1,140 & \texttt{ecumene/rust-sloth} & 18 & 52 & 374 \\
\texttt{ekzhang/bore} & 12 & 27 & 377 & \texttt{eliukblau/pixterm} & 20 & 55 & 423 \\
\texttt{elkowar/pipr} & 13 & 38 & 399 & \texttt{eradman/entr} & 18 & 54 & 571 \\
\texttt{esubaalew/run} & 38 & 106 & 1,194 & \texttt{eudoxia0/hashcards} & 20 & 60 & 1,014 \\
\texttt{facebook/zstd} & 37 & 111 & 2,023 & \texttt{facebookresearch/fasttext} & 26 & 74 & 308 \\
\texttt{filosottile/age} & 27 & 76 & 673 & \texttt{foriequal0/git-trim} & 24 & 67 & 479 \\
\texttt{gabotechs/dep-tree} & 35 & 105 & 785 & \texttt{git-bahn/git-graph} & 13 & 39 & 467 \\
\texttt{go-critic/go-critic} & 24 & 68 & 312 & \texttt{gromacs/gromacs} & 50 & 148 & 1,189 \\
\texttt{guumaster/hostctl} & 21 & 61 & 954 & \texttt{hairyhenderson/gomplate} & 36 & 104 & 2,873 \\
\texttt{halitechallenge/halite} & 18 & 51 & 224 & \texttt{hooklift/gowsdl} & 24 & 69 & 371 \\
\texttt{htop-dev/htop} & 20 & 56 & 686 & \texttt{hush-shell/hush} & 33 & 98 & 1,200 \\
\texttt{incu6us/goimports-reviser} & 24 & 71 & 505 & \texttt{ip7z/7zip} & 40 & 120 & 1,043 \\
\texttt{ismaelgv/rnr} & 25 & 72 & 675 & \texttt{isona/dirble} & 30 & 86 & 685 \\
\texttt{ivanceras/svgbob} & 26 & 75 & 470 & \texttt{jarun/nnn} & 29 & 51 & 477 \\
\texttt{jesseduffield/lazygit} & 42 & 112 & 796 & \texttt{jhspetersson/fselect} & 25 & 75 & 2,971 \\
\texttt{johanneskaufmann/html-to-markdown} & 27 & 81 & 855 & \texttt{jonas/tig} & 30 & 82 & 1,480 \\
\texttt{jrnxf/thokr} & 20 & 42 & 322 & \texttt{junegunn/fzf} & 32 & 96 & 1,856 \\
\texttt{kaushiksrini/parqeye} & 26 & 64 & 470 & \texttt{kisielk/errcheck} & 20 & 54 & 296 \\
\texttt{konradsz/igrep} & 20 & 58 & 354 & \texttt{ksxgithub/parallel-disk-usage} & 26 & 77 & 467 \\
\texttt{kyoheiu/felix} & 24 & 71 & 497 & \texttt{lfos/calcurse} & 22 & 64 & 640 \\
\texttt{lh3/seqtk} & 23 & 67 & 429 & \texttt{lua/lua} & 35 & 105 & 1,337 \\
\texttt{luajit/luajit} & 28 & 84 & 2,945 & \texttt{lymphatus/caesium-clt} & 25 & 75 & 561 \\
\texttt{madler/pigz} & 25 & 75 & 830 & \texttt{mfridman/tparse} & 23 & 68 & 413 \\
\texttt{mgdm/htmlq} & 21 & 63 & 1,453 & \texttt{mgechev/revive} & 50 & 105 & 450 \\
\texttt{mibk/dupl} & 18 & 54 & 373 & \texttt{mikefarah/yq} & 36 & 108 & 1,997 \\
\texttt{miserlou/loop} & 22 & 66 & 696 & \texttt{mookid/diffr} & 19 & 57 & 606 \\
\texttt{multiprocessio/dsq} & 30 & 89 & 542 & \texttt{nachoparker/dutree} & 18 & 53 & 497 \\
\texttt{naggie/dstask} & 16 & 48 & 1,270 & \texttt{nikolassv/bartib} & 20 & 60 & 716 \\
\texttt{noborus/ov} & 40 & 114 & 1,804 & \texttt{noborus/trdsql} & 36 & 108 & 1,306 \\
\texttt{nukesor/pueue} & 24 & 61 & 414 & \texttt{nuta/nsh} & 34 & 102 & 1,881 \\
\texttt{o2sh/onefetch} & 30 & 87 & 1,142 & \texttt{oppiliappan/eva} & 18 & 54 & 913 \\
\texttt{orf/gping} & 22 & 62 & 310 & \texttt{parcel-bundler/lightningcss} & 39 & 115 & 2,827 \\
\texttt{peco/peco} & 27 & 58 & 564 & \texttt{pemistahl/grex} & 24 & 71 & 1,311 \\
\texttt{pier-cli/pier} & 10 & 30 & 692 & \texttt{pls-rs/pls} & 20 & 60 & 316 \\
\texttt{psampaz/go-mod-outdated} & 15 & 45 & 284 & \texttt{quinn-rs/quinn} & 20 & 59 & 478 \\
\texttt{raviqqe/muffet} & 20 & 56 & 291 & \texttt{rbakbashev/elfcat} & 24 & 72 & 560 \\
\texttt{rcoh/angle-grinder} & 18 & 54 & 1,038 & \texttt{rhysd/kiro-editor} & 25 & 74 & 529 \\
\texttt{riquito/tuc} & 23 & 69 & 1,195 & \texttt{robertdavidgraham/masscan} & 39 & 116 & 2,533 \\
\texttt{rochacbruno/marmite} & 44 & 127 & 663 & \texttt{rs/curlie} & 20 & 55 & 687 \\
\texttt{rs/jplot} & 21 & 57 & 579 & \texttt{rust-ethereum/ethabi} & 13 & 39 & 997 \\
\texttt{rust-lang/mdbook} & 40 & 120 & 1,097 & \texttt{rvben/rumdl} & 40 & 119 & 3,263 \\
\texttt{sayanarijit/xplr} & 22 & 65 & 417 & \texttt{sclevine/yj} & 24 & 72 & 764 \\
\texttt{segmentio/chamber} & 25 & 72 & 1,691 & \texttt{sharkdp/fd} & 31 & 90 & 1,235 \\
\texttt{sharkdp/hexyl} & 19 & 57 & 906 & \texttt{sharkdp/hyperfine} & 22 & 66 & 289 \\
\texttt{sharkdp/pastel} & 30 & 87 & 1,114 & \texttt{shashwatah/jot} & 22 & 65 & 744 \\
\texttt{sibprogrammer/xq} & 28 & 83 & 772 & \texttt{sigoden/argc} & 36 & 108 & 995 \\
\texttt{simeg/eureka} & 18 & 54 & 344 & \texttt{sirwart/ripsecrets} & 18 & 54 & 610 \\
\texttt{sitkevij/hex} & 18 & 53 & 823 & \texttt{skeema/skeema} & 30 & 88 & 1,673 \\
\texttt{sqlite/sqlite} & 45 & 135 & 13,298 & \texttt{sstadick/hck} & 20 & 59 & 852 \\
\texttt{stacked-git/stgit} & 34 & 101 & 1,341 & \texttt{stathissideris/ditaa} & 25 & 75 & 608 \\
\texttt{stranger6667/jsonschema} & 27 & 81 & 1,860 & \texttt{svenstaro/genact} & 33 & 87 & 227 \\
\texttt{svenstaro/miniserve} & 21 & 60 & 304 & \texttt{thezoraiz/ascii-image-converter} & 20 & 60 & 462 \\
\texttt{tomarrell/wrapcheck} & 20 & 59 & 417 & \texttt{tomnomnom/gron} & 18 & 54 & 223 \\
\texttt{trasta298/keifu} & 32 & 81 & 255 & \texttt{tree-sitter/tree-sitter} & 37 & 110 & 1,203 \\
\texttt{typst/typst} & 35 & 105 & 1,622 & \texttt{unhappychoice/gittype} & 25 & 75 & 685 \\
\texttt{universal-ctags/ctags} & 15 & 42 & 544 & \texttt{wfxr/code-minimap} & 20 & 58 & 311 \\
\texttt{wfxr/csview} & 21 & 63 & 335 & \texttt{wgunderwood/tex-fmt} & 19 & 57 & 455 \\
\texttt{wintermute-cell/ngrrram} & 18 & 47 & 302 & \texttt{xampprocky/tokei} & 22 & 64 & 726 \\
\texttt{y2z/monolith} & 29 & 87 & 709 & \texttt{yaa110/nomino} & 22 & 66 & 312 \\
\texttt{yassinebridi/serpl} & 22 & 65 & 416 & \texttt{yoav-lavi/melody} & 35 & 104 & 1,205 \\
\texttt{ys-l/flamelens} & 12 & 29 & 224 & & & & \\
\end{longtable}
}

%% file: main.bbl
\begin{thebibliography}{23}
\providecommand{\natexlab}[1]{#1}
\providecommand{\url}[1]{\texttt{#1}}
\expandafter\ifx\csname urlstyle\endcsname\relax
  \providecommand{\doi}[1]{doi: #1}\else
  \providecommand{\doi}{doi: \begingroup \urlstyle{rm}\Url}\fi

\bibitem[{Anthropic}(2026)]{opus5}
{Anthropic}.
\newblock {Claude Opus 5 System Card}.
\newblock \url{https://www.anthropic.com/system-cards}, July 2026.

\bibitem[Carlini(2026)]{carliniBuildingCCompiler2026}
Nicholas Carlini.
\newblock Building a c compiler with a team of parallel claudes, 2 2026.
\newblock URL \url{https://www.anthropic.com/engineering/building-c-compiler}.

\bibitem[{DeepSeek-AI} et~al.(2026){DeepSeek-AI}, Xu, Lin, Xue, Wang, Xu, Wu, Zhang, Lin, Dong, Ling, et~al.]{deepseek}
{DeepSeek-AI}, Anyi Xu, Bangcai Lin, Bing Xue, Bingxuan Wang, Bingzheng Xu, Bochao Wu, Bowei Zhang, Chaofan Lin, Chen Dong, Chenchen Ling, et~al.
\newblock Deepseek-v4: Towards highly efficient million-token context intelligence.
\newblock \emph{arXiv preprint arXiv:2606.19348}, 2026.

\bibitem[Deng et~al.(2025)Deng, Da, Pan, He, Ide, Garg, Lauffer, Park, Pasari, Rane, Sampath, Krishnan, Kundurthy, Hendryx, Wang, Bharadwaj, Holm, Aluri, Zhang, Jacobson, Liu, and Kenstler]{swebenchpro}
Xiang Deng, Jeff Da, Edwin Pan, Yannis~Yiming He, Charles Ide, Kanak Garg, Niklas Lauffer, Andrew Park, Nitin Pasari, Chetan Rane, Karmini Sampath, Maya Krishnan, Srivatsa Kundurthy, Sean Hendryx, Zifan Wang, Vijay Bharadwaj, Jeff Holm, Raja Aluri, Chen Bo~Calvin Zhang, Noah Jacobson, Bing Liu, and Brad Kenstler.
\newblock Swe-bench pro: Can ai agents solve long-horizon software engineering tasks?, 2025.
\newblock URL \url{https://arxiv.org/abs/2509.16941}.

\bibitem[Ding et~al.(2025)Ding, Long, Pu, Zhou, Gao, Gao, He, Hou, Hu, Li, et~al.]{ding2025nl2repo}
Jingzhe Ding, Shengda Long, Changxin Pu, Huan Zhou, Hongwan Gao, Xiang Gao, Chao He, Yue Hou, Fei Hu, Zhaojian Li, et~al.
\newblock Nl2repo-bench: Towards long-horizon repository generation evaluation of coding agents.
\newblock \emph{arXiv preprint arXiv:2512.12730}, 2025.

\bibitem[{Google DeepMind}(2026)]{gemini38}
{Google DeepMind}.
\newblock {Gemini 3.8 Flash Model Card}.
\newblock \url{https://deepmind.google/models/model-cards/gemini-3-8-flash/}, September 2026.

\bibitem[Huang et~al.(2026)Huang, Reyna, Lerner, Xia, and Hempel]{huang2026professionalsoftwaredevelopersdont}
Ruanqianqian Huang, Avery Reyna, Sorin Lerner, Haijun Xia, and Brian Hempel.
\newblock Professional software developers don't vibe, they control: Ai agent use for coding in 2025, 2026.
\newblock URL \url{https://arxiv.org/abs/2512.14012}.

\bibitem[Jain et~al.(2025)Jain, Synnaeve, and Rozi{\`e}re]{jain2025testgeneval}
Kush Jain, Gabriel Synnaeve, and Baptiste Rozi{\`e}re.
\newblock Testgeneval: A real world unit test generation and test completion benchmark.
\newblock In \emph{International Conference on Learning Representations}, volume 2025, pp.\  14947--14999, 2025.

\bibitem[Jia \& Harman(2011)Jia and Harman]{jia2010analysis}
Yue Jia and Mark Harman.
\newblock An analysis and survey of the development of mutation testing.
\newblock \emph{IEEE transactions on software engineering}, 37\penalty0 (5):\penalty0 649--678, 2011.

\bibitem[Jimenez et~al.(2024)Jimenez, Yang, Wettig, Yao, Pei, Press, and Narasimhan]{jimenez2024swebench}
Carlos~E Jimenez, John Yang, Alexander Wettig, Shunyu Yao, Kexin Pei, Ofir Press, and Karthik Narasimhan.
\newblock Swe-bench: Can language models resolve real-world github issues?
\newblock In \emph{International Conference on Learning Representations}, volume 2024, pp.\  54107--54157, 2024.

\bibitem[Kogler et~al.(2026)Kogler, Hangler, Ehrhart, Dornauer, Wuersching, and Schrammel]{kogler2026restestbench}
Leon Kogler, Stefan Hangler, Maximilian Ehrhart, Benedikt Dornauer, Roland Wuersching, and Peter Schrammel.
\newblock Restestbench: A benchmark for evaluating the effectiveness of llm-generated rest api test cases from nl requirements.
\newblock \emph{arXiv preprint arXiv:2604.25862}, 2026.

\bibitem[Lin(2026)]{linScalingLongRunningAutonomousCoding2026}
Wilson Lin.
\newblock Scaling long-running autonomous coding, January 2026.
\newblock URL \url{https://cursor.com/blog/scaling-agents}.

\bibitem[Lu et~al.(2025)Lu, Yang, Ren, Hou, Xiao, Wang, Shi, Zhou, Zhan, and Li]{lu2025webgenbenchevaluatingllmsgenerating}
Zimu Lu, Yunqiao Yang, Houxing Ren, Haotian Hou, Han Xiao, Ke~Wang, Weikang Shi, Aojun Zhou, Mingjie Zhan, and Hongsheng Li.
\newblock Webgen-bench: Evaluating llms on generating interactive and functional websites from scratch, 2025.
\newblock URL \url{https://arxiv.org/abs/2505.03733}.

\bibitem[M{\"u}ndler et~al.(2024)M{\"u}ndler, Mueller, He, and Vechev]{mundler2024swtbench}
Niels M{\"u}ndler, Mark~Niklas Mueller, Jingxuan He, and Martin Vechev.
\newblock {SWT}-bench: Testing and validating real-world bug-fixes with code agents.
\newblock In \emph{The Thirty-eighth Annual Conference on Neural Information Processing Systems}, 2024.
\newblock URL \url{https://openreview.net/forum?id=9Y8zUO11EQ}.

\bibitem[Mustafa et~al.(2021)Mustafa, Wan-Kadir, Ibrahim, Shah, Younas, Khan, Zareei, and Alanazi]{mustafa2021automated}
Ahmad Mustafa, Wan~MN Wan-Kadir, Noraini Ibrahim, Muhammad~Arif Shah, Muhammad Younas, Atif Khan, Mahdi Zareei, and Faisal Alanazi.
\newblock Automated test case generation from requirements: A systematic literature review.
\newblock \emph{Computers, Materials and Continua}, 67\penalty0 (2):\penalty0 1819--1833, 2021.

\bibitem[Mürtz \& Müller(2025)Mürtz and Müller]{aitw}
Christian Mürtz and Mark~Niklas Müller.
\newblock Agents in the wild - dashboard.
\newblock \url{https://insights.logicstar.ai}, 2025.
\newblock URL \url{https://doi.org/10.5281/zenodo.15846865}.
\newblock Interactive web dashboard. Code available at \url{https://github.com/logic-star-ai/insights}.

\bibitem[{OpenAI}(2026{\natexlab{a}})]{gpt56}
{OpenAI}.
\newblock {GPT-5.6 System Card}.
\newblock \url{https://deploymentsafety.openai.com/gpt-5-6}, July 2026{\natexlab{a}}.

\bibitem[{OpenAI}(2026{\natexlab{b}})]{gpt6}
{OpenAI}.
\newblock {GPT-6 Astra System Card}.
\newblock \url{https://deploymentsafety.openai.com/gpt-6-astra}, September 2026{\natexlab{b}}.

\bibitem[Sun et~al.(2026)Sun, Zhao, Wang, Du, Ma, Wang, Zhang, Zhang, and Huang]{sun2026swemutation}
Yuxuan Sun, Yuze Zhao, Yufeng Wang, Yao Du, Zhiyuan Ma, Jinbo Wang, Mengdi Zhang, Kai Zhang, and Zhenya Huang.
\newblock Swe-mutation: Can llms generate reliable test suites in software engineering?
\newblock In \emph{Findings of the Association for Computational Linguistics: ACL 2026}, pp.\  39651--39674, 2026.

\bibitem[Vero et~al.(2025)Vero, Mündler, Chibotaru, Raychev, Baader, Jovanović, He, and Vechev]{baxbench}
Mark Vero, Niels Mündler, Victor Chibotaru, Veselin Raychev, Maximilian Baader, Nikola Jovanović, Jingxuan He, and Martin Vechev.
\newblock Baxbench: Can llms generate correct and secure backends?, 2025.
\newblock URL \url{https://arxiv.org/abs/2502.11844}.

\bibitem[Wang et~al.(2026)Wang, Xia, Zhao, Du, Miao, Deng, Yu, and Xing]{wang2026multifiletest}
Yibo Wang, Congying Xia, Wenting Zhao, Jiangshu Du, Chunyu Miao, Zhongfen Deng, Philip~S Yu, and Chen Xing.
\newblock Multifiletest: A multi-file-level llm unit test generation benchmark and impact of error fixing mechanisms.
\newblock In \emph{Findings of the Association for Computational Linguistics: ACL 2026}, pp.\  28151--28172, 2026.

\bibitem[Yang et~al.(2024)Yang, Jimenez, Wettig, Lieret, Yao, Narasimhan, and Press]{sweagent}
John Yang, Carlos~E. Jimenez, Alexander Wettig, Kilian Lieret, Shunyu Yao, Karthik Narasimhan, and Ofir Press.
\newblock Swe-agent: Agent-computer interfaces enable automated software engineering, 2024.
\newblock URL \url{https://arxiv.org/abs/2405.15793}.

\bibitem[Yang et~al.(2026)Yang, Lieret, Ma, Thakkar, Pedchenko, Sootla, McMilin, Yin, Hou, Synnaeve, et~al.]{yang2026programbench}
John Yang, Kilian Lieret, Jeffrey Ma, Parth Thakkar, Dmitrii Pedchenko, Sten Sootla, Emily McMilin, Pengcheng Yin, Rui Hou, Gabriel Synnaeve, et~al.
\newblock Programbench: Can language models rebuild programs from scratch?
\newblock \emph{arXiv preprint arXiv:2605.03546}, 2026.

\end{thebibliography}
